\documentclass{pasj00}
\draft
\usepackage{lscape}
\usepackage{subfigure}
\usepackage{multirow}
\usepackage{url}

\begin{document}
\SetRunningHead{M.~Mizumoto \etal}{A VDPC Model of 1H0707--495}

\title{Interpretation of the X-ray Spectral Variation of 1H0707--495 with a Variable Double Partial Covering Model}

\author{Misaki \textsc{Mizumoto}, \altaffilmark{1,2}
Ken \textsc{Ebisawa}, \altaffilmark{1,2}
and
Hiroaki \textsc{Sameshima}, \altaffilmark{1}
 }
\altaffiltext{1}{Institute of Space and Astronautical Science, Japan Aerospace Exploration Agency,
3-1-1 Yoshinodai, Chuo-ku, Sagamihara, Kanagawa 252-5210}
\email{mizumoto@astro.isas.jaxa.jp}
\altaffiltext{2}{Department of Astronomy, Graduate School of Science, The University of Tokyo, 7-3-1 Hongo, Bunkyo-ku, Tokyo 113-0033}


\maketitle

\begin{abstract}
The Narrow-line Seyfert 1 galaxy 1H0707--495 is known to exhibit significant X-ray spectral variations. 
Its X-ray energy spectrum is characterized by a strong soft excess emission, an extremely deep iron K-edge structure at $\sim7$ keV,
and a putative iron L-line/edge feature at $\sim1$ keV.
We have found that 
the energy spectrum of 1H0707--495 in 0.5--10 keV is successfully explained by a ``variable double partial covering model''
where the original continuum spectrum, 
which is composed of the soft multi-color disk blackbody component and the hard power-law component, 
is partially covered by two ionized absorption layers with different ionization states and the same partial covering fraction. 
The lower-ionized and thicker absorption layer primarily explains the iron K-edge feature, and
the higher-ionized and thinner absorption layer explains the L-edge feature.
We have discovered that 
the observed significant intensity/spectral variation within a $\sim$ day is mostly explained 
by only variation of the partial covering fraction.
In our model, the intrinsic luminosity and spectral shape are hardly variable within a $\sim$ day,
while some intrinsic variability above 3 keV is recognized.
This is consistent with the picture that the multi-color disk blackbody spectrum is almost invariable in this timescale,
and the hard power-law component is more variable.
We propose that the observed spectral variation of 1H0707--495 is caused by three physically independent variations with different timescales;
(1) intrinsic luminosity variation over days,
(2) variation of partial covering fraction at a timescale of hours, and
(3) small intrinsic hard component variation above 3 keV in a timescale of hours or less.
\end{abstract}


\section{Introduction}

Narrow-line Seyfert 1 galaxies (hereafter, NLS1s) are known to have narrower Balmer lines than usual Seyfert 1 galaxies \citep{ost85}, strong FeII emission \citep{hal87}, steep X-ray photon indices \citep{puc92}, and intense X-ray flux/spectral variation \citep{ulr97}.
The narrow line widths imply shallow gravitational potential around the central black holes, which is due to small black hole masses.
Strong FeII emission, steep X-ray photon indices, and intense X-ray time variation 
imply high accretion rates, which mean the high Eddington ratios \citep{sul00,bor02}.
The small black hole masses and high gas accretion rates suggest that 
the black holes are rapidly evolving young systems, where the elapsed time from the onset of the gas inflow is rather short \citep{mat00}.
Therefore, NLS1s are considered to be in an early phase of the AGN evolution \citep{gru04}.

Among NLS1s, 1H0707--495 is characterized by a sharp spectral drop at $\sim7$ keV, a strong soft excess emission, and a hint of iron L-edge feature \citep{bol02}.
Several different models have been proposed to explain its unique X-ray energy spectrum.
On one hand, some explain these features in terms of the ``relativistic disk-line model,'' 
where relativistically blurred inner-disk reflection around an extreme Kerr black hole is 
supposed to take place and be responsible for these spectral features (e.g.,~\cite{fab09}).
In this model, the spectral variation is interpreted to be mostly caused by changes in the geometry 
in the very vicinity of the black hole, such as the height of the continuum source above the black hole \citep{fab02,fab04}.
On the other hand, a partial covering model may also explain these spectral features.
\citet{gal04} attempted to explain the long-term spectral changes with the partial covering by a neutral single-layer absorber with iron-overabundance ($\sim3\times$solar).
\citet{tan04} used a {\it double-layer}\/ neutral partial covering model, 
and suggested that the partial covering clouds are produced presumably due to either disk instabilities or radiation-driven mass outflow.
In these models, the observed X-ray spectral variation is, at least to some extent, explained by change of the partial covering fraction.

Meanwhile, \citet{miy12} has shown that 
a characteristic broad iron-line feature and spectral variation of the Seyfert 1 galaxy MCG-6-30-15, which is often classed as an NLS1 \citep{mch05}, are successfully explained by a variable partial covering model.
In this model, intrinsic X-ray luminosity and spectral shape of the central X-ray source are not significantly variable; 
the observed flux and spectral variations are explained mostly by variation of the partial covering fraction 
by intervening absorbers, 
which are composed of two different ionization layers.
Here, we apply the variable partial covering model 
for the XMM-Newton and Suzaku observations of 1H0707--495, 
to see if this model is also valid in explaining spectral variations of another NLS1 
that is known to exhibit extreme spectral characteristics.
We examine in this paper whether spectral variation is explained 
even if we assume no intrinsic variation. 

We first describe the data we use in Section 2.
Next, in Section 3, we show that the spectral model used for MCG-6-30-15 by \citet{miy12} is
considered to be a ``Variable Double Partial Covering (VDPC) model'', 
and that the VDPC model can explain the X-ray spectral variation of 1H0707--495.
Section 4 gives further details on interpretation of the X-ray spectral variation.
Finally, we show our conclusions in Section 5.

\section{Observation and Data Reduction}

The XMM-Newton satellite \citep{jan01} observed 1H0707--495 fifteen times, and
the Suzaku satellite \citep{mit07} did once.
These observation IDs, start dates, and exposure times are listed in Table \ref{dataset}.

In the XMM-Newton data analysis, we used the European Photon Imaging Camera (EPIC)-pn data \citep{str01} in 0.5--10.0 keV
and the reflection grating spectrometer (RGS: \cite{den01}) in 0.5--1.5 keV.
For XMM-Newton data reduction, we used the XMM-Newton Software Analysis System ({\tt SAS}, v.12.0.1) and the latest calibration files as of March 2013.
The EPIC-pn spectra and light-curves were extracted with {\tt PATTERN$<=$12}\footnote{We also tried the selection {\tt PATTERN<=4} instead of {\tt PATTERN<=12},
but the results were almost the same within statistical error.}
from the circular regions with 20$^{\prime\prime}$ radius centered on the source, 
while background products were extracted from the annulus region of the inner and outer radius of 45$^{\prime\prime}$ and 75$^{\prime\prime}$, respectively.
High background period is excluded, when the count-rate of 10--12 keV with {\tt PATTERN$==$0} is higher than 0.4 cts/s.
The RGS spectra were processed with task {\tt rgsproc}.

In the Suzaku data analysis, we used the X-ray Imaging Spectrometer (XIS) data in 0.7--10.0 keV,
while we did not use the hard X-ray detector (HXD) PIN diode data because significant signals were hardly detected above 10 keV.
Since contamination calibration on the XIS window is not fully achieved, we did not use the XIS data below 0.7 keV.
For Suzaku data reduction we used the HEADAS 6.15.1 software package. 
The XIS data were screened with {\tt XSELECT} using the standard criterion \citep{koy07}. 
The XIS spectra and light-curves were extracted from circular regions of 3$^{\prime}$ radius centered on the source, 
while background products were extracted from the annulus region of the inner and outer radius of 4$^{\prime}$ and 7$^{\prime}$ respectively.
We generated XIS response matrices using the {\tt xisrmfgen} software, 
which takes into account the time-variation of the energy response. 
As for generating ancillary response files (ARFs), 
we used {\tt xissimarfgen} \citep{ish07}. 
This tool calculates ARFs through ray-tracing, 
and we selected the number of input photons as 400,000, with the ``estepfile'' parameter ``full''.
After extracting the products for the back-illuminated (BI) CCD (XIS 1) and for the three front-illuminated (FI) CCD XIS detectors (XIS 0, XIS 2, XIS 3) separately, 
we confirmed that the XIS 0, XIS 2, and XIS 3 products are almost identical to each other. 
We thus combined the XIS 0, XIS 2, and XIS 3 products in our spectral analysis, using 
 {\tt addascaspec} to create the merged XIS spectra and responses.

All the spectral fitting were made with {\tt xspec} v.12 \citep{arn96}.
In the following, the {\tt xspec} model names used in the spectral analysis are explicitly given.

\section{Data Analysis and Results}
\subsection{Variable Double Partial Covering Model}
First, we try to fit the time-average spectrum of each observation 
using a similar spectral model that was successful for MCG-6-30-15.
\citet{miy12} adopted the following spectral model to fit the MCG-6-30-15 (see their Equation (2)),
\begin{eqnarray}
F&=& W_H\, W_L\, (N_1 + N_2\,W_2) \,P + N_3 \,R \,P + I_{Fe},
\end{eqnarray}
where $P$ is the intrinsic power-law spectrum, 
$R$ is the outer disk reflection albedo, and 
$I_{Fe}$ is the fluorescent narrow iron emission line. 
$N_1$ is the normalization of the direct (non-absorbed) component, and 
$N_2$ is the normalization of the heavily absorbed component, 
where $W_2$ indicates the ionized thick partial absorber 
($N_H \approx 1.6 \times 10^{24}$cm$^{-2}$ and $\log \xi \approx 1.57$), and
$N_3$ is the normalization of the disk reflection component.
$W_H$ and $W_L$ are high-ionized and low-ionized warm absorbers, both of which are optically thin.

MCG-6-30-15 shows ionized iron K-absorption lines, which are explained by $W_H$, 
where $N_H \approx 2.4 \times 10^{23}$ cm$^{-2}$ and $\log \xi \approx 3.4$ (Table 1 in \cite{miy12}). 
However, in the case of 1H0707--495, these iron K-absorption lines are not observed, 
which indicates that $W_H$ is not necessary.
Also, a fluorescent narrow iron emission line, which is accompanied by outer disk reflection, 
is not recognized in 1H0707--495 either.
Hence, we consider the following simple form of the model,
\begin{eqnarray}
F&=& W_L\, (N_1 + N_2\,W_2) \,P \nonumber \\
 &=& W_L\, N\, (1 - \alpha + \alpha\,W_2) \,P, \label{vmpcmodel1}
\end{eqnarray}
where we have introduced the total normalization factor $N \equiv N_1+N_2$ and the
partial covering fraction $\alpha \equiv N_2/N$.
Here, $W_L$ can be written as
\begin{eqnarray}
W_L = \exp\left(- \sigma(E, \xi) N_{H,L}\right), \label{W_L}
\end{eqnarray}
where $\sigma(E, \xi)$
is the cross-section and $N_{H,L}$ is the column density of the low-ionized absorber.
\citet{miy12} discovered that, in the course of spectral variation of MCG-6-30-15, 
 $N_{H,L}$ and $\alpha$, which are in principle independently variable, are clearly correlated to each other. Namely, 
\begin{eqnarray}
N_{H,L} = \alpha N_{H,L}^{({\rm fixed})}, \label{N_H,L}
\end{eqnarray}
where $N_{H,L}^{({\rm fixed})}$ is invariable (Equation (8) in \cite{miy12}).
This finding, which shows that the two independent absorbers are correlated,
was unexpected and rather surprising.
Considering that $W_L$ is optically thin, from Equation (\ref{W_L}) and (\ref{N_H,L}), 
\begin{eqnarray}
W_L &\approx & 1- \sigma(E, \xi) N_{H,L}\nonumber\\
   &=& 1- \alpha \, \sigma(E, \xi) N_{H,L}^{(\rm fixed)}\nonumber\\
   &\approx & 1- \alpha + \alpha \exp\left( -\sigma(E, \xi) N_{H,L}^{(\rm fixed)}\right)\nonumber\\
   &=& 1- \alpha + \alpha W_L^{(\rm fixed)},
\end{eqnarray}
where we introduced $W_L^{(\rm fixed)} \equiv \exp\left( -\sigma(E, \xi) N_{H,L}^{(\rm fixed)}\right)$.
In the end, Equation (\ref{vmpcmodel1}) may be written as,
\begin{eqnarray}
F&=& N\, (1- \alpha + \alpha W_L^{(\rm fixed)}) \, \left(1 - \alpha + \alpha\,W_2\right) \,P.
\end{eqnarray}
Namely, the spectral model \citet{miy12} proposed is mathematically identical to a ``double partial covering model,''
which is similar to the one used by \citet{tan04}. 
Difference of our model from that by \citet{tan04} is that
we assume that the two absorption layers have the same covering fraction and different ionization states, 
while \citet{tan04} assumed that the two partial absorbers have different covering fractions 
and that both are neutral.
We again stress that, in this model, 
the two warm absorbers with different ionization degrees have the same covering fraction.
Presumably, the partial covering clouds have internal layers 
that contain the thick/low-ionized core and thin/high-ionized envelope \citep{miy12},
so that the two absorption layers have the same covering fraction.

Additionally, considering the soft excess component, 
interstellar absorption and strong Fe L- and K-shell edges, 
the model we use to fit 1H0707--495 spectra is 
\begin{eqnarray}
F&=&A_I\, N\, (1-\alpha + \alpha W_n \,e^{-\tau_1}) (1-\alpha + \alpha\, W_k \,e^{-\tau_2})(P+B)\label{VPC}
\end{eqnarray}
where $B$ is the spectrum from an accretion disk (we adopted {\tt diskbb} in {\tt xspec}), 
$A_I$ is the effect of interstellar absorption ({\tt wabs}),
and $W_n$ and $W_k$ are hotter/thinner and colder/thicker partial absorbers,
respectively.
The Fe L- and K-shell edges are mostly explained by $W_n$ and $W_k$, respectively, 
but additional edge components were introduced as $e^{- \tau_1}$ and $e^{- \tau_2}$ when needed.
It should be noted that the additional edges are put only to the spectral components that pass through the absorbers.

To model the warm absorbers, we used the absorption table calculated with XSTAR \citep{kal04} by \citet{miy09},
assuming the solar abundance and the photon index of the ionizing spectrum 2.0.
Hereafter, Equation (\ref{VPC}) is our baseline model, which we call the VDPC model, 
where the partial covering fraction $\alpha$ is mostly variable and
other spectral parameters are not significantly variable within an observation, 
as indicated for MCG-6-30-15 by \citet{miy09}.
We note that parameters besides the partial covering fraction $\alpha$ are variable 
for different observations ($=$ at timescales over several days.)

\subsection{Spectral Fitting to the Average Spectra}
\begin{figure*}
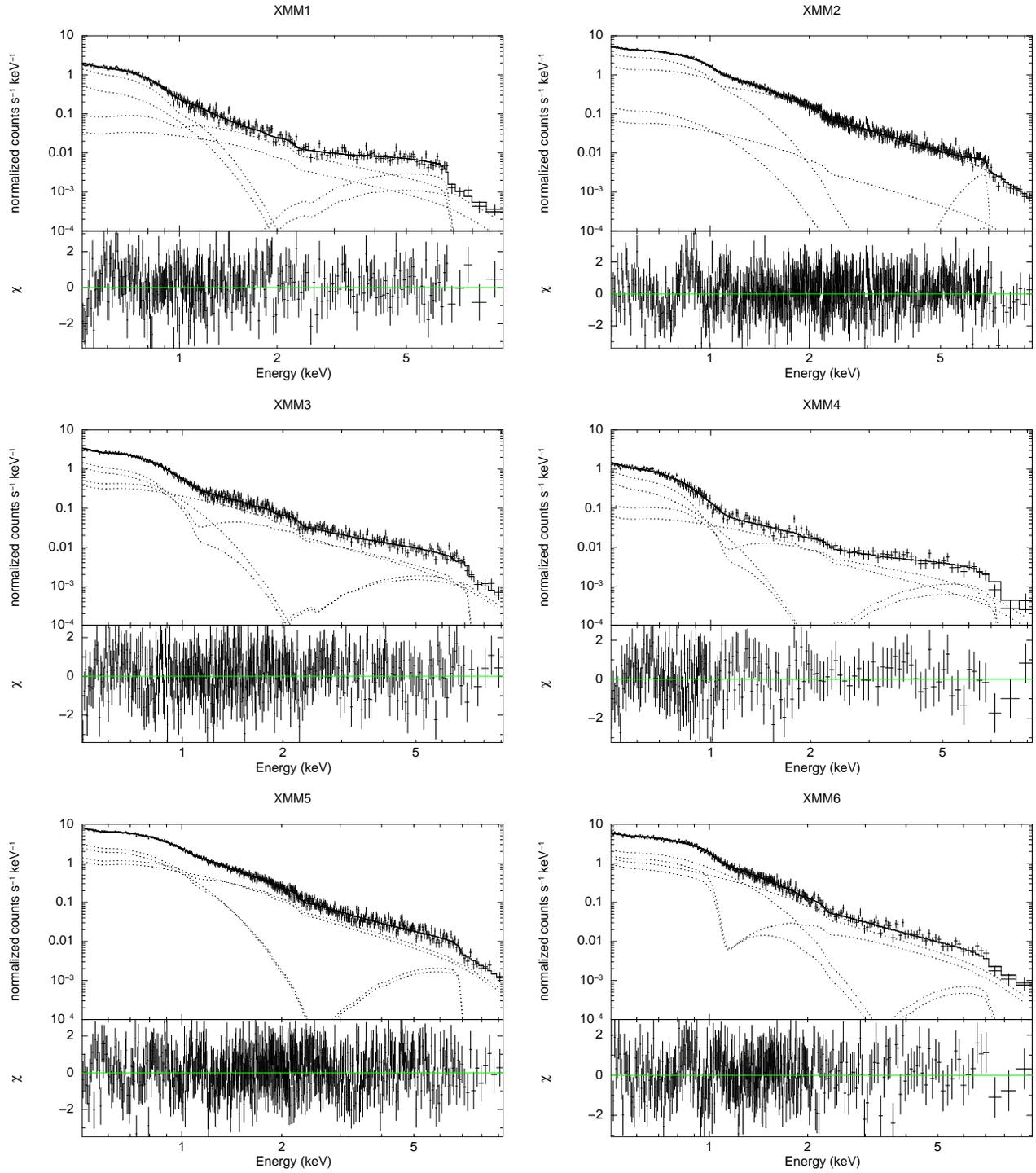

\centering
\subfigure{
        \resizebox{8.5cm}{!}{\includegraphics[angle=270]{1a.eps}}
        \resizebox{8.5cm}{!}{\includegraphics[angle=270]{1b.eps}}
}
\subfigure{
        \resizebox{8.5cm}{!}{\includegraphics[angle=270]{1c.eps}}
        \resizebox{8.5cm}{!}{\includegraphics[angle=270]{1d.eps}}
}
\subfigure{
        \resizebox{8.5cm}{!}{\includegraphics[angle=270]{1e.eps}}
        \resizebox{8.5cm}{!}{\includegraphics[angle=270]{1f.eps}}
}
\caption{The fitting results of the average spectra. 
Individual spectral components are shown with dotted lines.
In the Suzaku data, the black points and lines show the FI spectrum, 
and the red ones show the BI spectrum.}
\label{fig:average:a}
\end{figure*}

First, we consider the time-average spectra for the individual observations in Table 1.
The spectra are studied under the assumption of the VDPC model given in Equation (\ref{VPC}).
Figure \ref{fig:average:a} shows the fitting result for the total time-average spectra.
The fitting parameters and the reduced chi-squares are shown in Table \ref{average}.
Errors are quoted at the statistical 90\% level.
The foreground absorption was estimated as $N_H=4\times10^{20}$ cm$^{-2}$ from the Leiden-Argentice-Bonn 21 cm survey \citep{kal05}.
However, when we fix the interstellar absorption, 
there remain some residuals in the soft band,
suggesting existence of some excess absorption.
Thus, we allow the interstellar absorption to be variable, following \citet{dau12}.
This may be due to additional intrinsic absorption associated with 1H0707--495.

Most spectra are fitted reasonably well, 
where complex iron L- and K-features at $\sim 1$ keV and $\sim 7$ keV are primarily explained by partial covering by the hotter/thinner absorber ($W_n$) and colder/thicker absorber ($W_k$), respectively.
However, we note that there remain some residuals around 1 keV of the spectra with long exposure time (especially in XMM7, 11, 12, 13, and 14.)
These residuals may be explained by the fact that 
the time-averaged spectra of the VDPC model,
where partial covering fraction is the only variable parameter, 
is not mathematically identical to the VDPC model 
with the average partial covering fraction.
This situation is explained in the Appendix.
We note that the XMM15 observation was triggered after the source flux was significantly dropped 
compared with that of the 2008 observations \citep{fab12}.
Thus, parameters of the intrinsic X-ray spectrum were drastically changed 
compared with those of the 2008 observations.

\subsection{The RGS spectra}
In the VDPC model, the absorption edge/line features from the warm absorbers 
are expected in the X-ray spectra.
In fact, presence of some absorption-like features have been
pointed out by \citet{dau12}.
On the contrary, \citet{blu09} claimed that XMM-Newton RGS observations have few narrow absorption lines and
show little evidence for a line-of-sight ionized winds or warm absorbers.
To examine this claim, 
we analyze the RGS data to see if they are consistent with presence of the ionized warm absorbers
expected from the VDPC model.
Figure \ref{fig:rgs} shows the XMM7--10 combined RGS spectra, 
which is the same as what \citet{blu09} analyzed
with the best-fit VDPC model for the EPIC-pn time-average spectrum.
In the model spectrum, many absorption lines are clearly recognized.
However, even the strongest hydrogen-like oxygen Lyman $\alpha$ line is
not clearly recognized in the RGS spectrum,
because of the limited statistics.
This result shows that the high-resolution RGS spectra are consistent with the VDPC model 
with weak but plenty of absorption lines.
  
\subsection{Spectral Fitting to the Sliced Spectra}
Next, we apply the VDPC model to ``intensity-sliced spectra'' to investigate spectral variation
at time-scales shorter than a $\sim$ day.
The method for creating the intensity-sliced energy spectra is as follows:
(1) We create light-curves with a bin-width of 128 sec in the 0.2--12.0 keV band (see Figure \ref{fig:ltcrv:a}).
(2) We calculate all counts in each data and determine the four intensity ranges 
(separated by horizontal lines in Figure \ref{fig:ltcrv:a}) that contain almost the same counts.
(3) From the four time-periods corresponding to the different source flux levels,
we create the four intensity-sliced energy spectra.

First, we examine if spectral shape is truly variable for intensity-sliced spectra within an observation.
To that end, we first try to fit the intensity-sliced spectra only changing the normalization.
Figure \ref{fig:const} shows the result of XMM12,
which exhibits a wide spectral variation. 
Here, only normalizations of the intrinsic power-law and MCD are variable (ratio of the two components is fixed),
while all of the other spectral parameters are common.
The reduced chi-square is 3.24, which clearly indicates that the observed spectra are significantly variable.

So, next we examine if the observed spectral variation within an observations is explained
only varying the partial covering fraction.
During each observation, whose duration is from 40 ksec (XMM3 and 4) to 250 ksec (Suzaku), 
we assume that the intrinsic source luminosity and the spectral shape are invariable, 
and that the spectral variation is caused by only change of the partial covering fraction.
For each observation, all the spectral parameters are common to the four spectra, except for the covering fraction.
The fitting results are shown in Figure \ref{fig:slice:a} and Table \ref{tab:slice:a}.
In spite of the fact that 
our spectral variation model is extremely simple with {\it only}\/ a single variable parameter, 
we see that the 0.5--10 keV XMM and Suzaku spectral variations are explained reasonably well, 
where the reduced chi-square values (degree of freedom) are from 1.07 (1358) to 1.65 (451).
We also find that 
normalizations and other parameters of the disk blackbody component and the power-law component are significantly variable between observations, 
indicating that the intrinsic luminosity and spectra are variable in timescales longer than $\sim$ days.

\subsection{Time-Variation of the different energy bands}
We find, in the above subsection, that the VDPC model can explain the intensity-sliced spectra by only change of the covering fraction, 
where the luminosity is constant over the observation period for a $\sim$ day.
Next, we examine if shorter timescale variations are also explained with the VDPC model. 
To that end, we create simulated light-curves via the VDPC model 
for several different energy bands varying only the partial covering fraction,
and compared them with the observed light-curves in these energy bands. 

The method to create the simulated light-curves is as follows:
(1) For each observation, we fix all the model parameters besides the partial covering fraction at the best-fit values 
obtained from the intensity-sliced spectral analysis (Table \ref{tab:slice:a}).
(2) For each light-curve bin, which is 512 sec, the partial covering fraction value is calculated 
so that the observed counting rate in 0.5--10.0 keV and the model counting rate match.
(3) Given the partial covering fraction value thus determined for each bin, 
we create the simulated spectra using {\tt fakeit} command in {\tt xspec},
and calculate the simulated count-rates in 0.5--1.0 keV (soft), 1.0--3.0 keV (medium), 
and 3.0--10.0 keV (hard) respectively (Figure \ref{fig:cf}).
(4) We compare the simulated light-curves in the three bands with the observed ones.

In Figure \ref{fig:modelltcrv:a}, we show comparison of the observed light-curves and model light-curves,
as well as variation of the partial covering fraction.
We find that the observed light-curves in the soft band is almost fully reproducible by the VDPC model
with {\it only}\/ variation of the partial covering fraction, 
while the agreement becomes less satisfactory toward higher energy bands.
This may suggest that the VDPC model with a single variable parameter is 
too simple to fully explain the hard-band spectral variation;
in fact, some hard flare-like events have been reported independently of the soft-band variation (e.g., \cite{lei04}).
In any case, we stress that the VDPC model can reproduce the observed light-curves 
at the accuracy manifested in Figure \ref{fig:modelltcrv:a}
that demonstrates that the partial covering phenomenon {\it does}\/ take place in 1H0707--495 and 
plays a major role to cause apparently significant intensity/spectral variations.

\section{Discussion}

\subsection{Spectral features}
In addition to the cold/thick and hot/thin absorbers that respectively explain the iron K- and L- edges, 
we had to introduce the {\it ad-hoc} edge components to explain the extremely strong edge features.
We conjecture that the extremely strong iron edges might be due to 
either Fe overabundance and/or P Cygni profiles produced by winds.
When the iron is over-abundant as much as several times,
the strong spectral drops may be explained \citep{tan04}. 
Besides, if the velocity of the partial absorbers is $\sim 0.3 c$,
the iron line profiles would become {\it P Cygni profiles} and may explain the drops,
as discussed by \citet{don07} (see also \cite{tur14}\footnote{The proceeding (\cite{tur14}) is available at \url{http://xmm.esac.esa.int/external/xmm_science/workshops/2014symposium/presentations/TTurner_t.pdf}}.)
The ``P Cygni'' hypothesis is consistent with our model in Section 4.3, 
which suggests that the partial covering clouds are in the clumpy line-driven disk winds.
In this scenario, variation of the apparent edge energy (see Tables \ref{average} and \ref{tab:slice:a}) 
may be explained by change of an ionization degree or wind structures \citep{don07}.

\subsection{Timescale of variations}
In Section 3, we have shown that 
the VDPC model is successful to explain the observed light-curves 
in a timescale shorter than $\sim$ days, with only change of the partial covering fraction.
In particular, the simulated light-curves and observed ones below 3 keV are surprisingly identical.
This suggests that the intrinsic X-ray luminosity and spectral shape are almost invariant within a $\sim$ day, 
and variation of the partial covering fraction is the main cause of the observed spectral variation.
We see that the range of covering fraction variation is large, from null to $\sim90$ \% (Figure \ref{fig:modelltcrv:a}), 
which suggests that absorbing clouds in the line-of-sight are comparable in size to the
central X-ray emitting region, and 
fluctuating to cover/uncover the central source, 
independently of status of the central source.

While the covering fraction variation can almost reproduce the light-curves,
there remain some residuals above 3 keV.
When we assume there is no intrinsic variation in our model, the residuals
between the observation and the model indicates real intrinsic variation.
Figure \ref{fig:ratio} shows ratios of the observed light-curves to the simulated ones,
the energy band and time-bin of which are the same as those of Figure \ref{fig:modelltcrv:a}.
This figure shows intrinsic variations that cannot be explained by only variation of the covering fraction
do remain in the hard energy band.
To evaluate the residuals, 
we calculate the root-mean-square of the residuals between the observation and the model (Figure \ref{fig:rms}).
We defined the RMS of the residuals as
 $\sqrt{\sum_i^N((\mathrm{observed_i}-\mathrm{model_i})/\mathrm{model_i})^2/N}$
 in the three energy bands.
Figure \ref{fig:rms} clearly shows that the residuals are more significant in the hard energy band, 
where the power-law component is dominant.
This suggests that the power-law variation causes the intrinsic variation in the hard energy band,
which is consistent with the light-crossing time of X-ray emitting region, $\sim$ 500 sec,
when assuming that black hole mass is $ 10^{6.37} M_\odot$ and 
the Schwarzschild radius is $r_s \approx 6\times 10^{11}$ cm.
(See Section 4.3 for the validity of these values.)
On the contrary, little intrinsic variation is seen in the soft energy band, 
where the multi-color disk component is dominant.
This suggests that the viscous time of the disk is longer than the timescale of an observation. 

While the intrinsic luminosity seems almost invariant within a day,
comparing XMM1 with XMM2, for example,
variation of the intrinsic luminosity is about an order of magnitude over 2 years.
Even during a shorter interval, between XMM12 and XMM13, as an example,
variation of the intrinsic luminosity is about a factor of two over 2 days.
The standard disk may not explain such a large fluctuation in the soft X-ray emission
as well as the high disk temperature (Tables \ref{average} and \ref{tab:slice:a}) 
observed in 1H0707--495 \citep{min00, tan04}.
Assuming that 1H0707--495 has a slim-disk \citep{min00},
a large variation of the mass-accretion rate might explain these significant luminosity variations over days \citep{gal04}.

NLS1s have been known to exhibit significant intensity and spectral variations at
various timescales (see e.g.~\cite{lei99}.)
Our result indicates that this is due to combination of the variations in the intrinsic luminosity and 
in the covering fraction of the intervening absorbers at different timescales.
In summary, we propose that spectral variation of 1H0707--495 mainly consists of 
the two independent types of variations with different timescales;
(1) intrinsic luminosity variation over days, and 
(2) variation of partial covering fraction at a timescale of hours.

\subsection{Property of partial absorbers}
\citet{zog10} pointed out that 
the X-ray energy spectrum of 1H0707--495 is composed of at least two components, 
one dominating between 1 and 5 keV, the other dominating above the 1--5 keV range.
The VDPC model, in which the multi-color disk component and the high-ionized absorber determine the soft spectrum below $\sim3$ keV, 
while the power-law component and the low-ionized absorber determines the hard spectrum over $\sim3$ keV, 
is consistent with this spectral picture.
\citet{tan04} assumed that the covering fraction varies with the orbital motion of the clouds,
 and suggested that the clouds are rotating at around $r\sim 400 \; r_s$ $(\sim2\times10^{14})$ cm.
However, if the partial covering clouds are uniformly surrounding the central source, 
that would produce Fe-K fluorescence lines; this contradicts with absence of such lines \citep{zog10}. 
Namely, most of the partial covering clouds should exist in the line-of-sight 
not to produce the significant fluorescent iron emission line.
Therefore we suspect that the partial covering clouds are in the clumpy line-driven disk wind, 
which is thought to be an origin of the outflows \citep{ste90}.
Since the realistic radiation-hydrodynamic simulations show funnel-shaped disk winds \citep{pro00, pro04},
when the line-of sight is close to the tangential line of the disk wind,
we may observe the X-ray energy spectrum with hardly fluorescent iron emission line 
but significant absorption features.

Next, we estimate location of the warm absorber $W_n$, 
which is considered to be photoionized by the central X-ray source,   
in a similar manner suggested by \citet{miy12}.
We use the following relations:
\begin{eqnarray}
\xi&\equiv& L/nr^2=\frac{Ld}{N_Hr^2} {\rm,\,\, where \,\,}N_H=nd,\\
r&=&\left(\frac{Ld}{\xi N_H}\right)^{1/2}, 
\end{eqnarray}
where $r$ is distance from the central X-ray source to the absorber,
$d$ is a representative thickness of the absorbing region along the line-of-sight,
$n$ is density
and $L$ is the intrinsic luminosity.

We found the partial covering fraction varies from null to $\sim0.9$, 
that suggests that the partial absorber size is comparable to that of the central source, 
otherwise we would see a full covering (when the absorber size is much larger than the source size), 
or the variation amplitude should be much smaller (when opposite.)
When the source is not significantly variable, 
the covering fraction is not so variable for about 10$^4$ sec
(see, e.g.,~the top panel of Figure \ref{fig:modelltcrv:a}) 
which we take as a typical crossing timescale of the partial absorbers in front of the central X-ray source.
The crossing timescale may be expressed as $\sim d/v$, 
where $v$ is the velocity of partial absorbers.
Hence,
\begin{equation}
d\sim v\times10^4\sim10^{13}\left(\frac{v}{10^9\,\,\mathrm{cm\,\,s^{-1}}}\right)\,\,\mathrm{cm},
\end{equation}
where $v$ is normalized to that of a typical line-driven disk wind \citep{pro00}.
Here we take typical values of $N_H=5\times10^{23}$ cm$^{-2}$, $\log\xi=3$ for $W_n$, $L=10^{43}$ erg s$^{-1}$
expected by the parameters of power-law and multi-color disk in Table \ref{tab:slice:a}.
We thus estimate the distance from the central X-ray source to the absorber as follows:
\begin{equation}
r=\left(\frac{L}{\xi N_H}d\right)^{1/2}\simeq10^{14.5} \,\,\mathrm{cm}.
\end{equation}
Black hole mass of 1H0707--495 is estimated to be $ 10^{6.37} M_\odot$ 
from the empirical reverberation relation (\citet{zho05} and references therein), 
and the Schwarzschild radius is $r_s \approx 6\times 10^{11}$ cm.
Therefore, our analysis suggests that typical sizes of the X-ray emission region and the partial absorbers are
$\sim 20 \; r_s$, and the absorbers are typically located at $\sim500 \; r_s$ from the black hole. 
This is consistent with numerical simulations of the disk wind 
accelerated by a radiation force due to spectral lines (line force), 
which shows that the wind is launched within several $100 \; r_s$ radius (e.g., \cite{nom13}). 
In addition, this is also consistent with an X-ray time-lag analysis in a wide energy band of 1H0707--495,
which suggests that 
an absorbing medium that partially-covers the source at 20--100 $r_s$ from the black hole 
may be the cause of the observed energy dependent reverberation time-lags \citep{mil10}.

\section{Conclusion}
We have analyzed all the currently available XMM-Newton and Suzaku archival data of NLS1 1H0707--495, the total exposure time of which is $\sim1070$ ksec.  
We have found that the observed spectral variation is successfully explained by the VDPC model.
Our main conclusions are as follows:
\begin{enumerate}
\item We have shown that the spectral model \citet{miy12} proposed to explain the spectral variation of MCG-6-30-15 is identical to a ``variable double partial covering model'', 
which we denote the VDPC model in this paper, 
where two partial absorption layers have different ionization states and the same covering fraction.
\item The intensity-sliced spectra of 1H0707--495 within a $\sim$ day
are successfully fitted with the VDPC model only varying the partial covering fraction.
Light-curves within a $\sim$ day are also explained with the VDPC model 
only varying the partial covering fraction, 
where the agreement between the observation and the model is remarkable in the soft energy band (0.5--1 keV), while
less satisfactory in higher energy bands.
Some intrinsic variation is recognized above $\sim3$ keV.
\item Over timescales of $\sim$ days, intrinsic luminosity/spectral variations are significant, 
such as a factor of two over 2 days, or an order of magnitudes over 2 years.
\item We propose that the spectral variation of 1H0707--495 consists of the three types of variations with different timescales: 
(1) intrinsic luminosity variation over days, 
(2) variation of partial covering fraction at a timescale of hours, and
(3) small intrinsic hard component variation above 3 keV in a timescale of hours or less.
\item We suggest that the absorbing clouds are clumpy radiatively-driven disk winds localized in the line of sight.
We estimate the X-ray source size and typical cloud size $\sim 20 \; r_s$, 
and the clouds typically locate at $\sim 500 \; r_s$.
\end{enumerate}

Finally, we point out that significant flux/spectral variations similar to that of 1H0707--495 
are also observed from other NLS1s such as IRAS 13224--3809 \citep{bol03}, Mrk 335 \citep{lon13}, and Ark 564 \citep{bri07}.
It will be intriguing to apply the same analysis presented in this paper to these sources
to see if the VDPC model is also valid and 
their significant spectral variations are explained by change of the partial covering fractions.  
If this is the case, that will be an important step toward a systematic understanding of the NLS1s.

\bigskip
\bigskip

This research has made use of public Suzaku data obtained through the Data ARchives and Transmission System (DARTS), 
provided by the Institute of Space and Astronautical Science (ISAS) at Japan Aerospace Exploration Agency (JAXA).
This work is also based on observations obtained with XMM-Newton, an ESA science mission with instruments and contributions directly funded by ESA Member States and the USA (NASA), 
and the XMM-Newton data obtained through the High Energy Astrophysics Science Archive Research Center (HEASARC) at NASA/Goddard Space Flight Center.

For data reduction, we used software provided by HEASARC at NASA/Goddard Space Flight Center.


 
\begin{table}
  \caption{
Suzaku and XMM-Newton observations of 1H0707--495.
Observation IDs, start date, and good exposure times in sec are indicated.
The exposure time of Suzaku is that of XIS 0.
  }\label{dataset}
  \begin{center}
    \begin{tabular}{lllll}
      \hline
      Name &  Observation ID & Date & Exposure\\
      \hline
      XMM1 &0110890201&2000-10-21& 37.8 ks \\
      XMM2 &0148010301&2002-10-13& 68.1 ks\\
      XMM3 &0506200301&2007-05-14& 35.8 ks\\
      XMM4 &0506200201&2007-05-16& 26.9 ks\\
      XMM5 &0506200501&2007-06-20& 32.4 ks \\ 
      XMM6 &0506200401&2007-07-06& 14.7 ks \\
      XMM7 &0511580101&2008-01-29& 99.6 ks\\     
      XMM8 &0511580201&2008-01-31& 66.3 ks\\ 
      XMM9 &0511580301&2008-02-02& 59.7 ks\\ 
      XMM10&0511580401&2008-02-04& 66.6 ks\\     
      XMM11&0653510301&2010-09-13& 103.6 ks\\
      XMM12&0653510401&2010-09-15& 101.9 ks\\
      XMM13&0653510501&2010-09-17& 95.8 ks\\
      XMM14&0653510601&2010-09-19& 97.6 ks\\
      XMM15&0554710801&2011-01-12& 64.4 ks\\  
      \hline
      Suzaku & 700008010 &2005-12-03 & 97.9 ks\\
      \hline
    \end{tabular}
  \end{center}
\end{table}

 \begin{landscape}
 \begin{table}
  \caption{
Results of the average spectral fitting. 
Normalization of the power-law means the X-ray flux after removing the absorption in units of $10^{-12}$ erg cm$^{-2}$ s$^{-1}$ between 0.4--10 keV. 
The normalization of MCD means {\tt diskbb} normalization, 
$\left((R_\mathrm{in})/(\mathrm{km})/(D/10\,\mathrm{kpc})\right)^2\cos\theta$.
Errors are quoted at the statistical 90\% level.
Often upper-limits of the edge depths are not determined, and only the lower-limits are shown.
  }\label{average}
  \begin{center}
    \scalebox{0.8}{
    \begin{tabular}{llllllllll}
      \hline
      &&XMM1&XMM2&XMM3&XMM4&XMM5&XMM6&XMM7&XMM8 \\
      \hline
      Interstellar absorption &$\mathrm{N_H}$ ($10^{21}$cm$^{-2}$)&
1.0 $\pm0.3$ &
0.9 $\pm0.1$ &
0.9 $\pm0.3$ &
0.8 $_{-0.5}^{+0.6}$ &
1.11 $\pm0.18$ &
0.28 $\pm0.02$ &
0.90 $\pm0.09$ &
0.73 $_{-0.07}^{+0.08}$ \\
      \hline
      Thick absorber &$\mathrm{N_H}$ ($10^{23}$cm$^{-2}$)&
6 $_{-2}^{+1}$ &
8 $\pm2 \times10^1$&
8 $_{-2}^{+3}$ &
7 $_{-2}^{+3}$ &
15 $_{-7}^{+4}$ &
16 $_{-6}^{+8}$ &
42 $_{-14}^{+2}$ &
50 $_{-15}^{+7}$  \\
      &$\mathrm{\log \xi}$&
0.4 $_{-0.4}^{+1.6}$ &
1.1 $\pm1.1$ &
0.3 $_{-0.3}^{+1.7}$ &
0.4 $_{-0.4}^{+1.5}$ &
0.6 $_{-0.6}^{+0.5}$ &
0.7 $\pm0.7$ &
0.1 $_{-0.1}^{+0.2}$ &
0.7 $_{-0.7}^{+1.5}$  \\
      Edge &$\mathrm{E_{cut}}$ (keV)&
6.74 $_{-0.09}^{+0.07}$ &
6.9 $_{-0.1}^{+0.1}$ &
7.1 $_{-0.2}^{+0.4}$ &
6.9 $\pm0.3$ &
6.86 $_{-0.17}^{+0.14}$ &
---&
6.86 $\pm0.07$ &
7.0 $\pm0.1$\\
      &$\tau$&$>10$&$>10$&$>10$&$>10$&5 $_{-4}$&---&$>10$&$>10$\\
      \hline
      Thin Absorber & $\mathrm{N_H}$ ($10^{23}$cm$^{-2}$)&
0.9 $_{-0.4}^{+0.6}$ &
55.4 $_{-52.0}^{}$ &
1.4 $\pm0.5$ &
1.5 $_{-0.7}^{+1.3}$ &
1.4 $_{-1.1}^{+3.6}$&
1.1 $_{-0.9}^{+1.7}$ &
0.13 $_{-0.06}^{+0.36}$ &
0.11 $_{-0.03}^{+0.14}$     \\
      &$\log\xi$&
2.81 $_{-0.10}^{+0.07}$ &
4.2 $_{-0.9}^{+0.5}$ &
2.78 $_{-0.08}^{+0.06}$ &
2.8 $\pm0.1$ &
3.21 $_{-0.10}^{+0.18}$ &
3.21 $_{-0.05}^{+0.15}$ &
3.00 $_{-0.09}^{+0.14}$ &
2.95 $_{-0.02}^{+0.12}$  \\
      Edge &$\mathrm{E_{cut}}$ (keV)&
--&
1.030 $_{-0.007}^{+0.006}$ &
1.06 $\pm0.01$ &
1.02 $\pm0.03$ &
1.03 $\pm0.02$ &
1.019 $_{-0.014}^{+0.033}$ &
1.018 $\pm0.007$ &
1.067 $\pm0.006$ \\
      &$\tau$&--&
0.48 $\pm0.02$ &
2.2 $\pm0.2$ &
2.0 $\pm0.4$ &
0.52 $_{-0.12}^{+0.19}$ &
6 $_{-5}$ &
0.43 $\pm0.02$ &
0.48 $\pm0.02$ \\ 
      \hline
      Power law&Photon index&
2.1 $\pm0.1$ &
2.70 $_{-0.05}^{+0.05}$ &
2.7 $\pm0.1$ &
2.3 $\pm0.2$ &
2.81 $\pm0.07$ &
2.69 $\pm0.15$ &
2.66 $\pm0.04$ &
2.78 $\pm0.05$\\
      &norm&
2.5 $_{-0.4}^{+0.6}$ &
1.1 $\pm0.3 \times 10^2$&
6 $\pm1$ &
1.9 $_{-0.6}^{+1.0}$ &
17 $\pm4$ &
6.4 $_{-1.0}^{+1.3}$ &
40 $\pm7$ &
6 $_{-1}^{+2}\times10^1$ \\
      \hline
      MCD&$\mathrm{T_{in}}$ (eV)&
126 $_{-6}^{+7}$ &
170 $\pm5$ &
135 $_{-9}^{+11}$ &
121 $_{-11}^{+14}$ &
150 $\pm8$ &
2.3 $\pm0.2$ $\times 10^2$ &
160 $\pm4$ &
190 $\pm5$  \\
   &norm&
8.0 $_{-3.3}^{+6.0}\times10^3$ &
1.9 $_{-0.6}^{+0.7}\times10^4$ &
4.2 $_{-2.0}^{+3.7}\times10^3$ &
4.4 $_{-2.7}^{+7.4}\times10^3$ &
6 $_{-3}^{+4}\times10^3$ &
2.5 $_{-0.9}^{+1.7}\times10^2$ &
9.7 $_{-2.2}^{+2.8}\times10^3$ &
4.5 $_{-1.0}^{+1.3}\times10^3$ \\
      \hline
      Partial covering&covering factor&
0.74 $\pm0.05$ &
0.960 $^{+0.008}_{-0.013}$ &
0.59 $^{+0.05}_{-0.06}$ &
0.67 $^{+0.08}_{-0.10}$ &
0.56 $_{-0.07}^{+0.10}$ &
0.42 $^{+0.03}_{-0.04}$ &
0.90 $^{+0.01}_{-0.02}$ &
0.90 $^{+0.02}_{-0.03}$\\
      \hline
      \multicolumn{2}{r}{Reduced chisq (d.o.f)}&
      1.09 (288) & 1.13 (653) & 1.04 (399) & 1.10 (183) & 0.99 (568) &  0.99 (338) & 1.19 (756) & 1.21 (717)    \\
      \hline
    \end{tabular}
    }
  \end{center}
\end{table}
\end{landscape}
\addtocounter{table}{-1}

\begin{landscape}
 \begin{table}
  \caption{
  {\it Continued.}
  }\label{average:b}
  \begin{center}
    \scalebox{0.8}{
    \begin{tabular}{llllllllll}
      \hline
      &&XMM9&XMM10&XMM11&XMM12&XMM13&XMM14&XMM15&Suzaku \\
      \hline
      Interstellar absorption &$\mathrm{N_H}$ ($10^{21}$cm$^{-2}$)&
0.81 $\pm0.02$ &
0.86 $\pm0.12$ &
0.60 $\pm0.09$ &
0.69 $\pm0.08$ &
0.75 $\pm0.09$ &
0.61 $\pm0.08$ &
0.73 $_{-0.32}^{+0.33}$ &
2.0 $\pm0.8$   \\
      \hline
      Thick absorber &$\mathrm{N_H}$ ($10^{23}$cm$^{-2}$)&
45 $_{-13}^{+7}$ &
47 $_{-8}^{+6}$ &
7.9 $_{-1.4}^{+2.3}$ &
7.9 $_{-1.4}^{+2.4}$ &
5.6 $_{-1.4}^{+1.6}$ &
8.9 $_{-2.1}^{+1.6}$ &
3.2 $_{-1.4}^{+5.2}$ &
6.5 $\pm0.9$  \\
      &$\mathrm{\log \xi}$& 
0.5 $\pm0.5$ &
0.5 $_{-0.5}^{+0.3}$ &
0.3 $_{-0.3}^{+1.6}$ &
0.2 $_{-0.2}^{+1.6}$ &
0.3 $_{-0.3}^{+2.0}$ &
0.4 $_{-0.4}^{+1.7}$ &
0.4 $_{-0.4}^{+1.8}$  &
0.6 $\pm0.3$  \\
      Edge &$\mathrm{E_{cut}}$ (keV)&
6.91 $_{-0.10}^{+0.08}$ &
6.88 $_{-0.10}^{+0.09}$ &
7.06 $_{-0.09}^{+0.12}$ &
7.3 $\pm0.2$ &
7.22 $\pm 0.10$ &
7.22 $_{-0.13}^{+0.25}$ &
6.78 $_{-0.10}^{+0.11}$&
7.02 $\pm 0.08$\\
      &$\tau$&
      $>10$ & $>10$ & $>10$ & $>10$ & $>10$ & $>10$ & $>10$ & $>10$ \\
      \hline
      Thin Absorber& $\mathrm{N_H}$ ($10^{23}$cm$^{-2}$)&
0.18 $_{-0.05}^{+0.04}$ &
0.15 $_{-0.06}^{+0.33}$ &
2.60 $_{-0.14}^{+0.10}$ &
1.52 $_{-0.07}^{+0.28}$ &
1.38 $_{-0.22}^{+0.11}$ &
1.47 $_{-0.10}^{+0.20}$ &
10 $\pm10$ & 
0.3 $^{+3.0}_{-0.1}$  \\
      &$\log\xi$& 
2.95 $_{-0.03}^{+0.15}$ &
2.95 $_{-0.03}^{+0.15}$ &
2.940 $\pm0.006$ &
2.937 $_{-0.004}^{+0.025}$ &
2.95 $\pm0.01$ &
2.936 $_{-0.006}^{+0.019}$ &
3.5$_{-0.4}^{+1.0}$ &
2.95 $^{+0.26}_{-0.09}$\\
      Edge &$\mathrm{E_{cut}}$ (keV)&
1.054 $_{-0.008}^{+0.007}$ &
1.039 $\pm0.006$ &
1.050 $_{-0.003}^{+0.005}$ &
1.053 $_{-0.003}^{+0.004}$ &
1.044 $\pm0.003$ &
1.048 $_{-0.002}^{+0.003}$ & --- & --- \\
      &$\tau$&
0.54 $_{-0.04}^{+0.05}$ &
0.68 $_{-0.05}^{+0.06}$ &
7.0 $_{-0.2}^{+0.4}$ &
6.1 $\pm0.2$ &
2.59 $_{-0.08}^{+0.09}$ &
5.5 $\pm0.2$ &
---&---  \\ 
      \hline
      Power law&Photon index&
2.80 $_{-0.05}^{+0.06}$ &
2.69 $\pm0.06$ &
3.18 $\pm0.06$ &
3.20 $\pm0.05$ &
3.12 $\pm0.06$ &
3.22 $\pm0.06$ &
1.53 $_{-0.16}^{+0.18}$ &
2.4 $\pm0.2$  \\
      &norm&
62 $_{-16}^{+19}$ &
47 $_{-14}^{+16}$ &
16.5 $_{-1.5}^{+1.6}$ &
18.3 $_{-1.4}^{+1.5}$ &
12.1 $_{-1.0}^{+1.2}$ &
14.6 $_{-1.2}^{+1.3}$ &
0.93 $_{-0.10}^{+0.13}$ &
2.5 $^{+1.0}_{-0.7}$ \\
      \hline
      MCD&$\mathrm{T_{in}}$ (eV)&
178 $\pm7$ &
170 $\pm7$ &
262 $_{-12}^{+11}$ &
245 $\pm9$ &
185 $\pm7$ &
222 $\pm9$ &
148 $_{-9}^{+10}$ &
103 $_{-7}^{+8}$ \\
     &norm&
8 $_{-3}^{+4} \times10^3$&
1.1 $_{-0.4}^{+0.6} \times10^4$&
1.4 $_{-0.2}^{+0.3} \times10^2$&
2.2 $_{-0.3}^{+0.4} \times10^2$&
1.2 $_{-0.2}^{+0.3} \times10^3$&
3.7 $_{-0.6}^{+0.8} \times10^2$&
8.6 $_{-3.8}^{+6.9} \times10^2$&
6 $^{+12}_{-4}\times10^4$ \\
      \hline
      Partial covering&covering factor&
0.92 $^{+0.02}_{-0.03}$ &
0.92 $^{+0.02}_{-0.03}$ &
0.60 $\pm0.01$ &
0.53 $\pm0.01$ &
0.62 $^{+0.02}_{-0.02}$ &
0.55 $\pm0.01$ &
0.82 $\pm0.04$ &
0.72 $^{+0.08}_{-0.07}$ \\
      \hline
      \multicolumn{2}{r}{Reduced chisq (d.o.f)}&
      1.10 (620) & 1.51 (586) & 1.26 (702) & 1.36 (761) & 1.41 (654) & 1.09 (669) & 1.10 (259) & 1.11 (582)  \\
      \hline
    \end{tabular}
    }
  \end{center}
\end{table}
\end{landscape}

 \begin{landscape}
 \begin{table}
  \caption{
Results of the intensity-sliced spectral fitting. See the caption of Table 2 for explanation of the parameters.
  }\label{tab:slice:a}
  \begin{center}
  \scalebox{0.8}{
    \begin{tabular}{llllllllll}
      \hline
      &&XMM1&XMM2&XMM3&XMM4&XMM5&XMM6&XMM7&XMM8 \\
      \hline
      Interstellar absorption &$\mathrm{N_H}$ ($10^{21}$ cm$^{-2}$)&
1.0 $\pm0.3$ &
0.7 $\pm0.1$ &
0.7 $\pm0.3$ &
0.4 $_{-0.4}^{+0.5}$ &
0.9 $\pm0.1$ &
0.3 $\pm0.2$&
0.61 $_{-0.07}^{+0.13}$ &
0.68 $\pm0.08$\\
      \hline
      Thick absorber &$\mathrm{N_H}$ ($10^{23}$cm$^{-2}$)&
13.6 $^{+2.7}_{-0.5}$&
25 $_{-11}^{+6}$ &
14.7 $_{-0.5}^{+9.9}$ &
5 $_{-2}^{+5}$ $\times10$&
9.0$_{-1.3}^{+8.5}$ &
30 $_{-15}$ &
17 $_{-2}^{+3}$ &
13.4 $_{-0.9}^{+9.0}$ 
  \\
      &$\mathrm{\log \xi}$&
0.35 $^{+0.06}_{-0.05}$&
0.1 $_{-0.1}^{+2.3}$ &
0.36 $_{-0.08}^{+0.09}$ &
0.2 $_{-0.2}^{+1.3}$ &
1.5 $_{-1.5}^{+0.7}$ &
0.3 $_{-0.3}^{+2.4}$ &
0.6 $_{-0.1}^{+1.0}$ &
0.4 $_{-0.4}^{+1.3}$ 
\\
      Edge &$\mathrm{E_{cut}}$ (keV)&
6.6 $\pm0.1 $&
7.0 $_{-0.2}^{+0.5} $&
7.1 $\pm0.3$&
---&
6.8 $\pm0.2$&
---&
6.93 $_{-0.09}^{+0.08}$ &
7.0 $_{-0.1}^{+0.3}$ \\
      &$\tau$&$>10$&$>10$&$>10$&---&$>10$&---&$>10$&$>10$\\
      \hline
      Thin Absorber & $\mathrm{N_H}$ ($10^{23}$cm$^{-2}$)&
0.8 $_{-0.3}^{+0.4}$&
23 $_{-20}$ &
1.9 $_{-0.4}^{+0.5}$ &
2.2 $\pm1.0$&
7 $_{-5}$ $\times10$ &
4 $_{-2}^{+5}$ $\times10$&
2.6 $_{-0.7}^{+0.5}$ &
0.5 $_{-0.2}^{+0.7}$
    \\
      &$\log\xi$&
2.82 $_{-0.09}^{+0.07}$&
3.4 $_{-0.2}^{+0.3}$&
2.83 $\pm0.04$ &
2.82 $_{-0.07}^{+0.06}$&
3.40 $_{-0.16}^{+0.04}$ &
3.3 $_{-0.2}^{0.1}$ &
3.01 $_{-0.04}^{+0.02}$&
2.97 $_{-0.05}^{0.11}$ 
 \\
      Edge &$\mathrm{E_{cut}}$ (keV)&
---&
1.019 $_{-0.006}^{+0.009} $&
1.073 $_{-0.009}^{+0.016}$ &
1.05 $\pm0.03$&
1.05 $\pm0.01$ &
1.03 $\pm0.01$ &
1.045 $_{-0.007}^{+0.006}$ &
1.063 $\pm0.006$ \\
      &$\tau$&
      ---  &
2.1 $_{-0.3}^{+0.4} $&
1.6 $\pm0.2 $&
1.2 $\pm0.4$&
1.8 $_{-0.3}^{+0.4}$ &
4.2 $_{-0.6}^{+0.8}$ &
1.9 $\pm0.2$&
1.5 $\pm0.2$ \\ 
      \hline
      Power law&Photon index&
1.98 $_{-0.10}^{+0.11} $&
2.65 $\pm0.05$ &
2.55 $\pm0.09$ &
1.85 $\pm0.13$&
2.70 $\pm0.05$ &
2.42 $_{-0.16}^{+0.15}$ &
2.69 $\pm0.04$&
2.78 $\pm0.04$\\
      &norm&
4.3 $_{-0.8}^{+1.0}$ &
8.5 $_{-0.5}^{+0.6} $&
11.1 $_{-1.9}^{+2.4} $&
4.7 $_{-1.5}^{+2.6}$&
12.1 $_{-1.1}^{+1.3} $&
6.2 $_{-0.9}^{+1.1} $&
10.6 $\pm0.9$&
12.6 $_{-0.8}^{+0.9}$ \\
      \hline
      MCD&$\mathrm{T_{in}}$ (eV)&
125 $_{-6}^{+7}$ &
189 $\pm8$&
145 $_{-10}^{+12} $&
138 $_{-14}^{+17}$&
166 $_{-9}^{+10}$ &
225 $\pm17$&
192 $_{-13}^{+8}$&
198 $\pm7$   \\
   &norm&
1.4 $_{-0.3}^{+0.4}$ $\times10^4$&
7.4 $\pm0.4$ $\times10^2$&
4.3 $_{-0.5}^{+0.6}$ $\times10^3$&
4 $_{-2}^{+6}$  $\times10^3$&
2.0 $\pm0.1$ $\times10^3$&
3.5 $_{-0.3}^{+0.4}$ $\times10^2$&
5.9 $_{-1.2}^{+3.1}$ $\times10^2$&
7.3 $\pm0.4$ $\times10^2$\\
      \hline
      Partial covering&covering factor&
0.90 $\pm0.02 $ &
0.63 $\pm0.02 $ &
0.90 $\pm0.01$ &
0.93 $^{+0.02}_{-0.03} $ &
0.57 $\pm0.03 $ &
0.60 $\pm0.03 $ &
0.69 $\pm0.02 $ &
0.65 $\pm0.02$\\
&&
0.86 $\pm0.03$ &
0.51 $\pm0.02$ &
0.71 $\pm0.04 $ &
0.88 $^{+0.04}_{-0.05} $ &
0.43 $\pm0.04$ &
0.53 $^{+0.03}_{-0.03} $ &
0.55 $^{+0.02}_{-0.03} $ &
0.50 $\pm0.02$ \\
&&
0.80 $^{+0.04}_{-0.05} $ &
0.40 $\pm0.03$ &
0.63 $\pm0.05$ &
0.80 $^{+0.07}_{-0.08} $ &
0.31 $\pm0.05$ &
0.46 $\pm0.04 $ &
0.44 $\pm0.03$ &
0.39 $\pm0.03$\\
&&
0.69 $^{+0.06}_{-0.07} $ &
0.26 $^{+0.03}_{-0.04} $ &
0.52 $^{+0.06}_{-0.07} $ &
0.65 $^{+0.11}_{-0.14} $ &
0.14 $^{+0.05}_{-0.06} $ &
0.18 $^{+0.05}_{-0.06} $ &
0.27 $^{+0.03}_{-0.04} $ &
0.26 $\pm0.03 $\\      \hline
      \multicolumn{2}{r}{Reduced chisq (d.o.f.)}&
1.15 (582) &
1.12 (1557) &
1.18 (804) &
1.28 (365) &
1.07 (1358) &
1.10 (768) &
1.18 (1755) &
1.10 (1720)     \\
      \hline
    \end{tabular}
    }
  \end{center}
\end{table}
\end{landscape}
\addtocounter{table}{-1}

\begin{landscape}
 \begin{table}
  \caption{
  {\it Continued.}
  }\label{tab:slice:b}
  \begin{center}
    \scalebox{0.8}{
    \begin{tabular}{llllllllll}
      \hline
      &&XMM9&XMM10&XMM11&XMM12&XMM13&XMM14&XMM15&Suzaku \\
      \hline
      Interstellar absorption &$\mathrm{N_H}$ ($10^{21}$cm$^{-2}$)&
0.69 $_{-0.09}^{0.10}$ &
0.64 $_{-0.10}^{0.11}$ &
0.47 $_{-0.09}^{0.09}$ &
0.58 $ \pm 0.08 $&
0.53 $\pm0.09 $&
0.55 $\pm0.09$&
1.0 $\pm0.3$&
1.3 $_{-0.7}^{+0.8}$\\
      \hline
      Thick absorber&$\mathrm{N_H}$ ($10^{23}$cm$^{-2}$)&
16 $_{-4}^{10}$ &
21 $_{-9}^{6}$ &
19 $_{-4}^{3}$ &
16 $_{-1}^{12}$ &
19 $_{-4}^{2}$ &
16.3 $_{-0.9}^{3.9}$ &
13.7 $_{-0.6}^{2.0}$ &
23 $_{-10}^{+6}$      \\
&$\mathrm{\log \xi}$&
0.14 $_{-0.14}^{0.12}$ &
0.8 $_{-0.8}^{1.6}$ &
0.10$_{-0.10}^{0.17}$ &
1.5 $_{-1.5}^{0.5}$ &
0.19 $_{-0.19}^{0.11}$ &
0.13$_{-0.13}^{1.38}$ &
0.36 $_{-0.07}^{0.08}$ &
0.1 $_{-0.1}^{+1.0}$ \\
      Edge&$\mathrm{E_{cut}}$ (keV)&
6.95 $_{-0.13}^{+0.12}$ &
6.98 $_{-0.17}^{+0.12} $&
6.95 $_{-0.08}^{+0.09} $&
7.07 $\pm0.13 $&
7.1 $_{-0.1}^{+0.2} $&
7.1 $_{-0.1}^{+0.3} $&
6.38 $_{-0.13}^{+0.16} $&
6.6  $\pm0.2 $\\
      &$\tau$&$>10$&$>10$&$>10$&$>10$&$>10$&$>10$&$>10$&$>10$\\
      \hline
      Thin absorber & $\mathrm{N_H}$ ($10^{23}$cm$^{-2}$)&
5.9 $_{-2.1}^{2.3}$ &
2.6 $_{-0.9}^{0.6}$ &
2.8 $\pm0.1$ &
2.6 $_{-0.5}^{0.1}$ &
3.2 $_{-0.4}^{0.5}$ &
2.6 $_{-0.5}^{0.3}$ &
10 $\pm10$ &
0.3 $_{-0.1}^{3.0}$      \\
&$\log\xi$&
3.12 $_{-0.04}^{0.02}$  &
3.08 $_{-0.04}^{0.03}$ &
2.937 $\pm0.005$ &
2.97 $_{-0.03}^{0.01}$ &
2.99 $\pm0.02$ &
2.98 $_{-0.03}^{0.02}$ &
3.5 $_{-0.4}^{+1.0}$&
2.95 $_{-0.09}^{+0.26}$ \\
      Edge &$\mathrm{E_{cut}}$ (keV)&
1.045 $_{-0.007}^{+0.005}$ &
1.041 $_{-0.008}^{+0.005} $&
1.058 $_{-0.005}^{+0.003} $&
1.062 $_{-0.003}^{+0.002}$ &
1.040 $_{-0.009}^{+0.003}$ &
1.052 $_{-0.003}^{+0.004} $&
---&
--- \\
      &$\tau$&
2.4 $_{-0.3}^{+0.4}$ &
2.5 $_{-0.3}^{+0.4} $&
$>$10 &
$>$10 &
$>$10 &
$>$10 &
---&
--- \\ 
      \hline
      Power law&Photon index&
2.78 $\pm0.05$ &
2.69 $\pm0.06$ &
3.03 $\pm0.06 $ &
3.11 $_{-0.06}^{+0.05} $ &
3.08 $\pm0.06$ &
3.17 $\pm0.06 $ &
1.10 $\pm0.10 $ &
1.89 $_{-0.10}^{+0.11}$\\
      &norm&
10.6 $\pm0.8 $ &
8.6 $_{-0.6}^{+0.7} $ &
18.0 $_{-1.6}^{+1.7} $ &
22.9 $_{-1.8}^{+1.9} $ &
17.8 $_{-1.5}^{+1.7} $ &
18.2 $_{-1.6}^{+1.7} $ &
1.7 $\pm0.3$ &
3.5 $_{-0.9}^{+1.2}$\\
\hline
      MCD&$\mathrm{T_{in}}$ (eV)&
194 $\pm8$ &
188 $\pm8 $ &
275 $_{-13}^{+12} $ &
272 $\pm11$ &
236 $\pm12$ &
248 $\pm11$ &
139 $_{-8}^{+9} $ &
109 $\pm7$    \\
&norm&
7.9 $\pm0.5$ $\times10^2$&
10.0 $\pm0.5$ $\times10^2$&
1.2 $\pm0.2$ $\times10^2$&
1.5 $\pm0.2$ $\times10^2$&
2.7 $_{-0.4}^{+0.5}$ $\times10^2$&
2.3 $_{-0.3}^{+0.4}$ $\times10^2$&
1.5 $_{-0.3}^{+0.4}$ $\times10^3$&
4 $_{-2}^{+7}$ $\times10^4$   \\
 \hline
      Partial covering&covering factor&
0.66 $\pm0.02$ &
0.730 $^{+0.013}_{-0.015} $ &
0.772 $\pm0.007 $ &
0.768 $\pm0.006 $ &
0.787 $\pm0.007 $ &
0.721 $\pm0.008$ &
0.89 $^{+0.02}_{-0.03} $ &
0.90 $_{-0.04}^{+0.03}$\\
&&
0.54 $\pm0.02 $ &
0.58 $\pm0.02$ &
0.663 $^{+0.009}_{-0.010} $ &
0.615 $^{+0.009}_{-0.010} $ &
0.677 $^{+0.010}_{-0.011} $ &
0.628 $^{+0.010}_{-0.011} $ &
0.84 $^{+0.03}_{-0.04} $ &
0.82 $_{-0.07}^{+0.08}$\\
&&
0.40 $\pm0.03$ &
0.47 $^{+0.02}_{-0.03} $ &
0.584 $^{+0.011}_{-0.012} $ &
0.500 $^{+0.012}_{-0.013} $ &
0.586 $\pm0.013 $ &
0.548 $^{+0.012}_{-0.013} $ &
0.78 $^{+0.04}_{-0.06} $ &
0.71 $_{-0.10}^{+0.07}$\\
&&
0.20 $^{+0.03}_{-0.04} $ &
0.33 $\pm0.03 $ &
0.429 $^{+0.015}_{-0.016} $ &
0.311 $^{+0.016}_{-0.017} $ &
0.409 $^{+0.017}_{-0.019} $ &
0.389 $^{+0.016}_{-0.017} $ &
0.62 $^{+0.07}_{-0.10} $ &
0.53$_{-0.17}^{+0.11}$\\  
  \hline
      \multicolumn{2}{r}{Reduced chisq (d.o.f)}&
1.24 (1497) &
1.32 (1387) &
1.33 (1641) &
1.49 (1823) &
1.50 (1515) &
1.20 (1616) &
1.65 (451) &
1.27 (629)    \\
      \hline
    \end{tabular}
    }
  \end{center}
\end{table}
\end{landscape}


\addtocounter{figure}{-1}
\begin{figure*}
\addtocounter{subfigure}{1}
\centering
\subfigure{
        \resizebox{8.5cm}{!}{\includegraphics[angle=270]{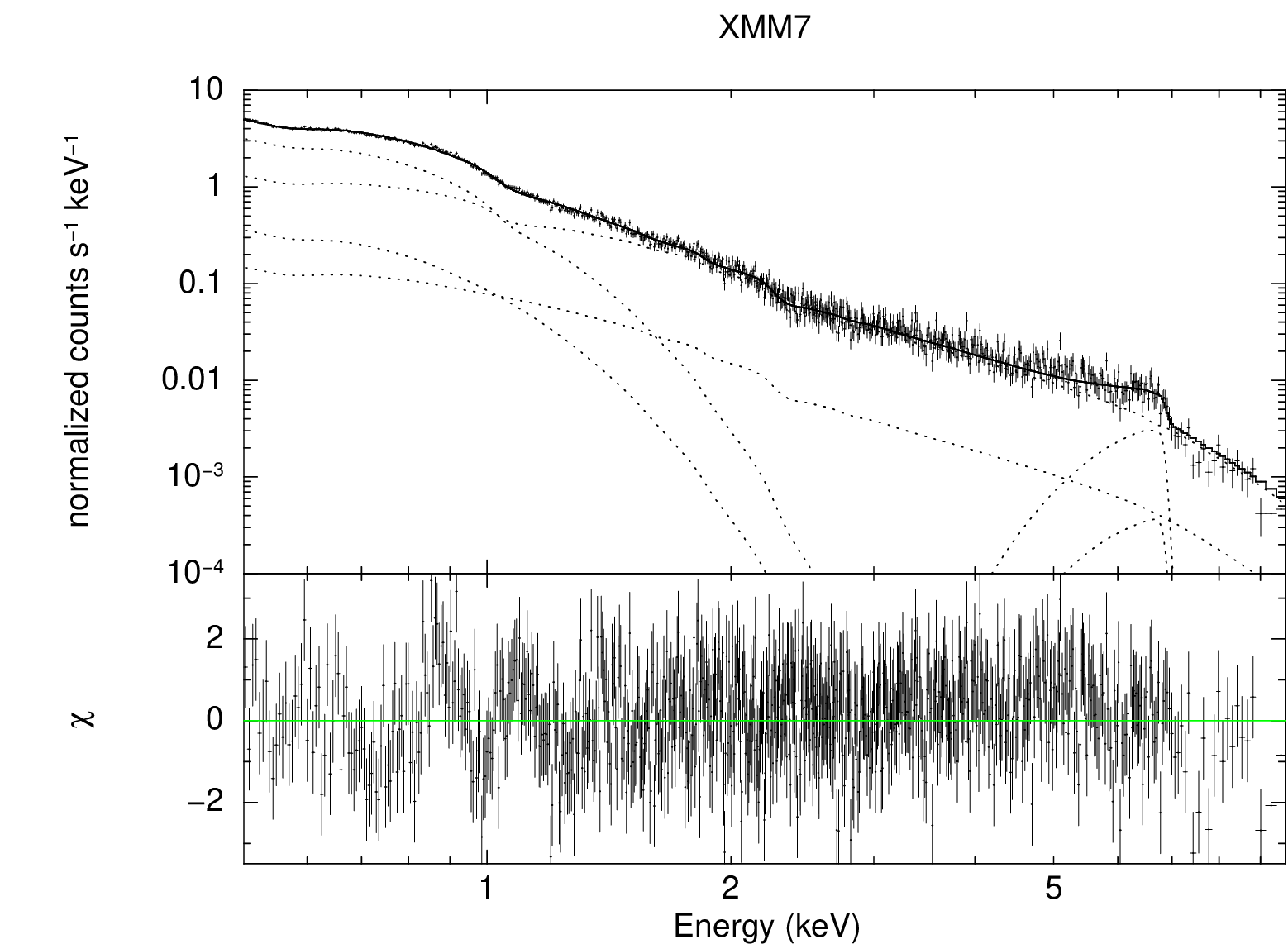}}
        \resizebox{8.5cm}{!}{\includegraphics[angle=270]{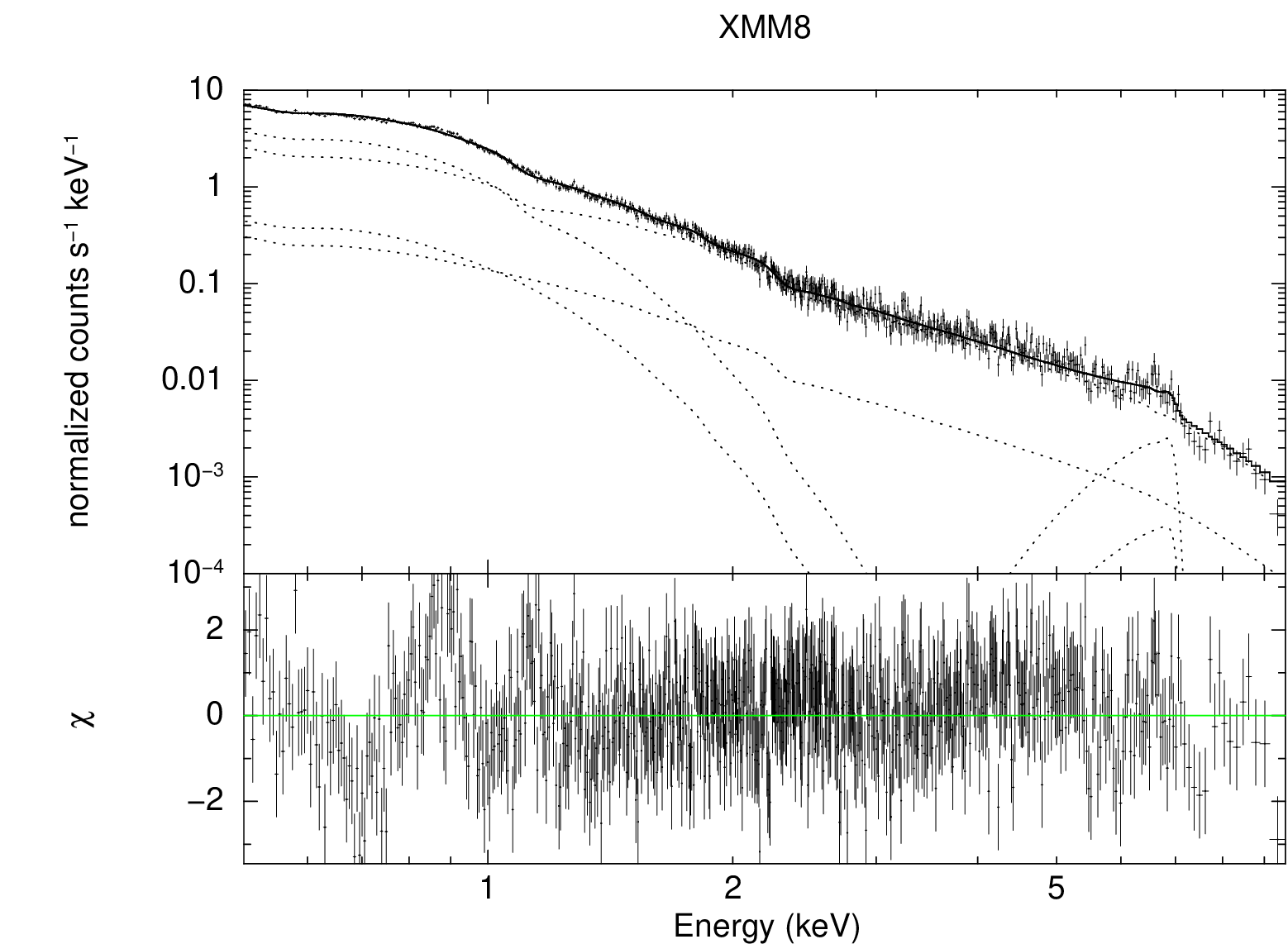}}
}
\subfigure{
        \resizebox{8.5cm}{!}{\includegraphics[angle=270]{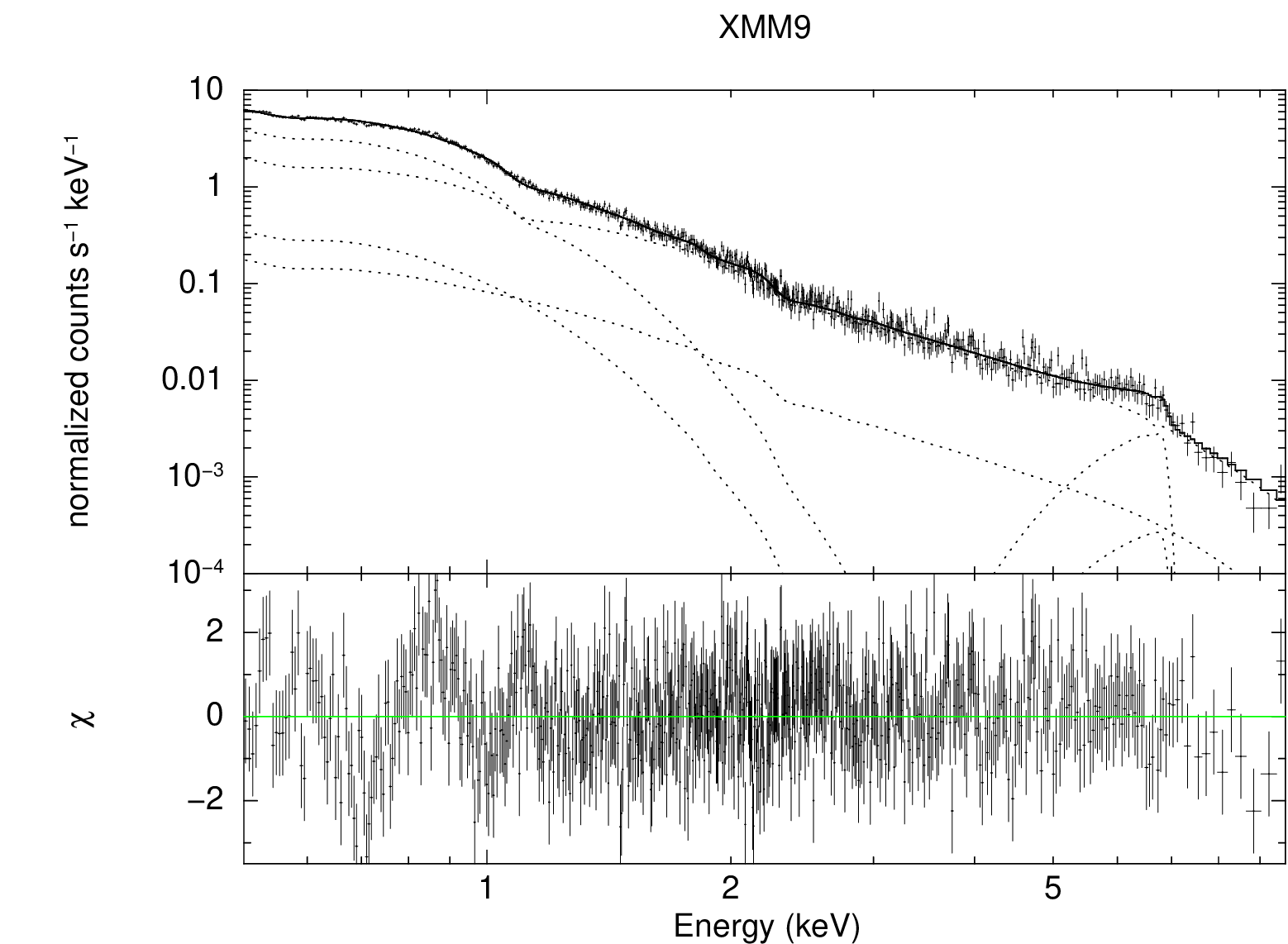}}
        \resizebox{8.5cm}{!}{\includegraphics[angle=270]{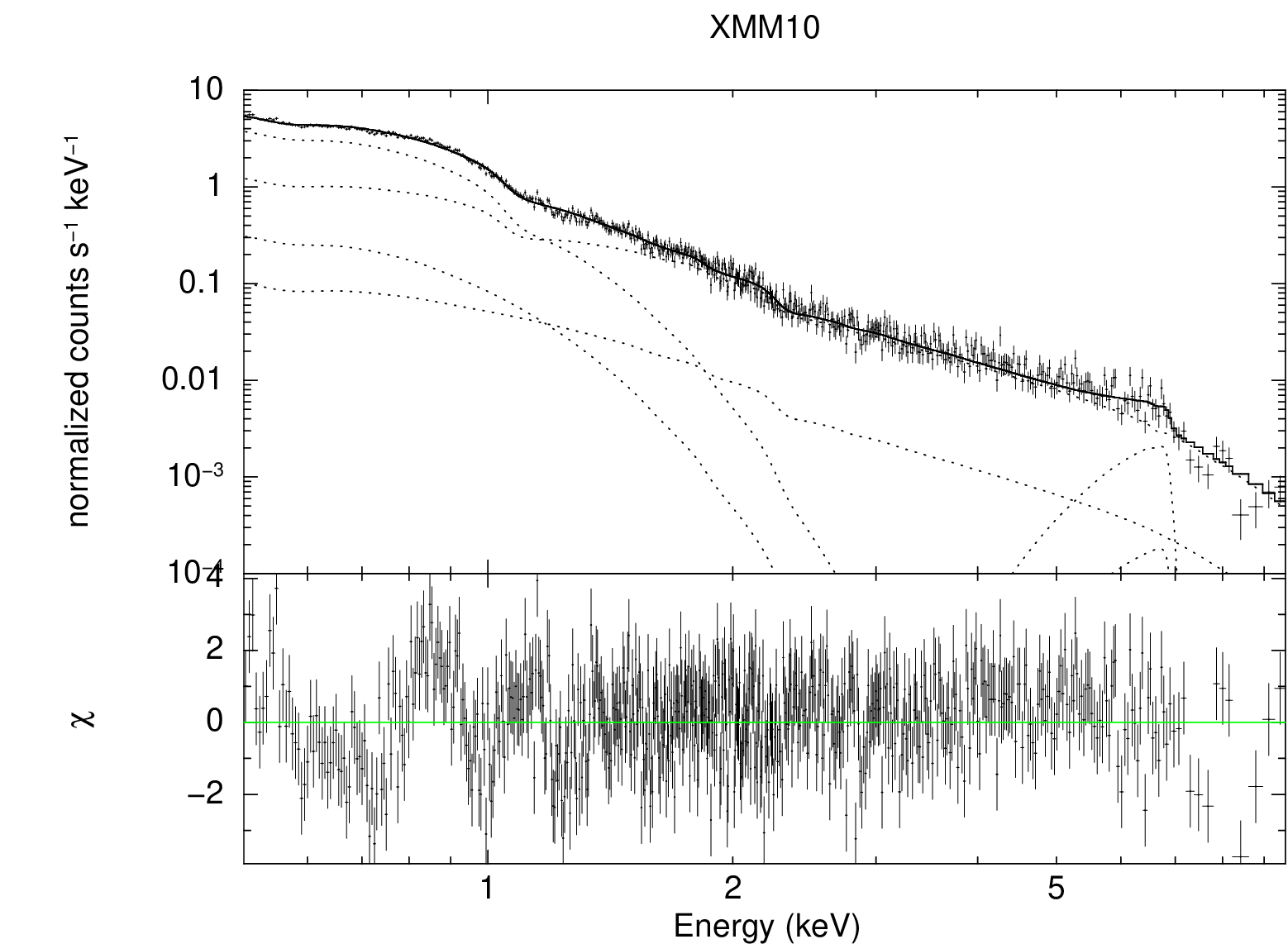}}
}
\subfigure{
        \resizebox{8.5cm}{!}{\includegraphics[angle=270]{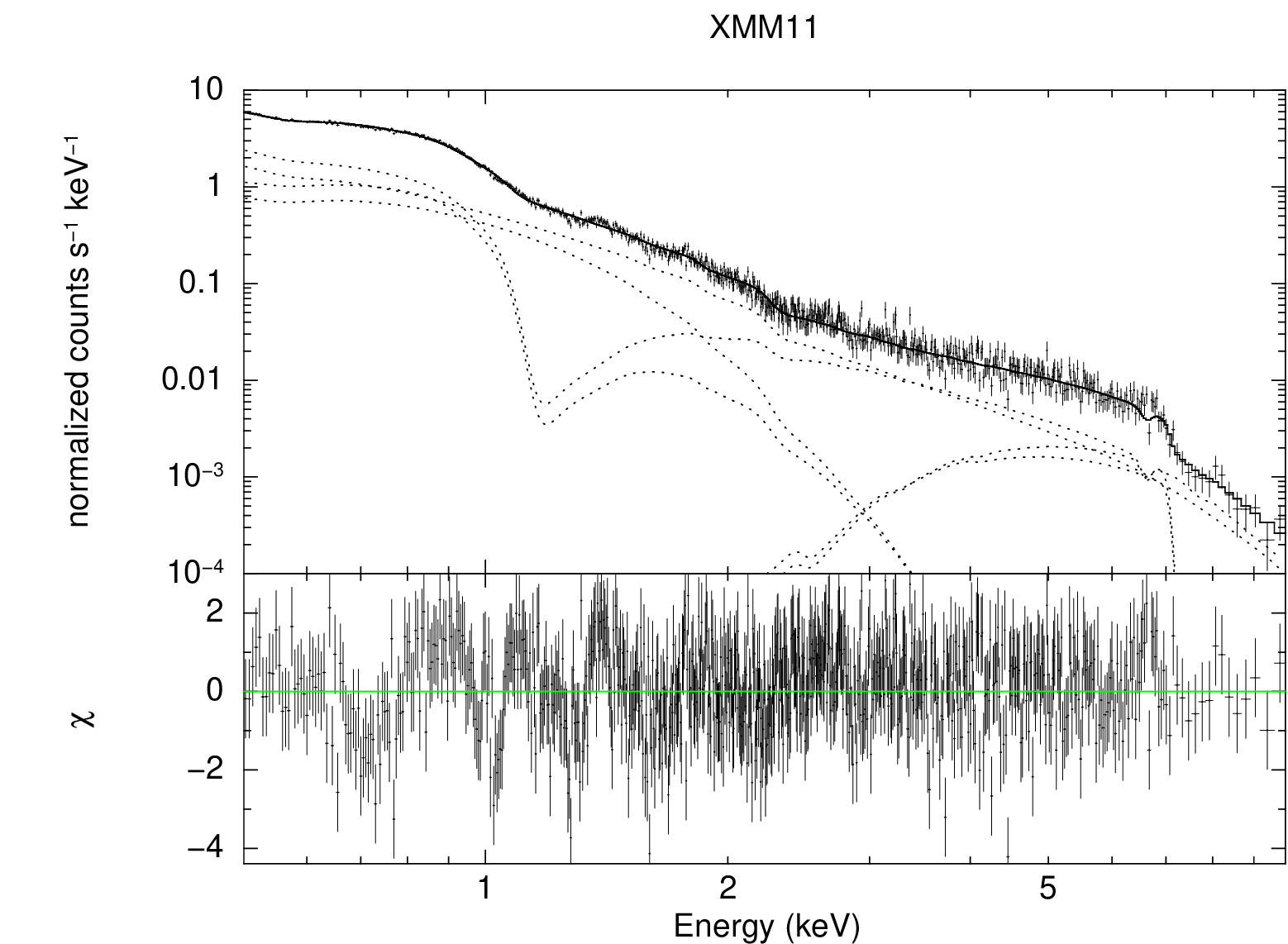}}
        \resizebox{8.5cm}{!}{\includegraphics[angle=270]{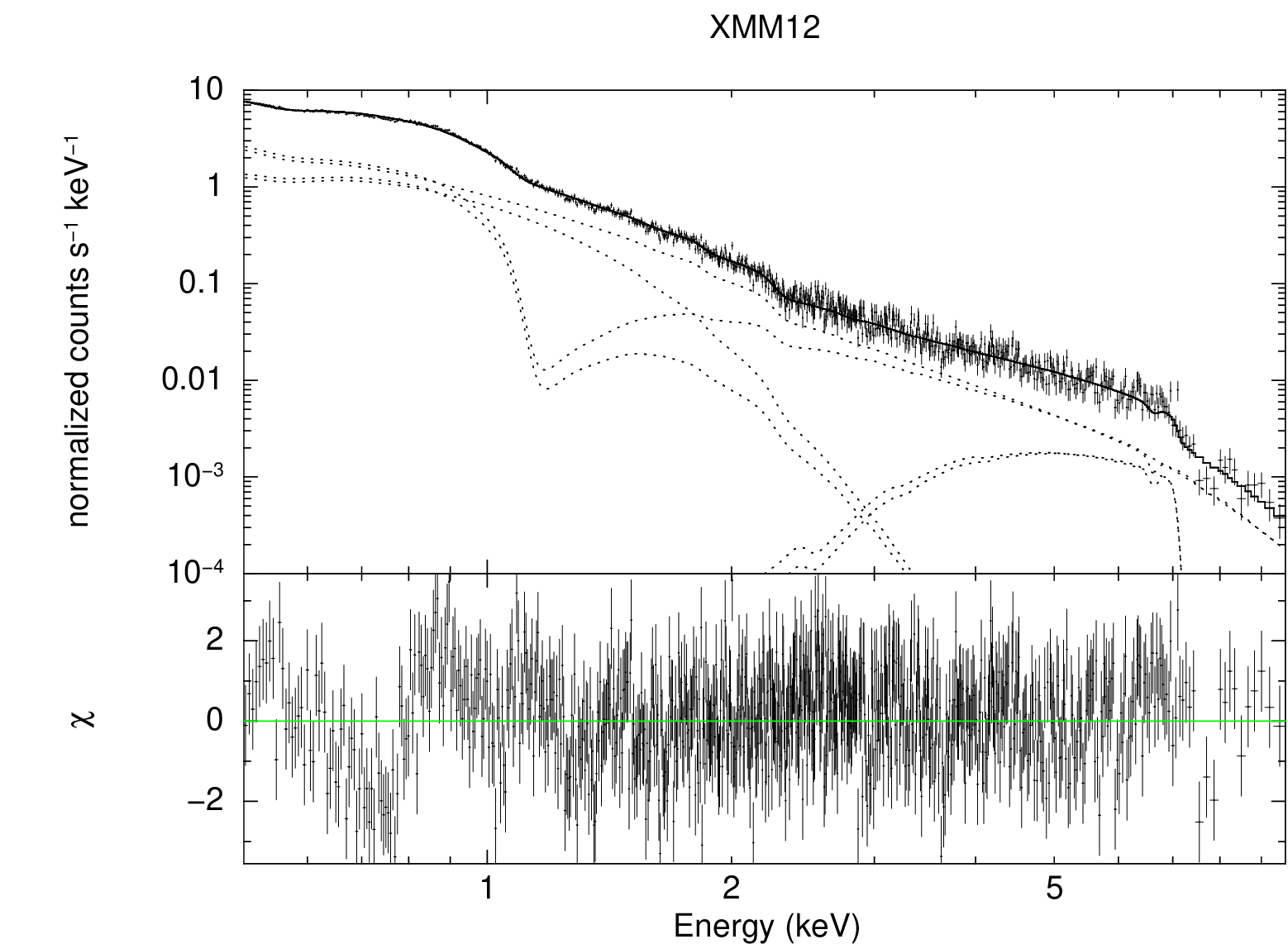}}
}
\caption{
        {\it Continued.}
}
\label{fig:average:b}
\end{figure*}
\addtocounter{figure}{-1}
\begin{figure*}
\addtocounter{subfigure}{1}
\centering
\subfigure{
        \resizebox{8.5cm}{!}{\includegraphics[angle=270]{1m.eps}}
        \resizebox{8.5cm}{!}{\includegraphics[angle=270]{1n.eps}}
}
\subfigure{
        \resizebox{8.5cm}{!}{\includegraphics[angle=270]{1o.eps}}
        \resizebox{8.5cm}{!}{\includegraphics[angle=270]{1p.eps}}
}
\subfigure{
}
\caption{
        {\it Continued.}
}
\label{fig:average:c}
\end{figure*}

\begin{figure*}
\centering
\subfigure{
        \resizebox{8cm}{!}{\includegraphics[angle=270]{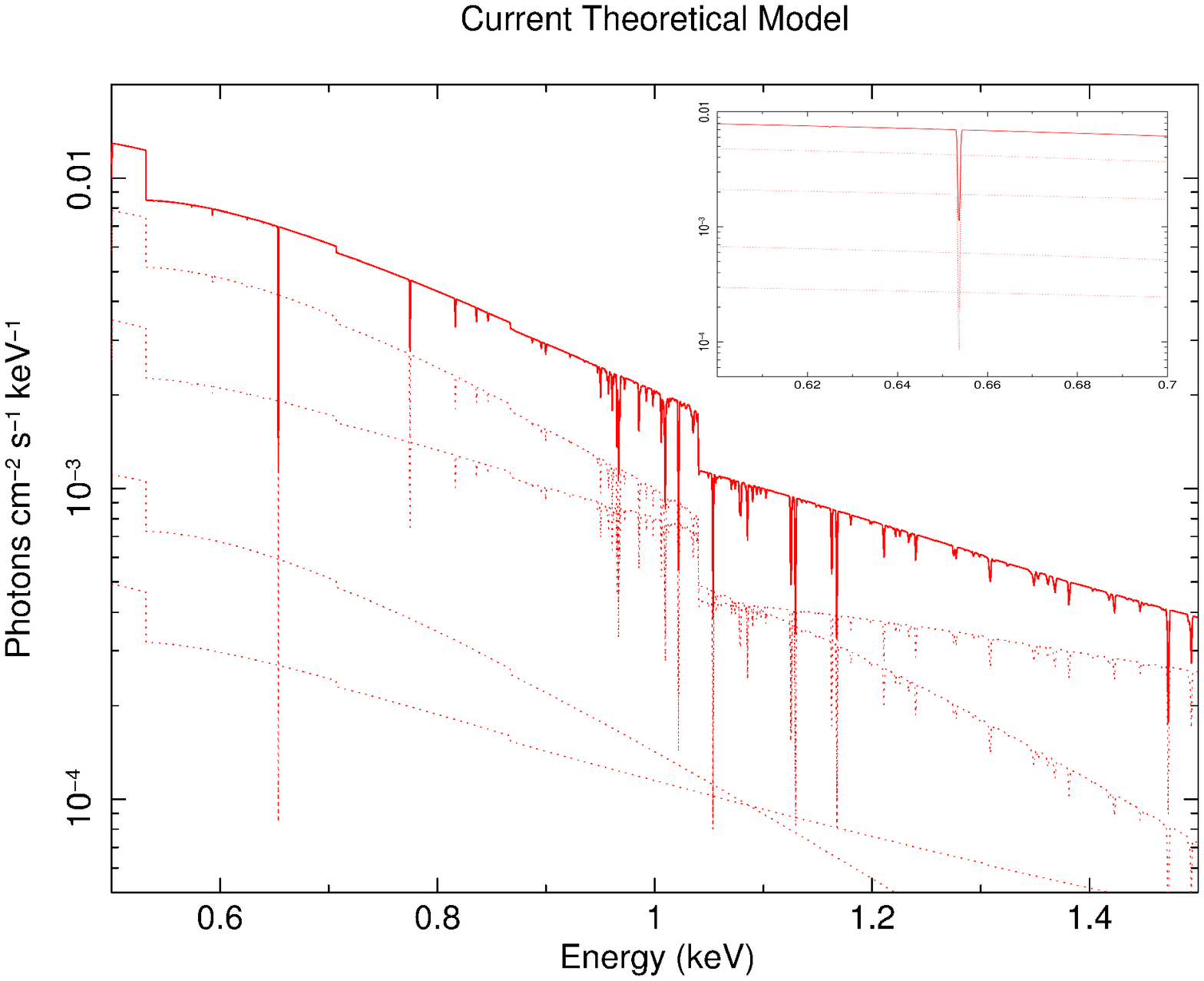}}
        \resizebox{8cm}{!}{\includegraphics[angle=270]{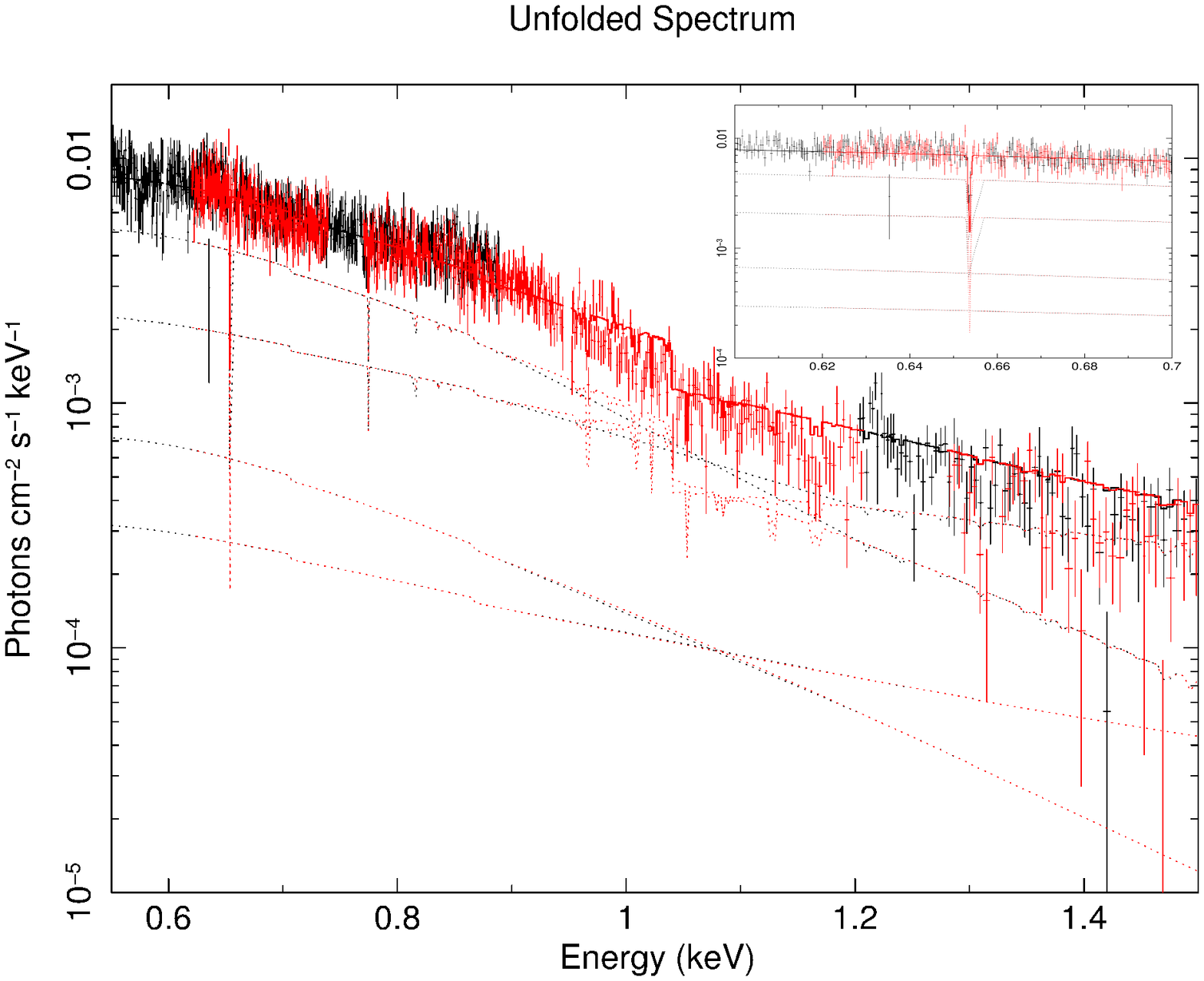}}
}
\caption
{RGS spectra combined in XMM7--10 and the best-fit VDPC model. 
The reduced chi-squares (and degree of freedoms) are 1.28 (1369).
The left figure shows the VDPC model, and the right figure shows the unfold spectra with best-fit model.
The black points and lines show the RGS1 spectrum and models, and the red ones show the RGS2 ones.
The insets are expansion of the region where the absorption line (the O VIII Lyman $\alpha$ line) is most prominent.}
\label{fig:rgs}
\end{figure*}


\begin{figure*}
\centering
\subfigure{
        \resizebox{14cm}{!}{\includegraphics[angle=270]{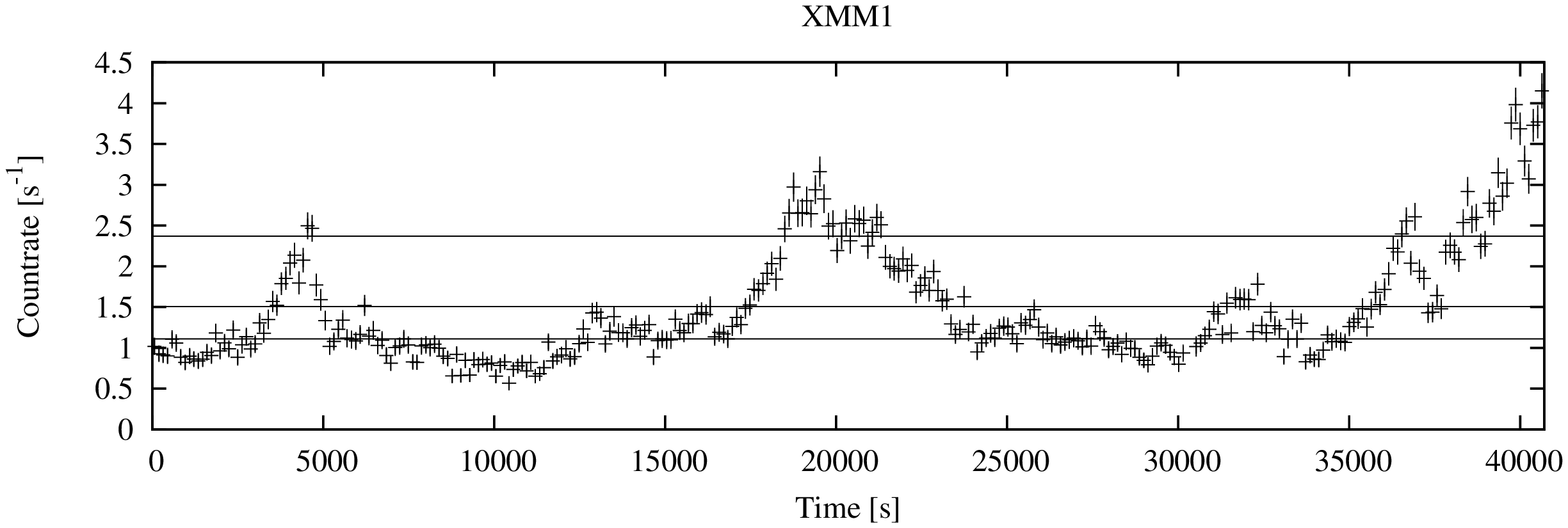}}
}
\subfigure{        
        \resizebox{14cm}{!}{\includegraphics[angle=270]{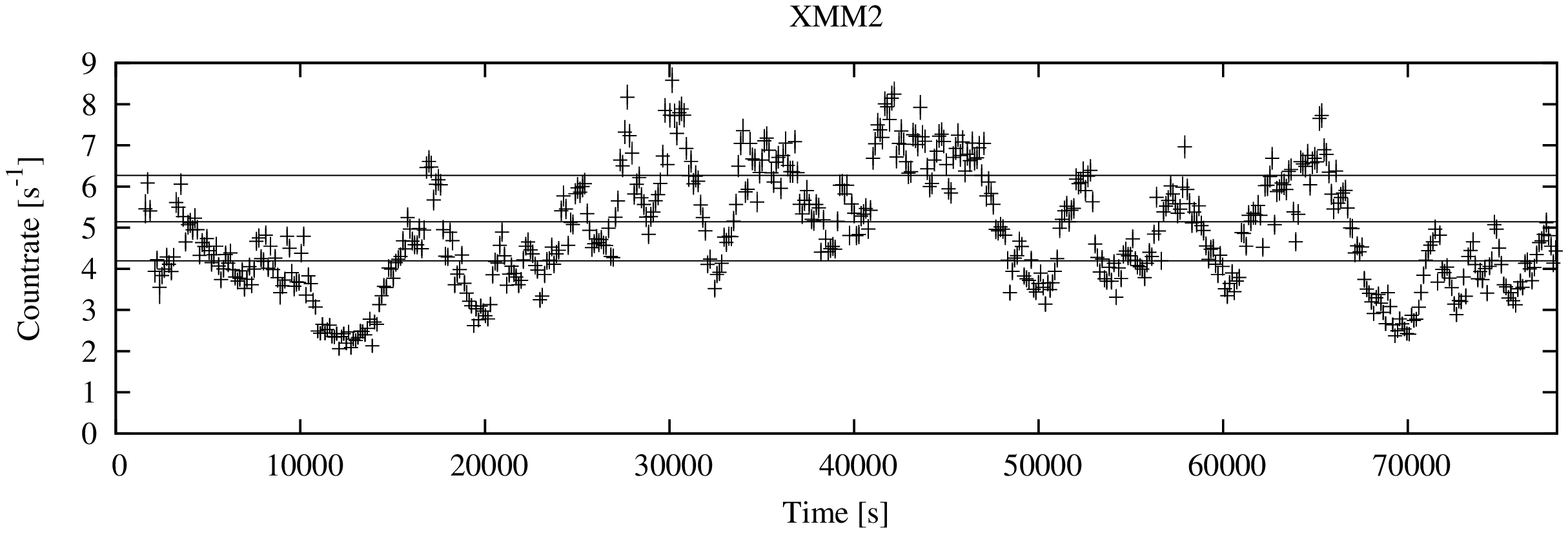}}
}
\subfigure{        
        \resizebox{14cm}{!}{\includegraphics[angle=270]{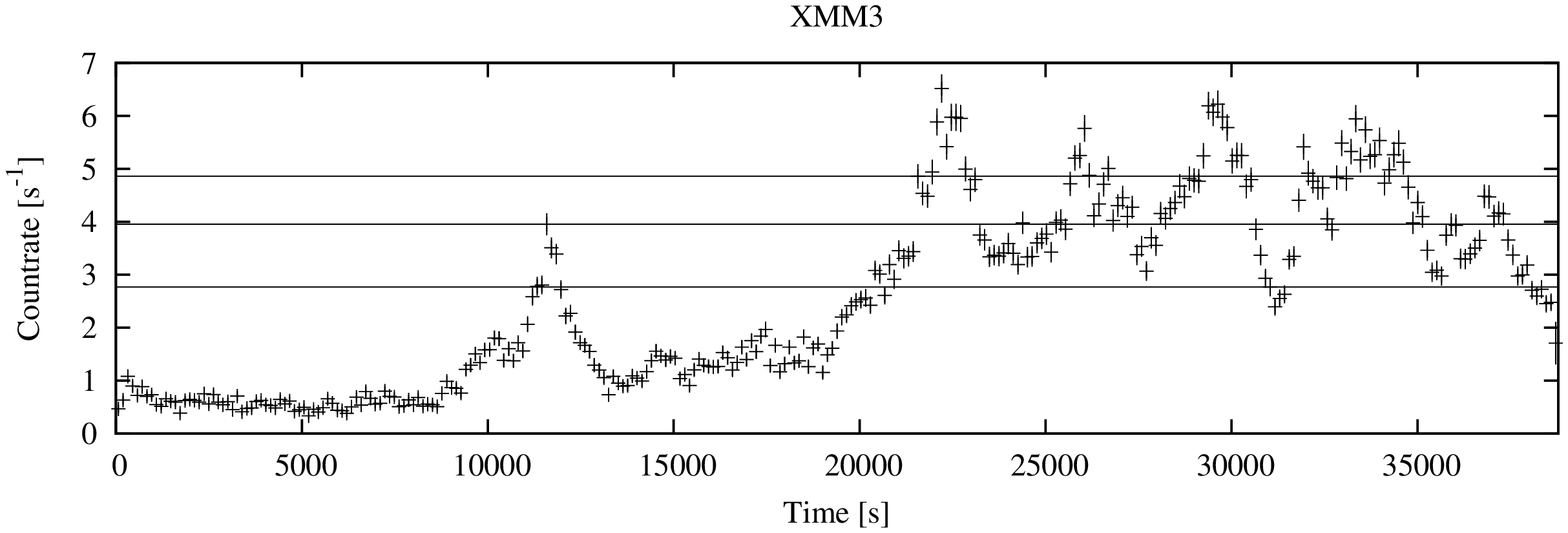}}
}
\subfigure{        
        \resizebox{14cm}{!}{\includegraphics[angle=270]{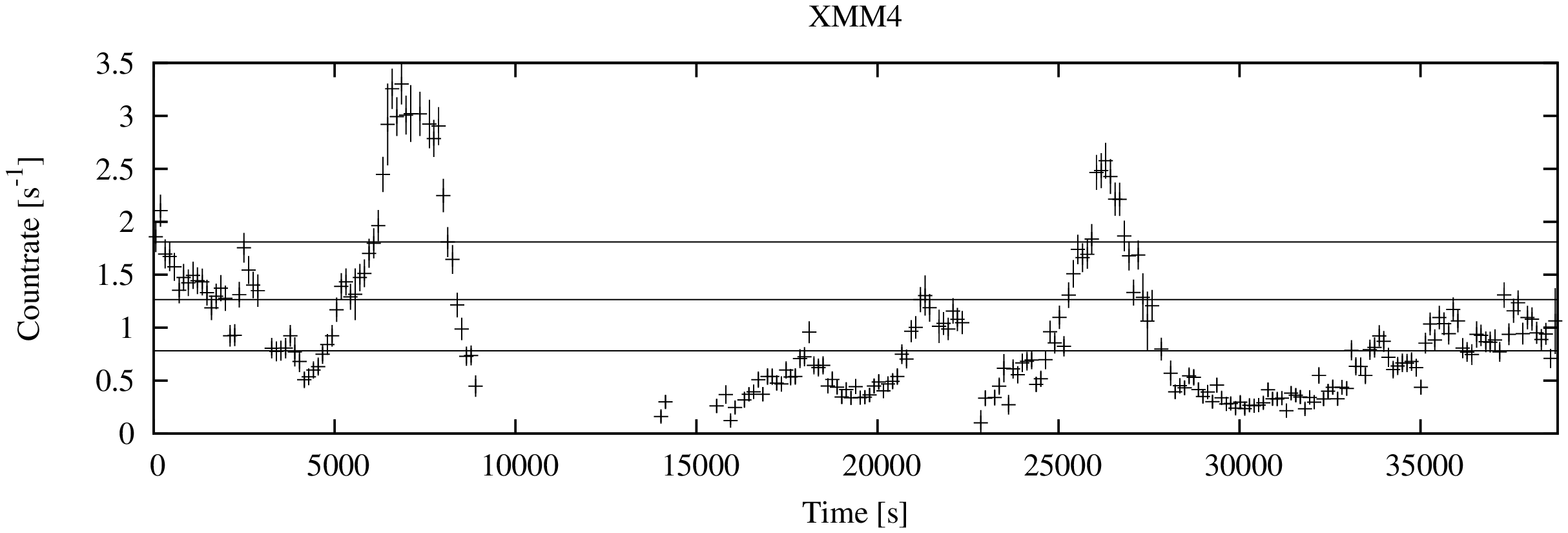}}
}
\caption{The light-curves in the 0.2--12.0 keV band and the thresholds of intensity-sliced spectra. 
The blanked regions show excluded high background period (in the XMM data) or unobserved period (in the Suzaku data.)
See the main text for the details.}
\label{fig:ltcrv:a}
\end{figure*}
\addtocounter{figure}{-1}
\begin{figure*}
\addtocounter{subfigure}{1}
\centering
\subfigure{        
        \resizebox{14cm}{!}{\includegraphics[angle=270]{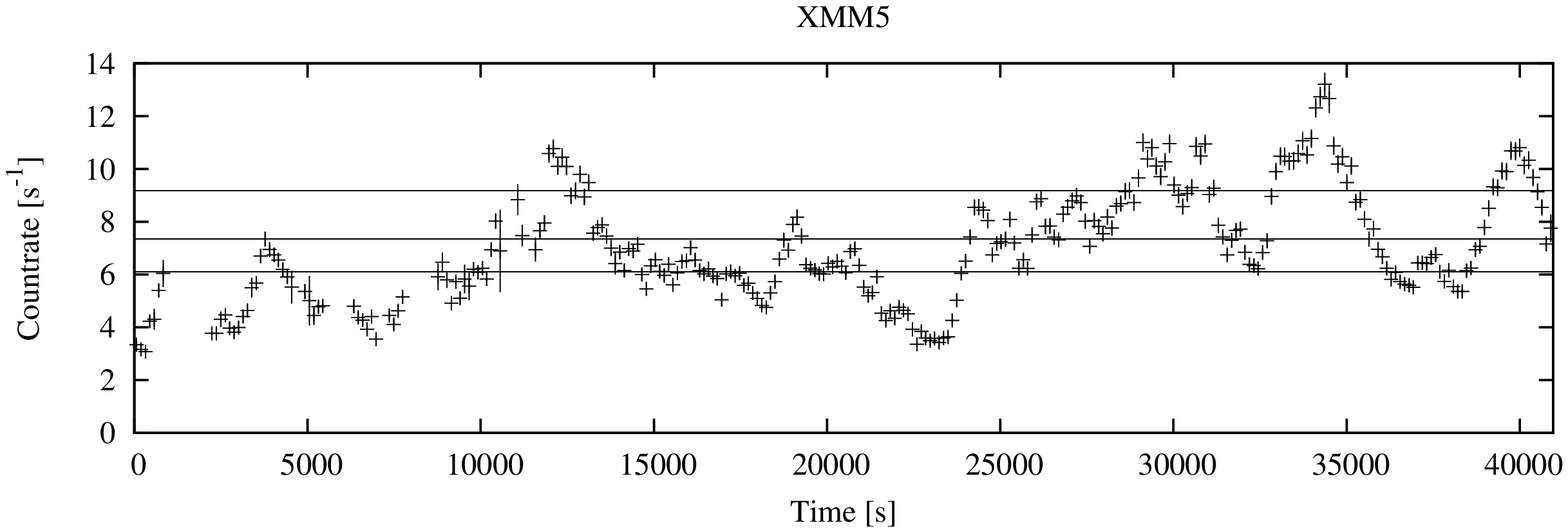}}
}
\subfigure{        
        \resizebox{14cm}{!}{\includegraphics[angle=270]{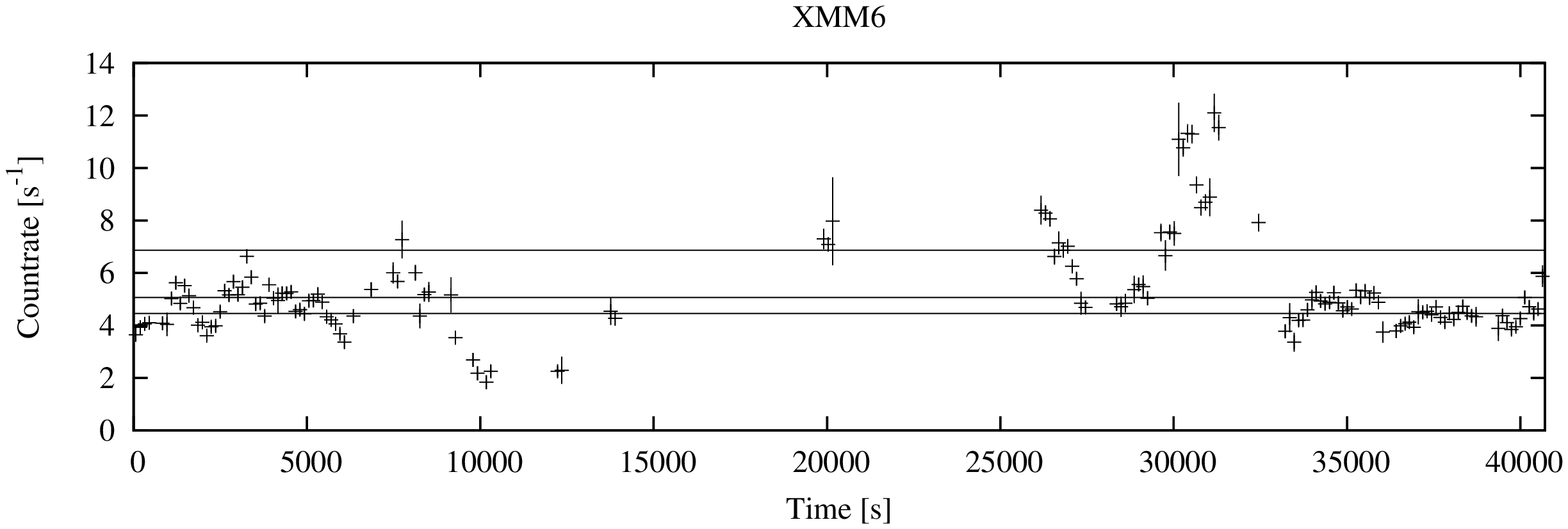}}
}
\subfigure{        
        \resizebox{14cm}{!}{\includegraphics[angle=270]{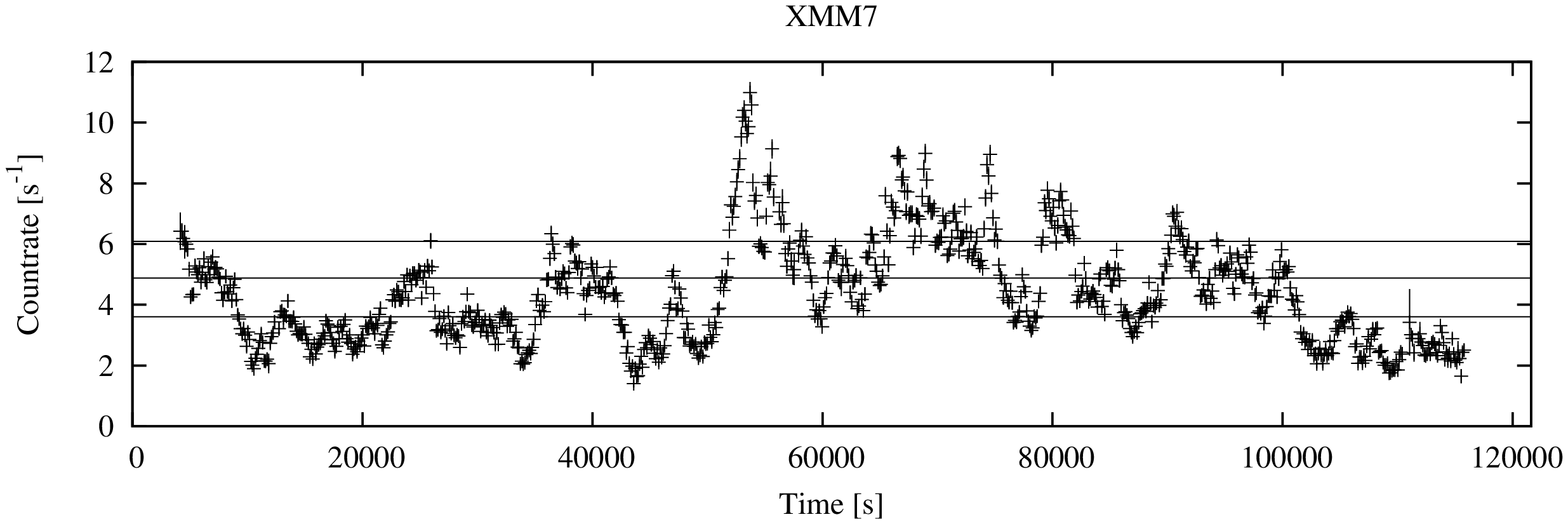}}
}
\subfigure{        
        \resizebox{14cm}{!}{\includegraphics[angle=270]{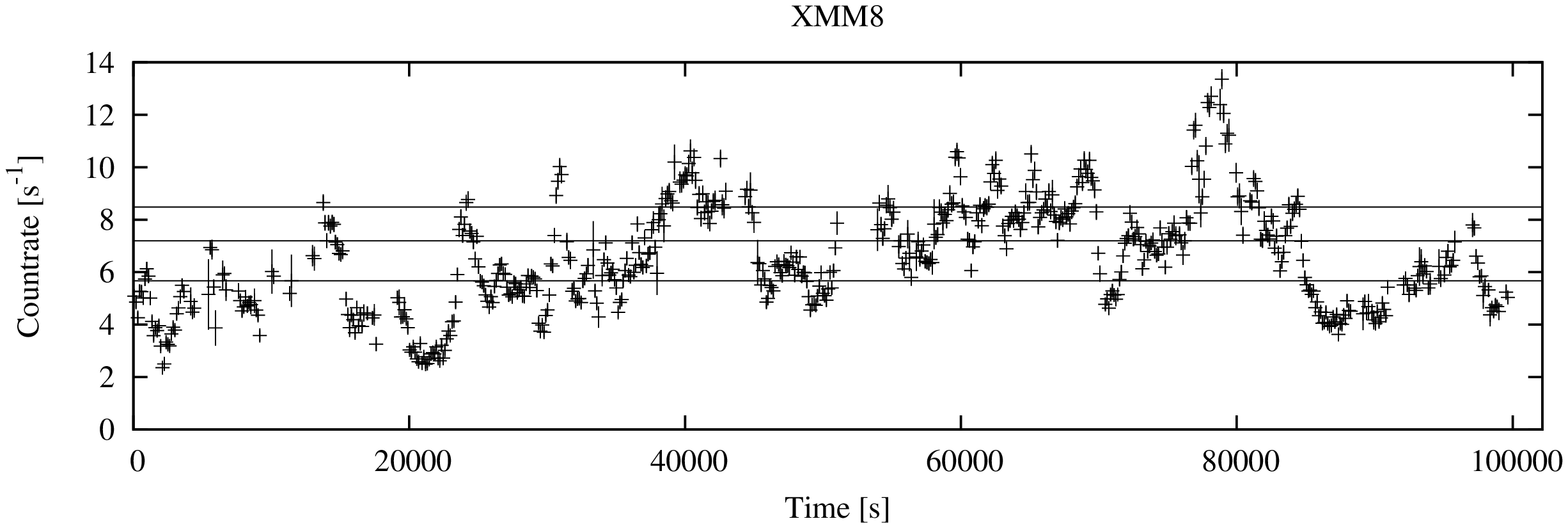}}
}
\caption{
        {\it Continued.}
}
\label{fig:ltcrv:b}
\end{figure*}
\addtocounter{figure}{-1}
\begin{figure*}
\addtocounter{subfigure}{1}
\centering
\subfigure{        
        \resizebox{14cm}{!}{\includegraphics[angle=270]{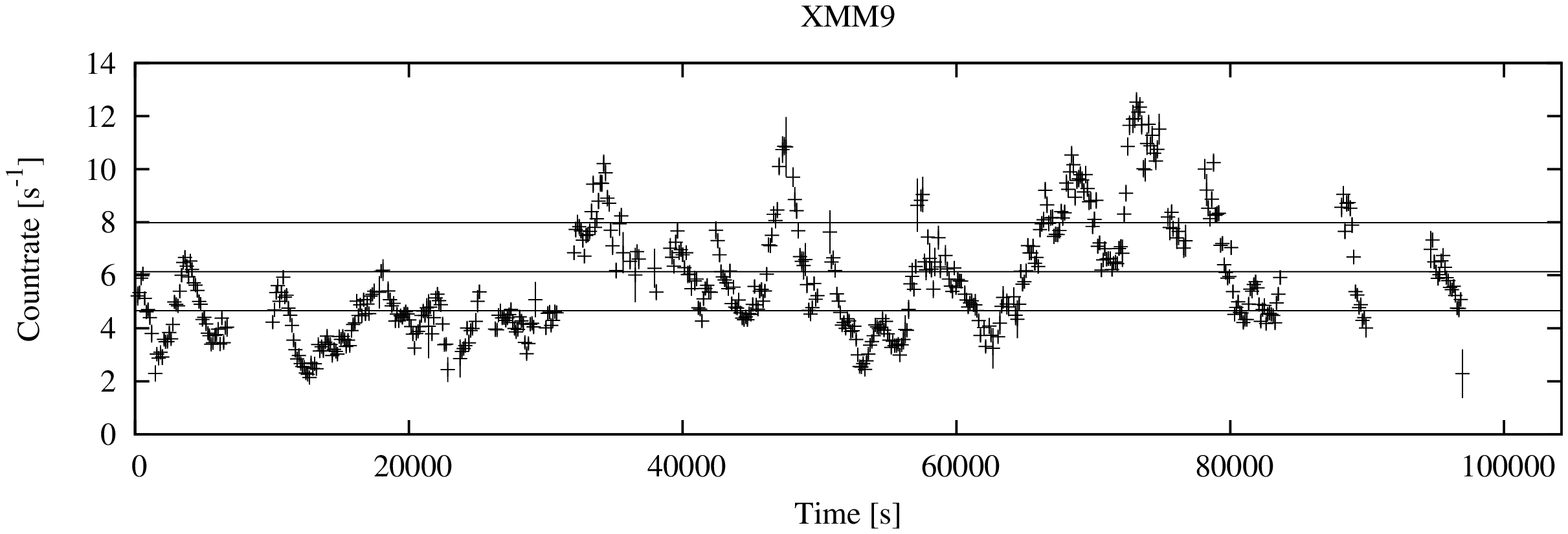}}
}
\subfigure{        
        \resizebox{14cm}{!}{\includegraphics[angle=270]{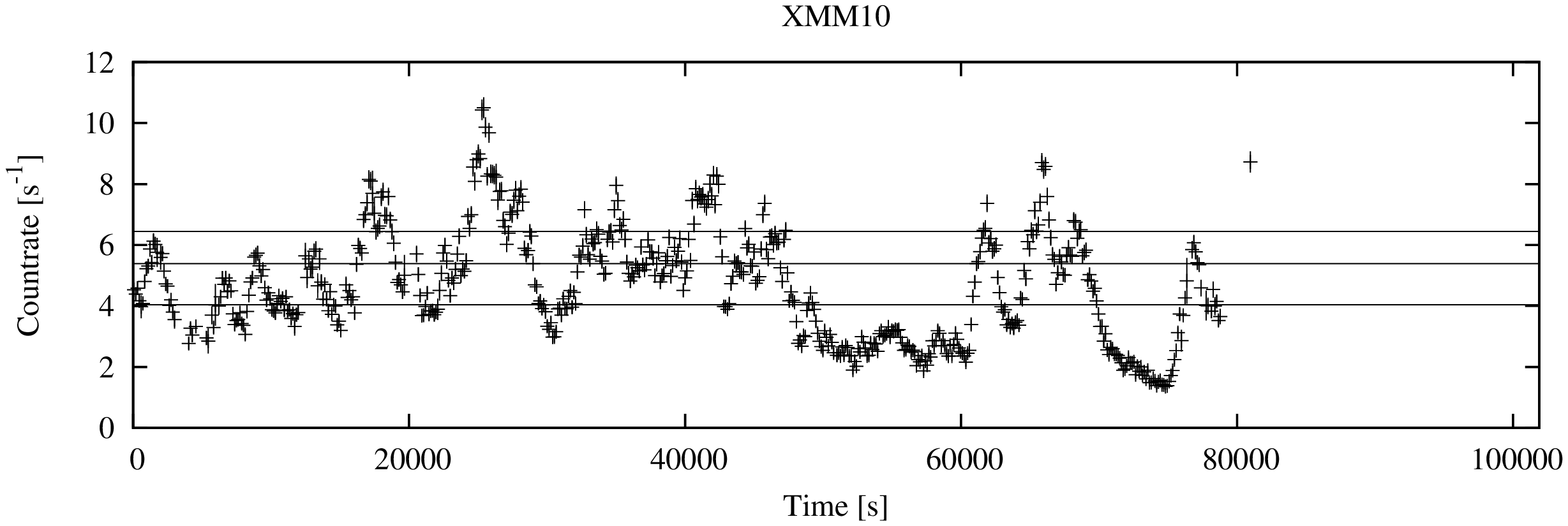}}
}
\subfigure{        
        \resizebox{14cm}{!}{\includegraphics[angle=270]{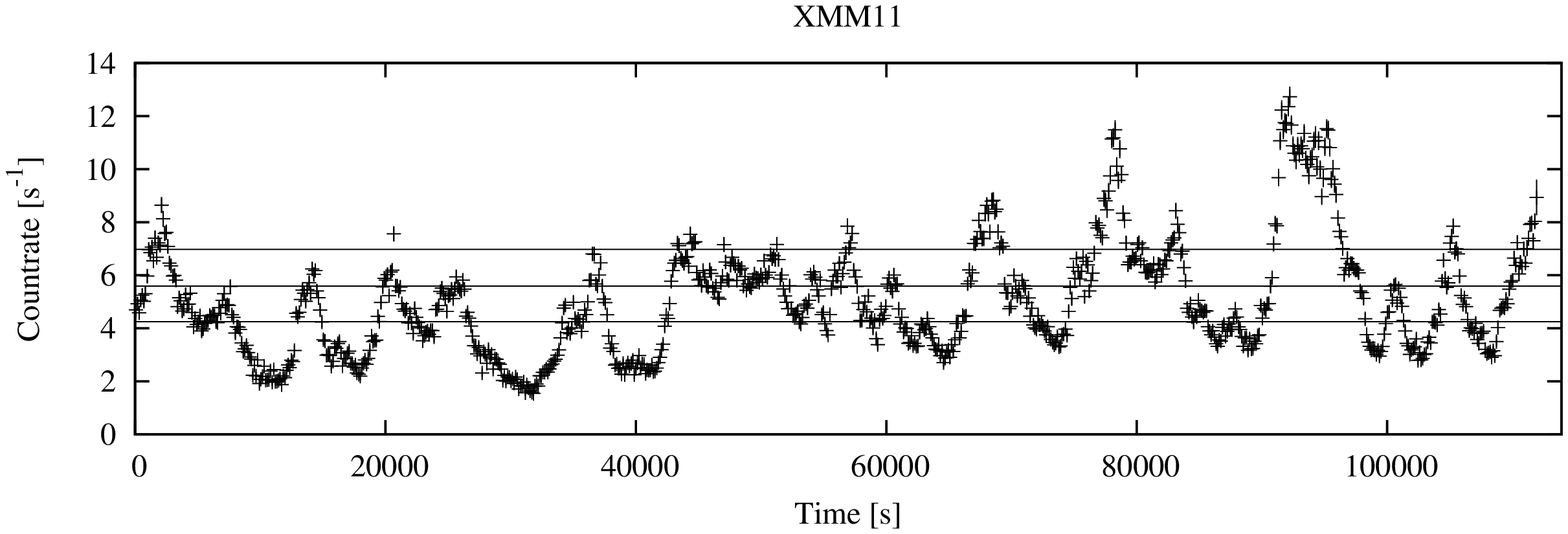}}
}
\subfigure{        
        \resizebox{14cm}{!}{\includegraphics[angle=270]{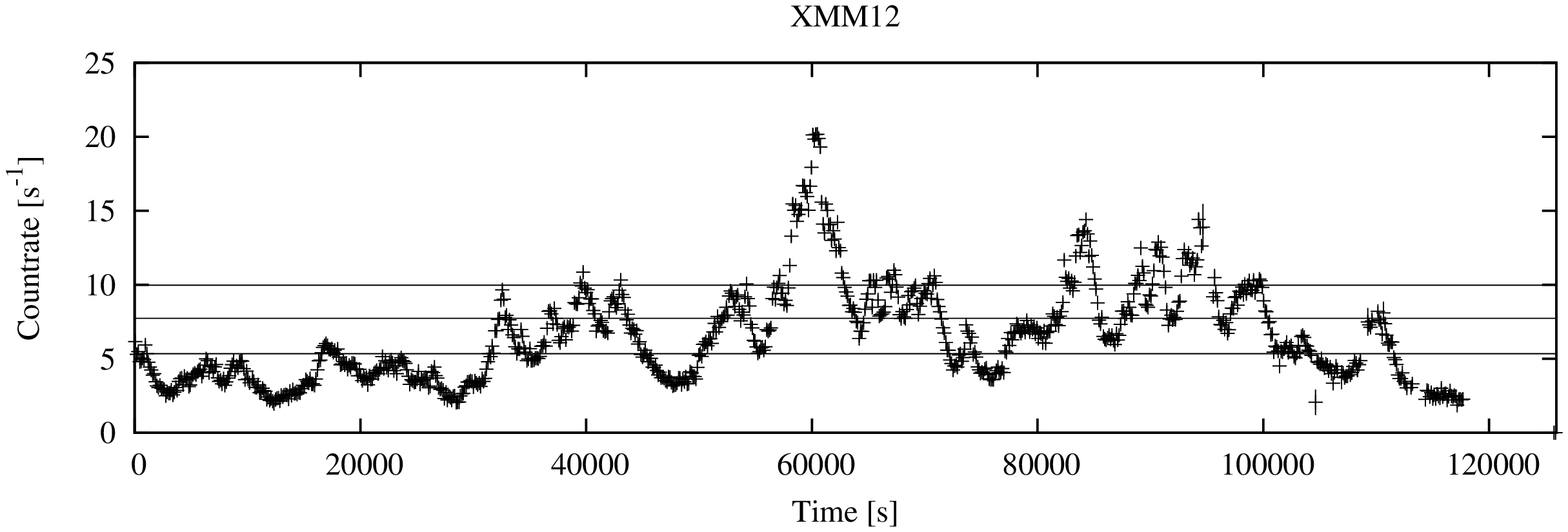}}
}
\caption{
        {\it Continued.}
}
\label{fig:ltcrv:c}
\end{figure*}
\addtocounter{figure}{-1}
\begin{figure*}
\addtocounter{subfigure}{1}
\centering
\subfigure{        
        \resizebox{14cm}{!}{\includegraphics[angle=270]{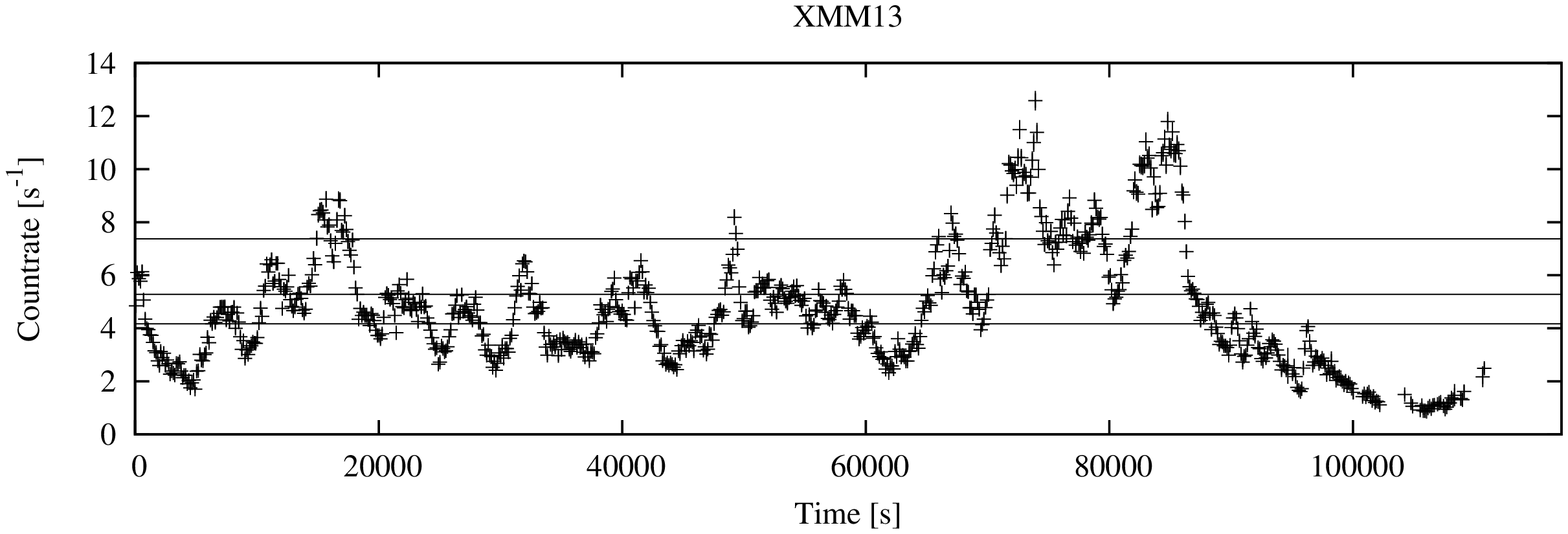}}
}
\subfigure{        
        \resizebox{14cm}{!}{\includegraphics[angle=270]{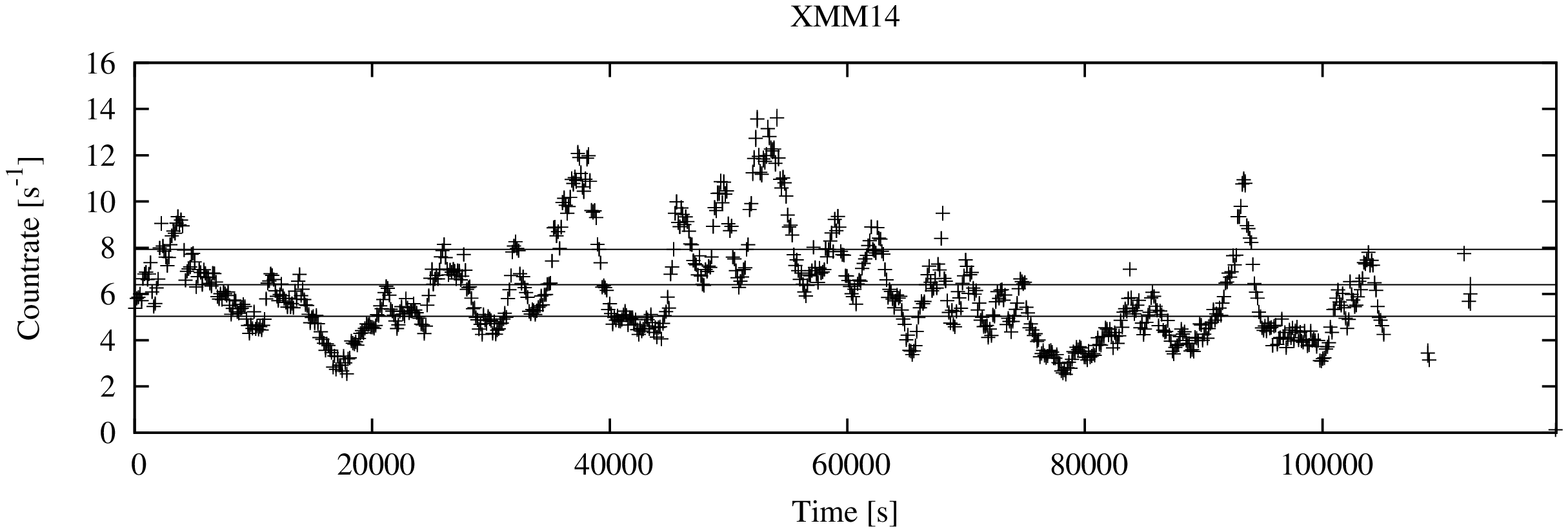}}
}
\subfigure{        
        \resizebox{14cm}{!}{\includegraphics[angle=270]{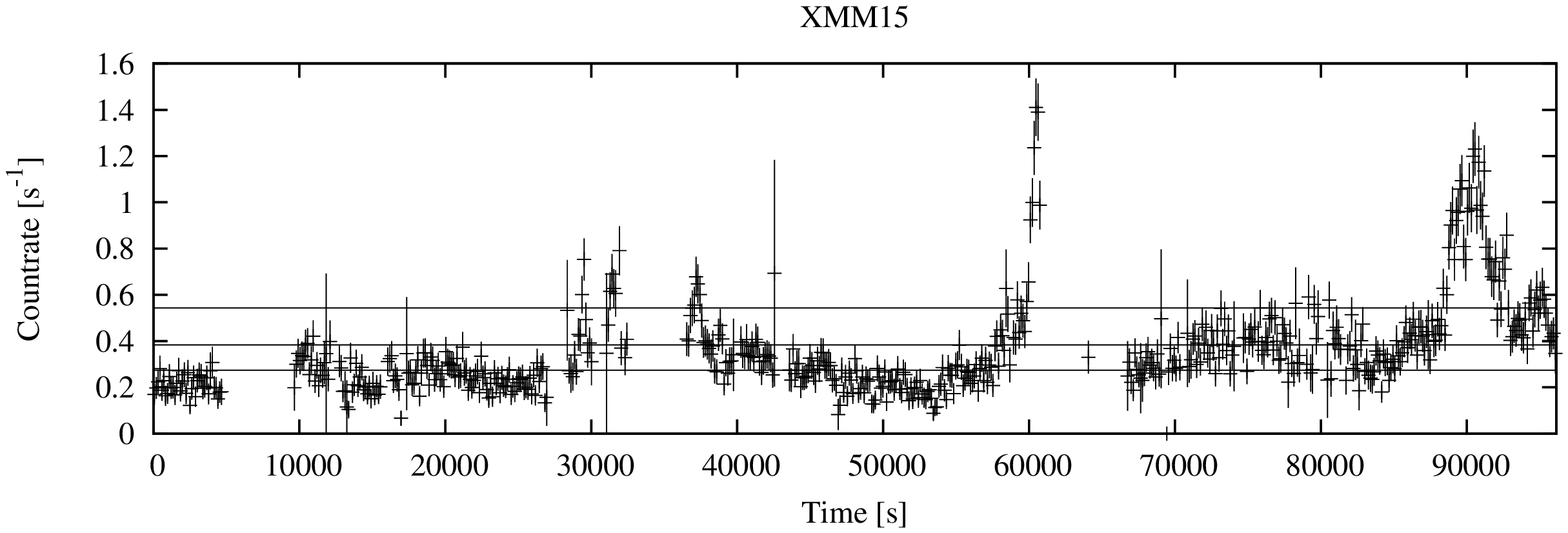}}
}
\subfigure{        
        \resizebox{14cm}{!}{\includegraphics[angle=270]{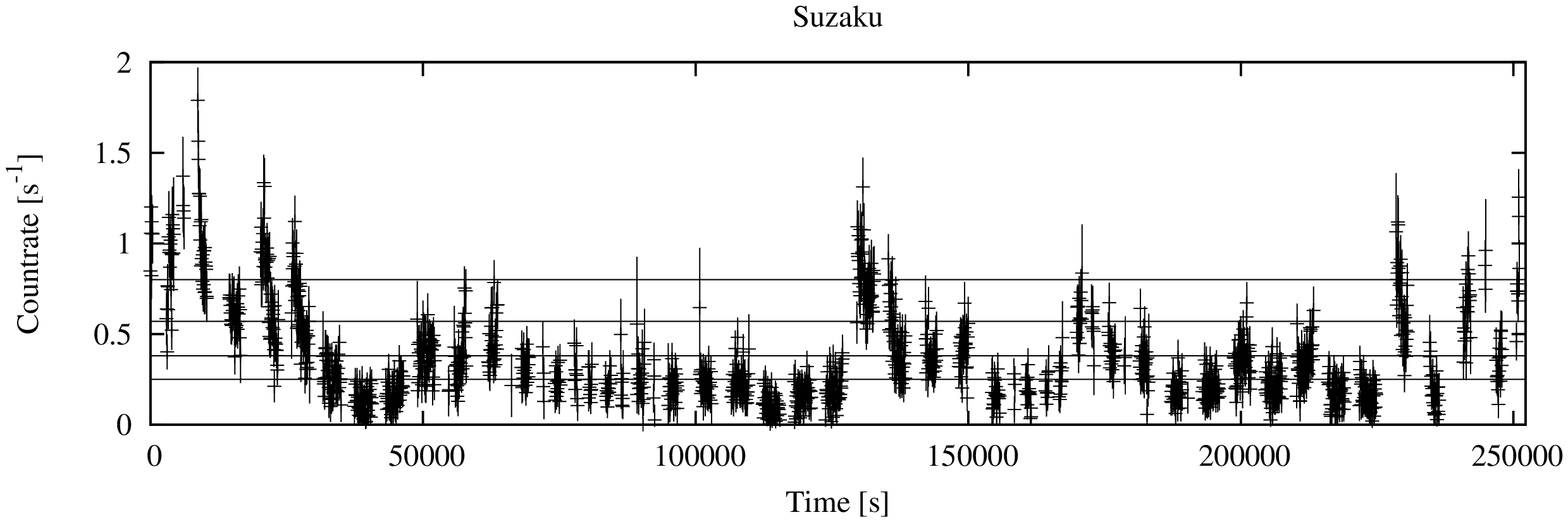}}
}
\caption{
        {\it Continued.}
}
\label{fig:ltcrv:d}
\end{figure*}

\begin{figure}
 \begin{center}
  \includegraphics[width=10cm,angle=270]{4.eps}
 \end{center}
\caption{The best-fit model of XMM12 when all parameters except for normalizations are invariant.}
\label{fig:const}
\end{figure}


\begin{figure*}
\centering
\subfigure{
        \resizebox{8cm}{!}{\includegraphics[angle=270]{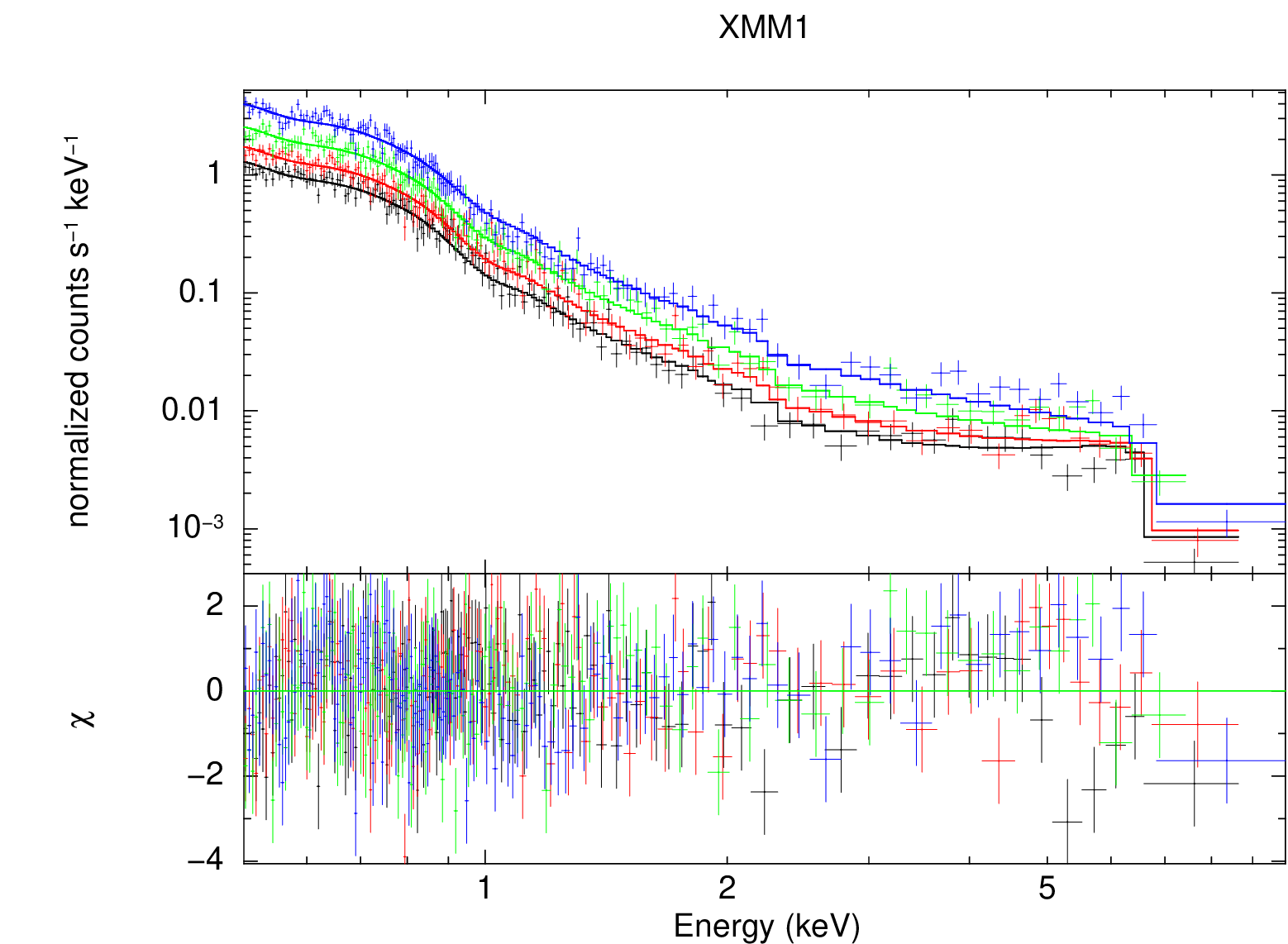}}
        \resizebox{8cm}{!}{\includegraphics[angle=270]{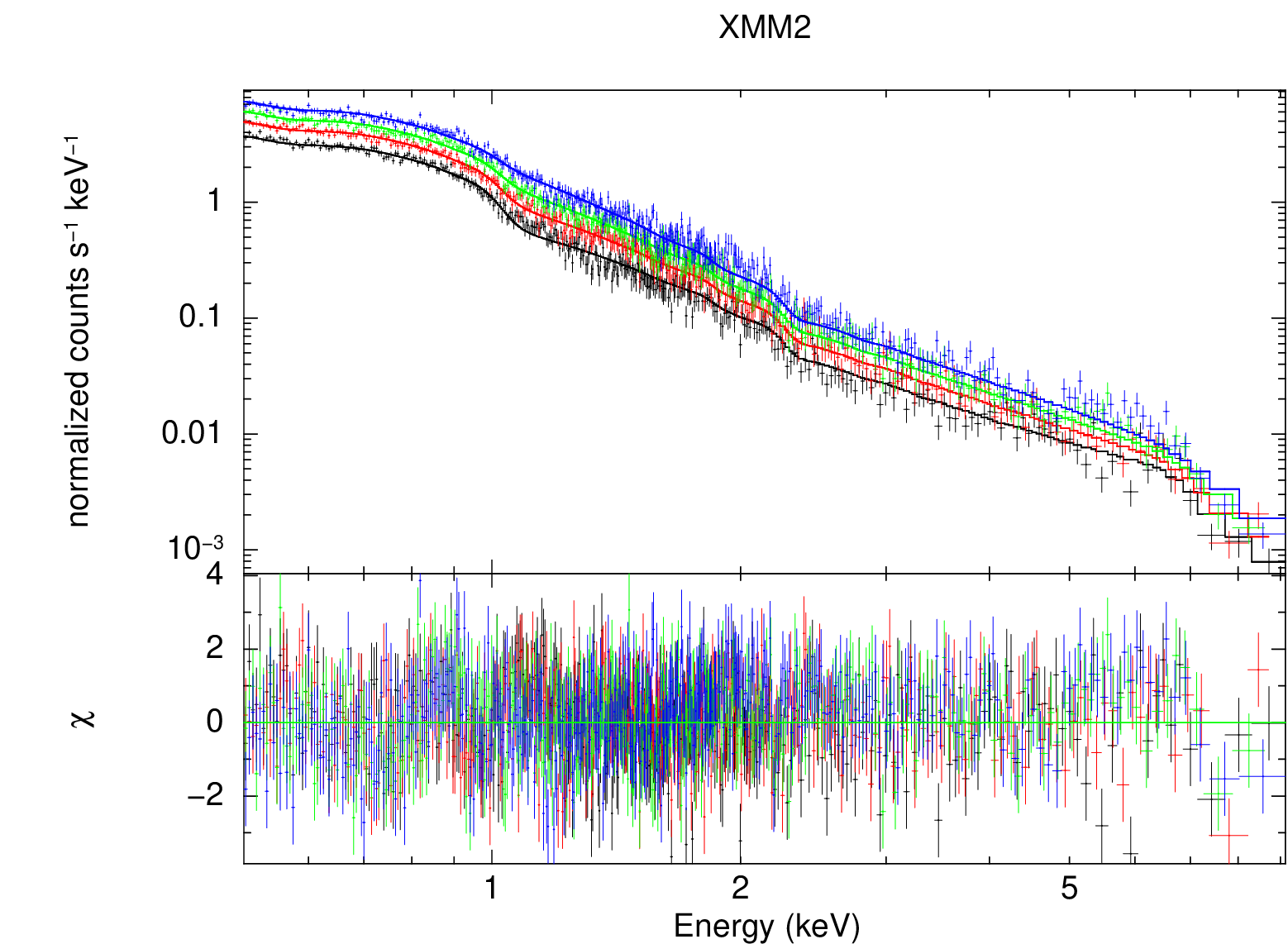}}
}
\subfigure{
        \resizebox{8cm}{!}{\includegraphics[angle=270]{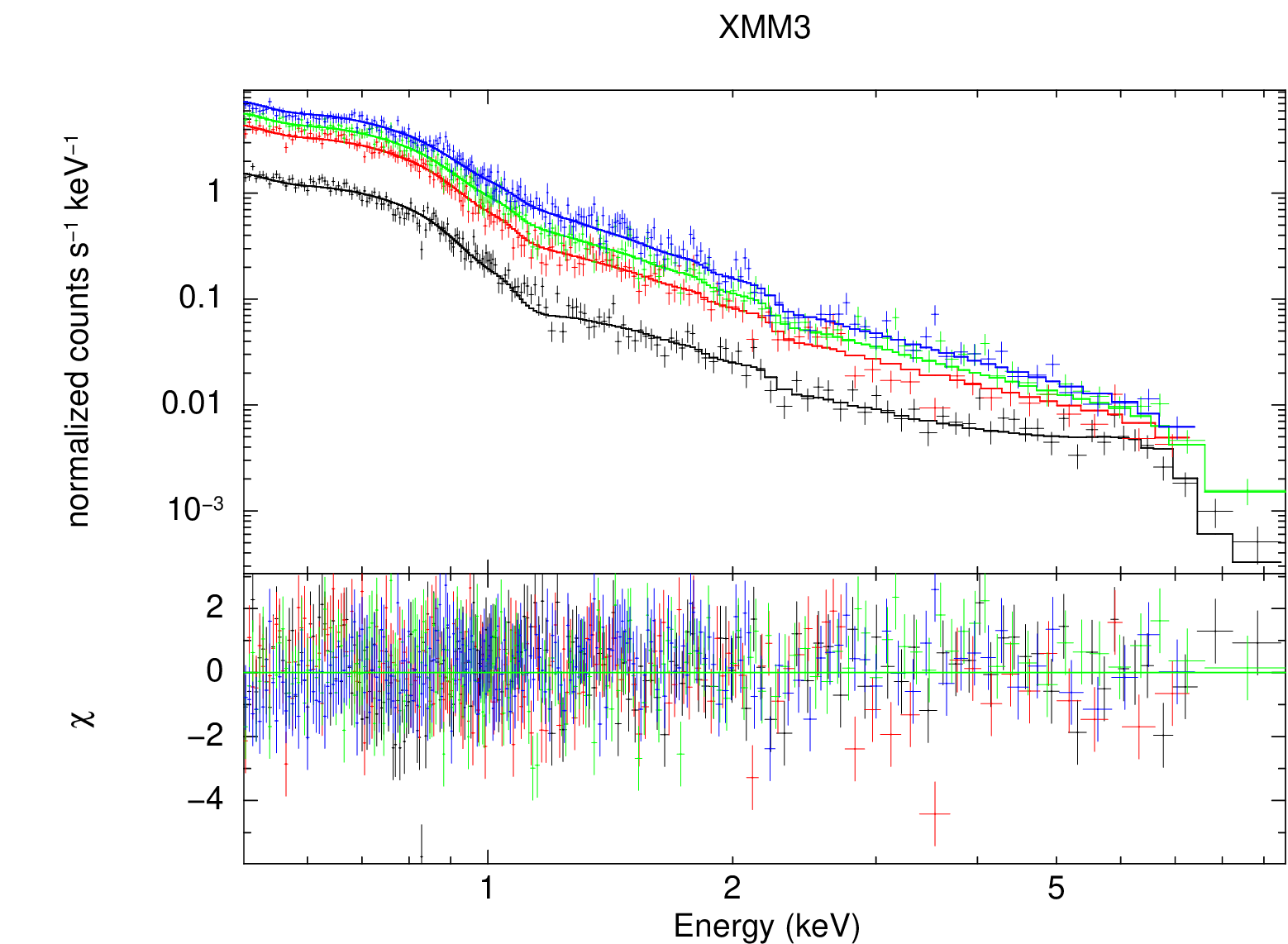}}
        \resizebox{8cm}{!}{\includegraphics[angle=270]{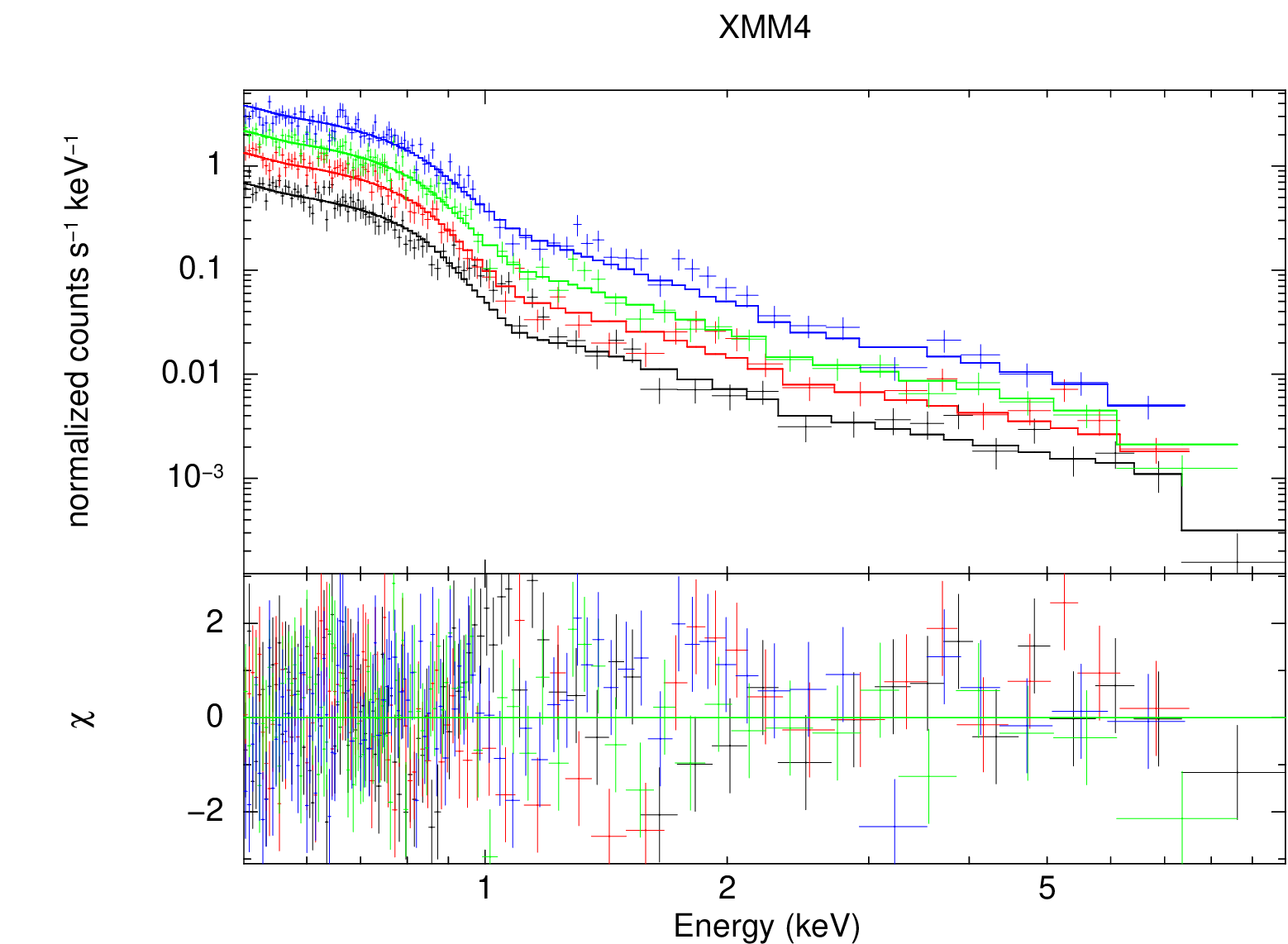}}
}
\subfigure{
        \resizebox{8cm}{!}{\includegraphics[angle=270]{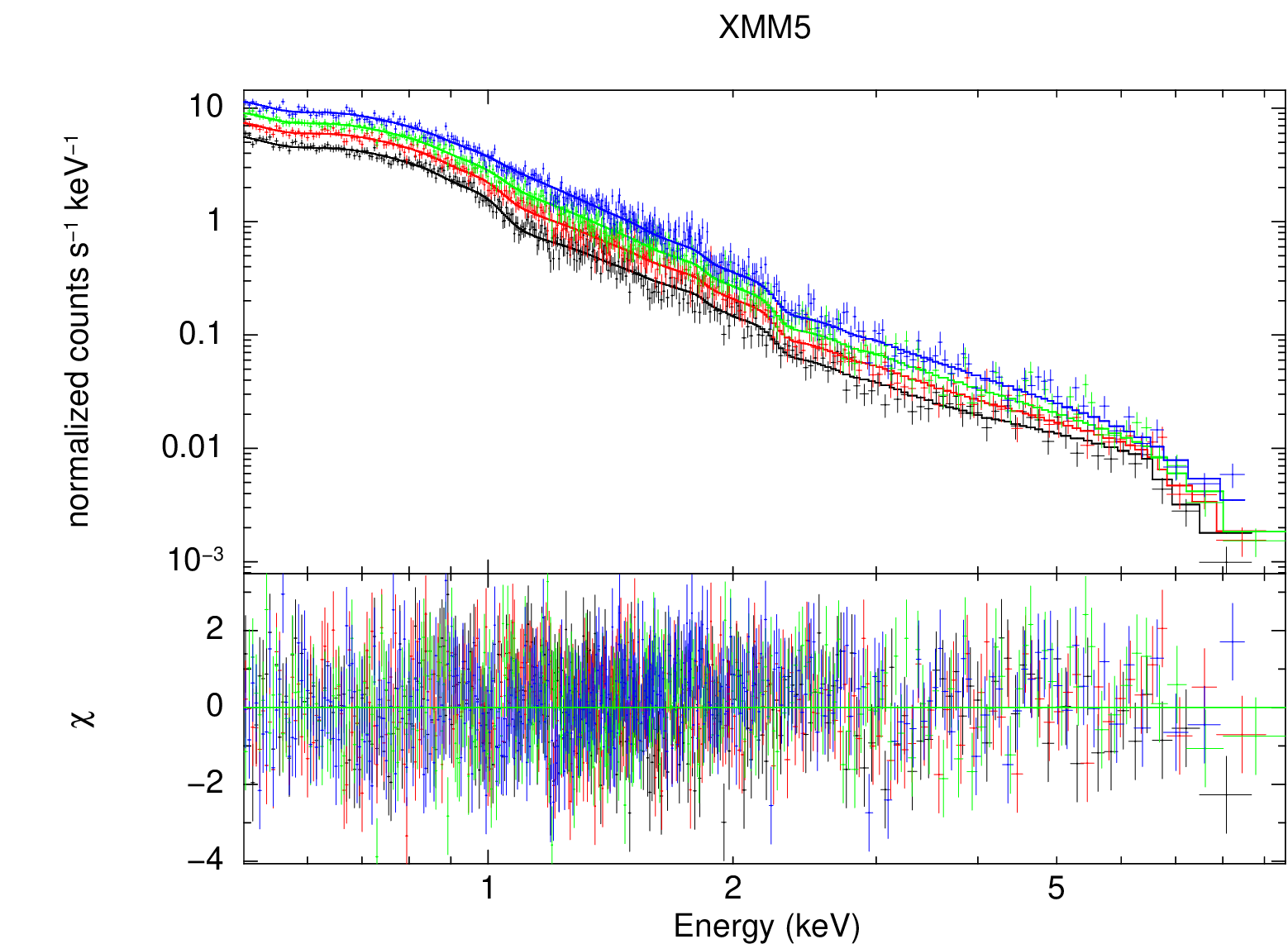}}
        \resizebox{8cm}{!}{\includegraphics[angle=270]{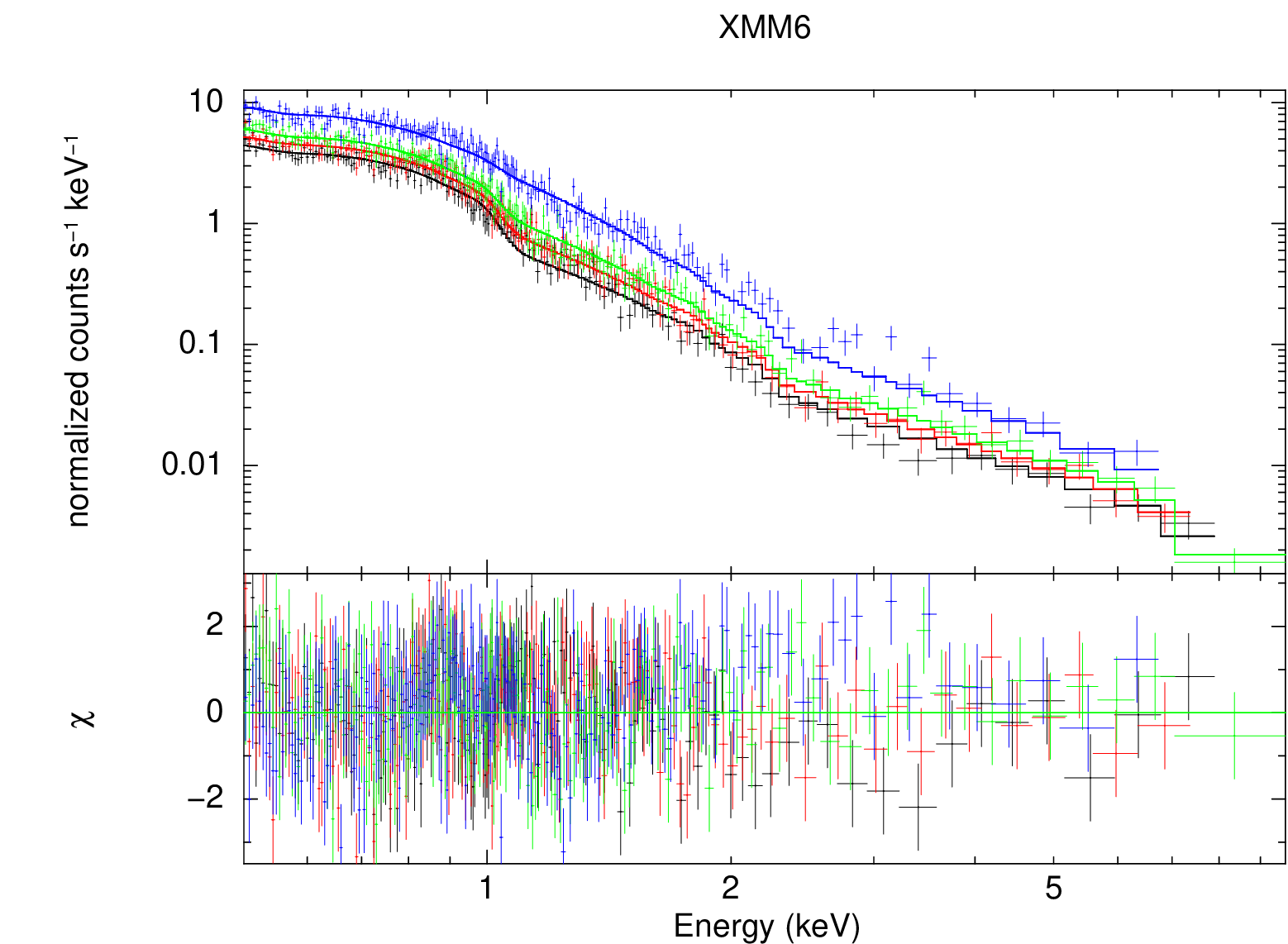}}
}
\caption{The fitting results of the intensity-sliced spectra. In the Suzaku data, the FI spectra are only shown.}
\label{fig:slice:a}
\end{figure*}
\addtocounter{figure}{-1}
\begin{figure*}
\addtocounter{subfigure}{1}
\centering
\subfigure{
        \resizebox{8cm}{!}{\includegraphics[angle=270]{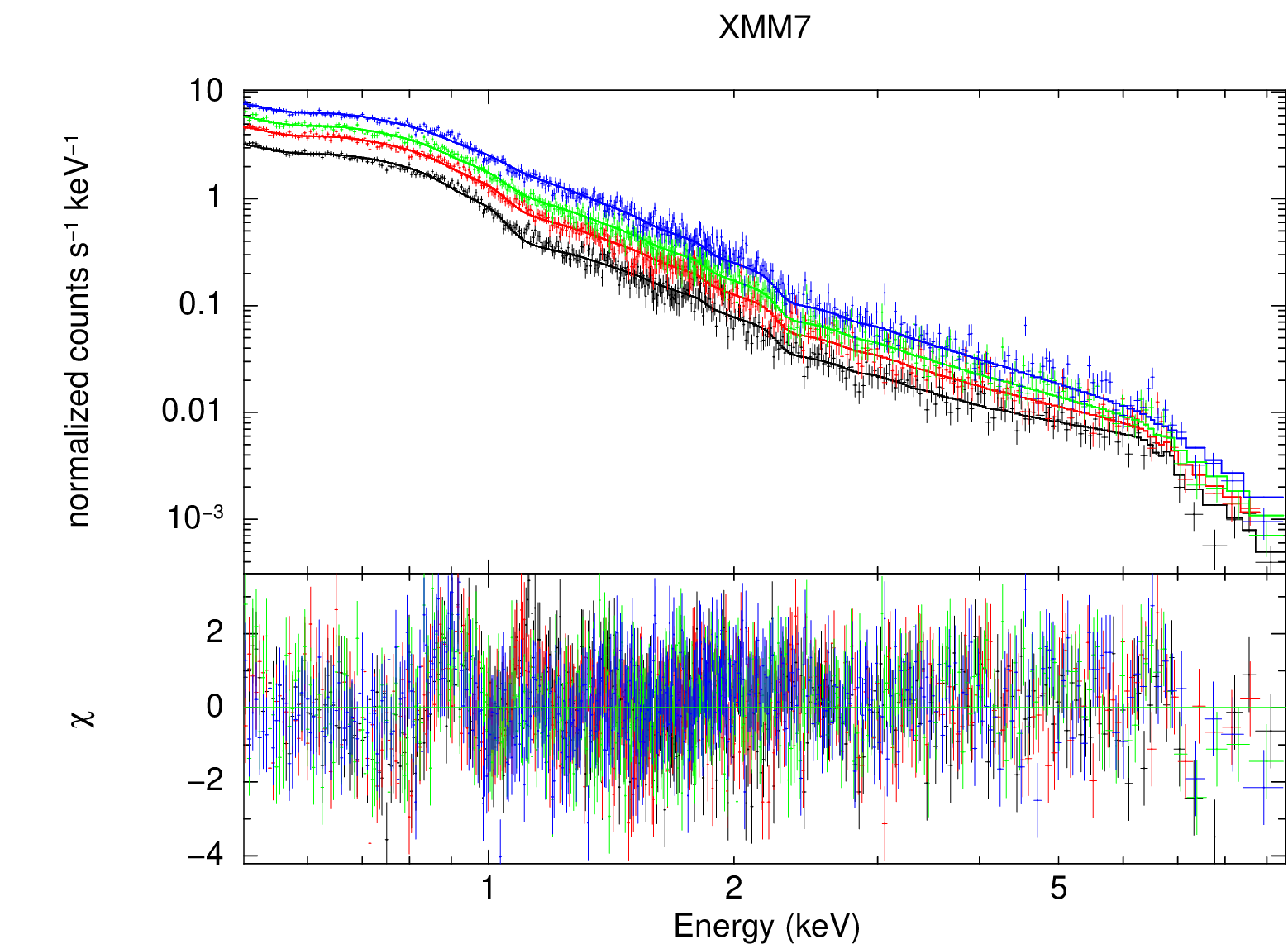}}
        \resizebox{8cm}{!}{\includegraphics[angle=270]{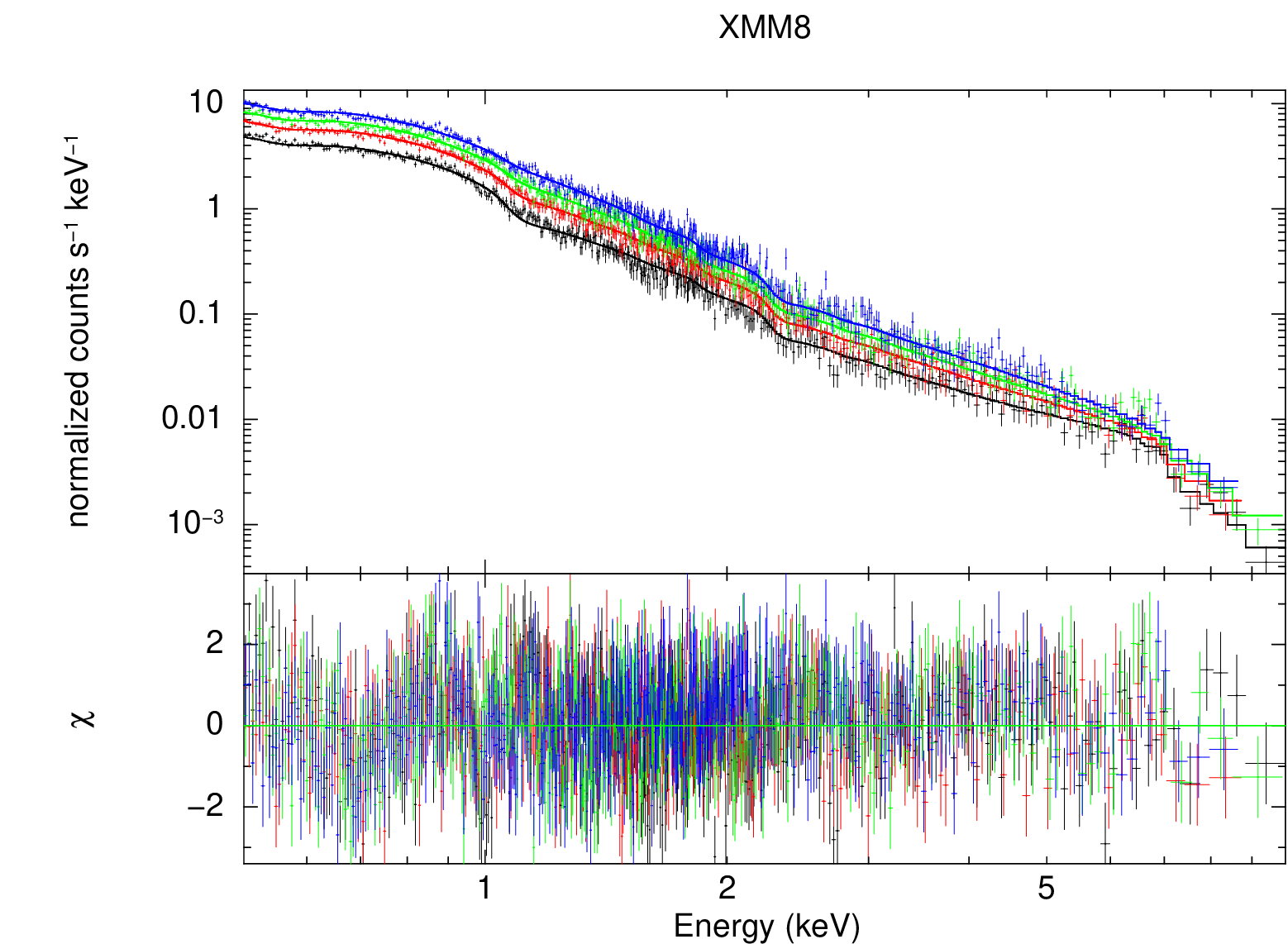}}
}
\subfigure{
        \resizebox{8cm}{!}{\includegraphics[angle=270]{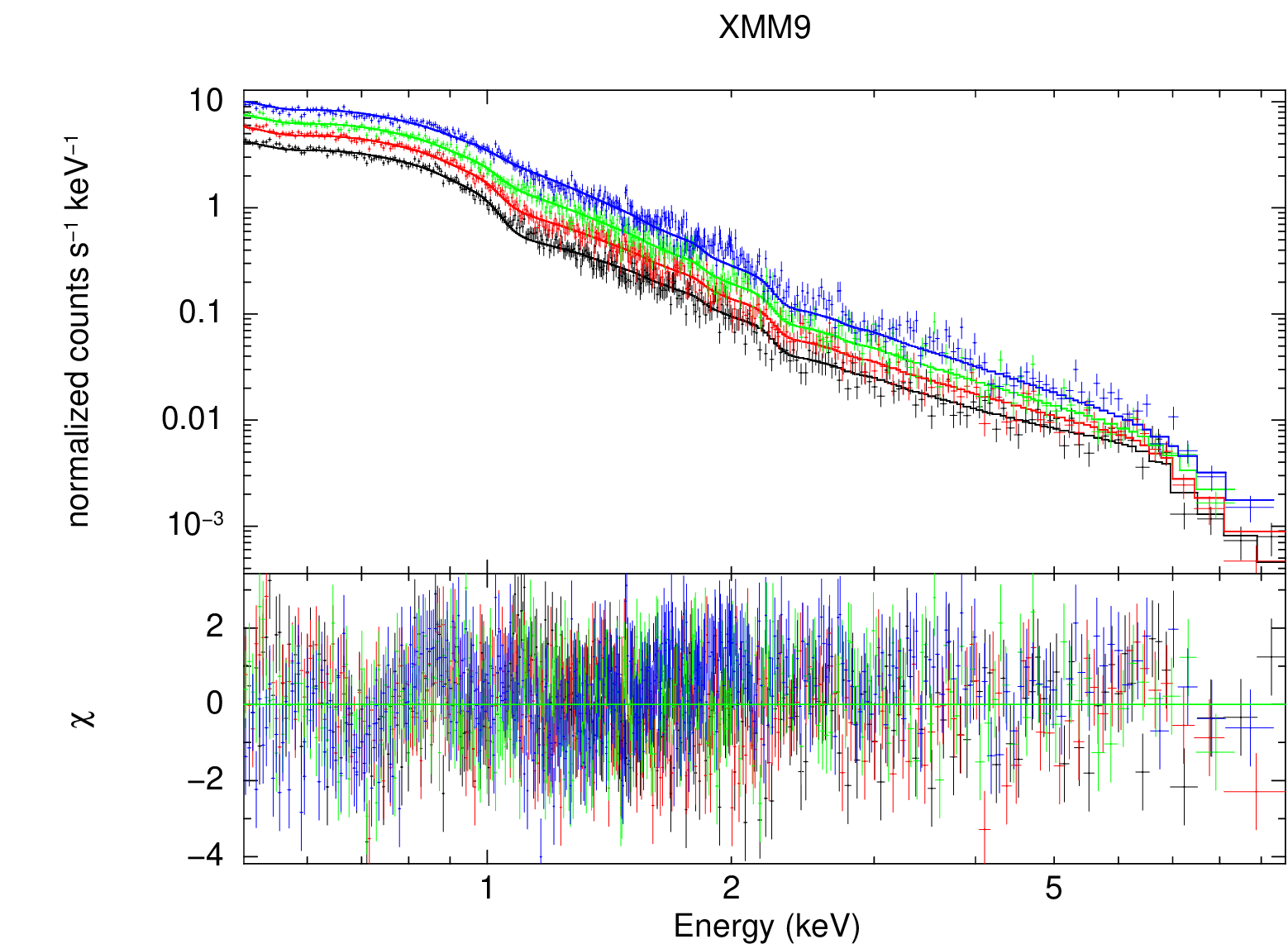}}
        \resizebox{8cm}{!}{\includegraphics[angle=270]{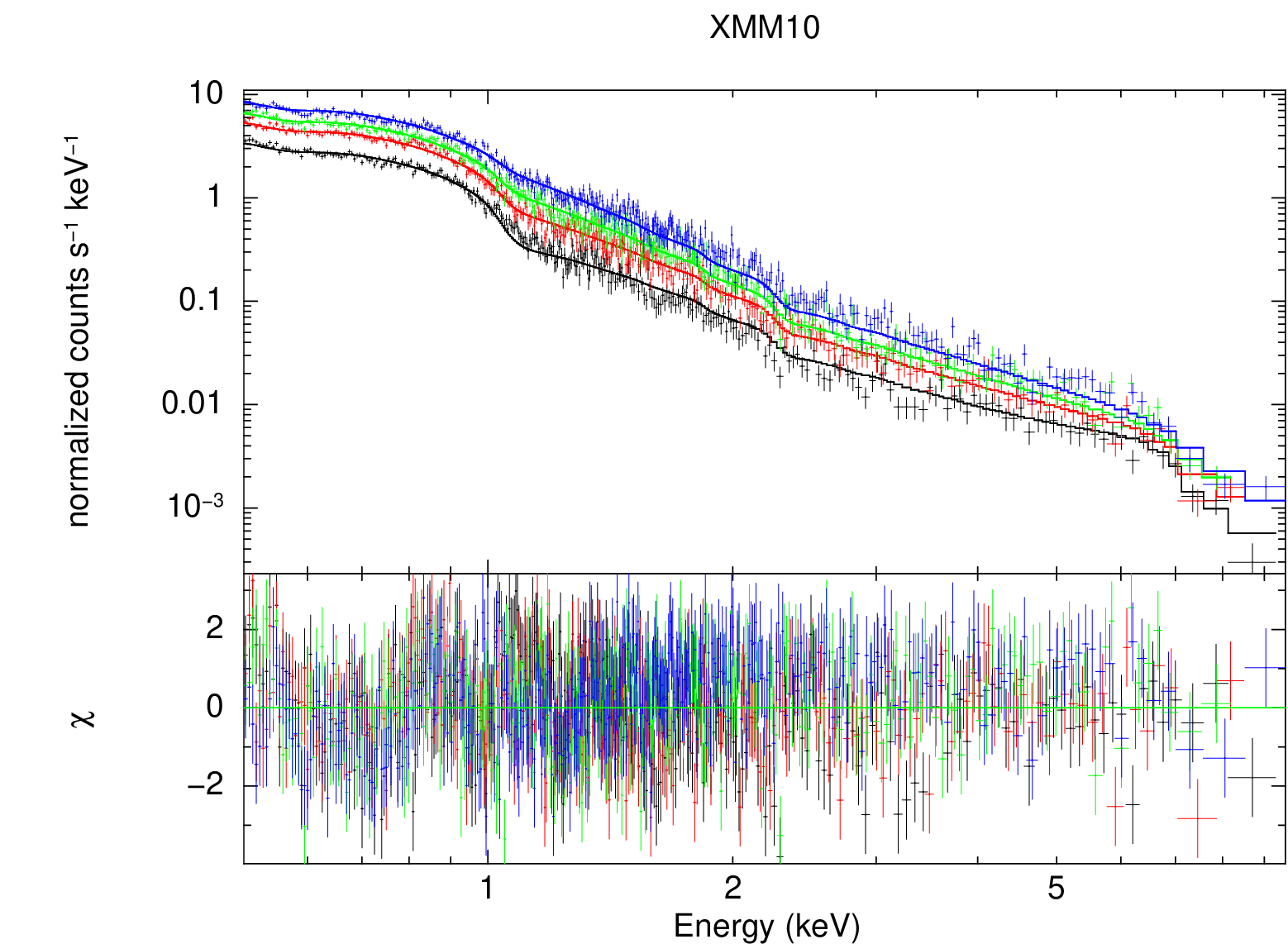}}
}
\subfigure{
        \resizebox{8cm}{!}{\includegraphics[angle=270]{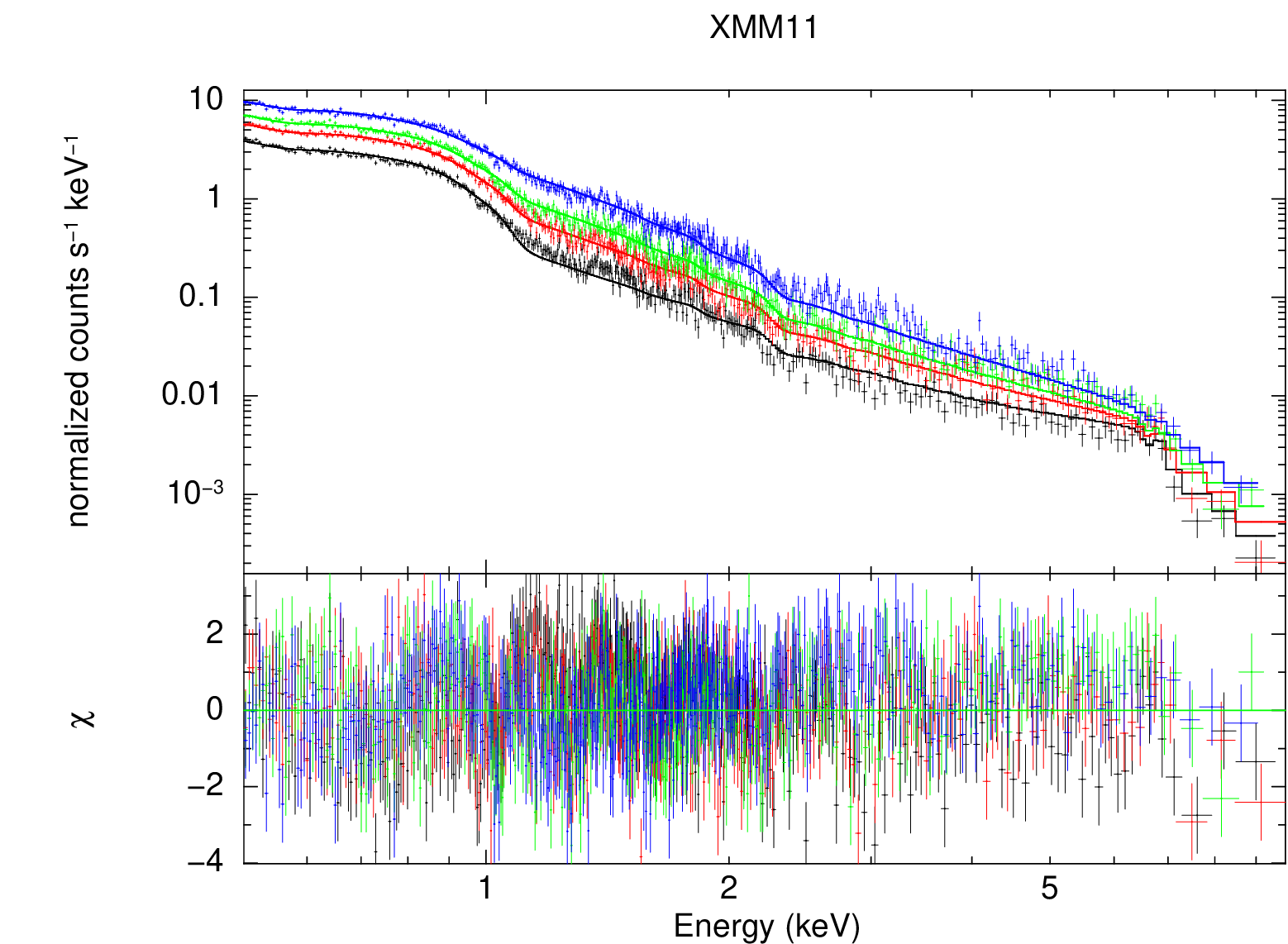}}
        \resizebox{8cm}{!}{\includegraphics[angle=270]{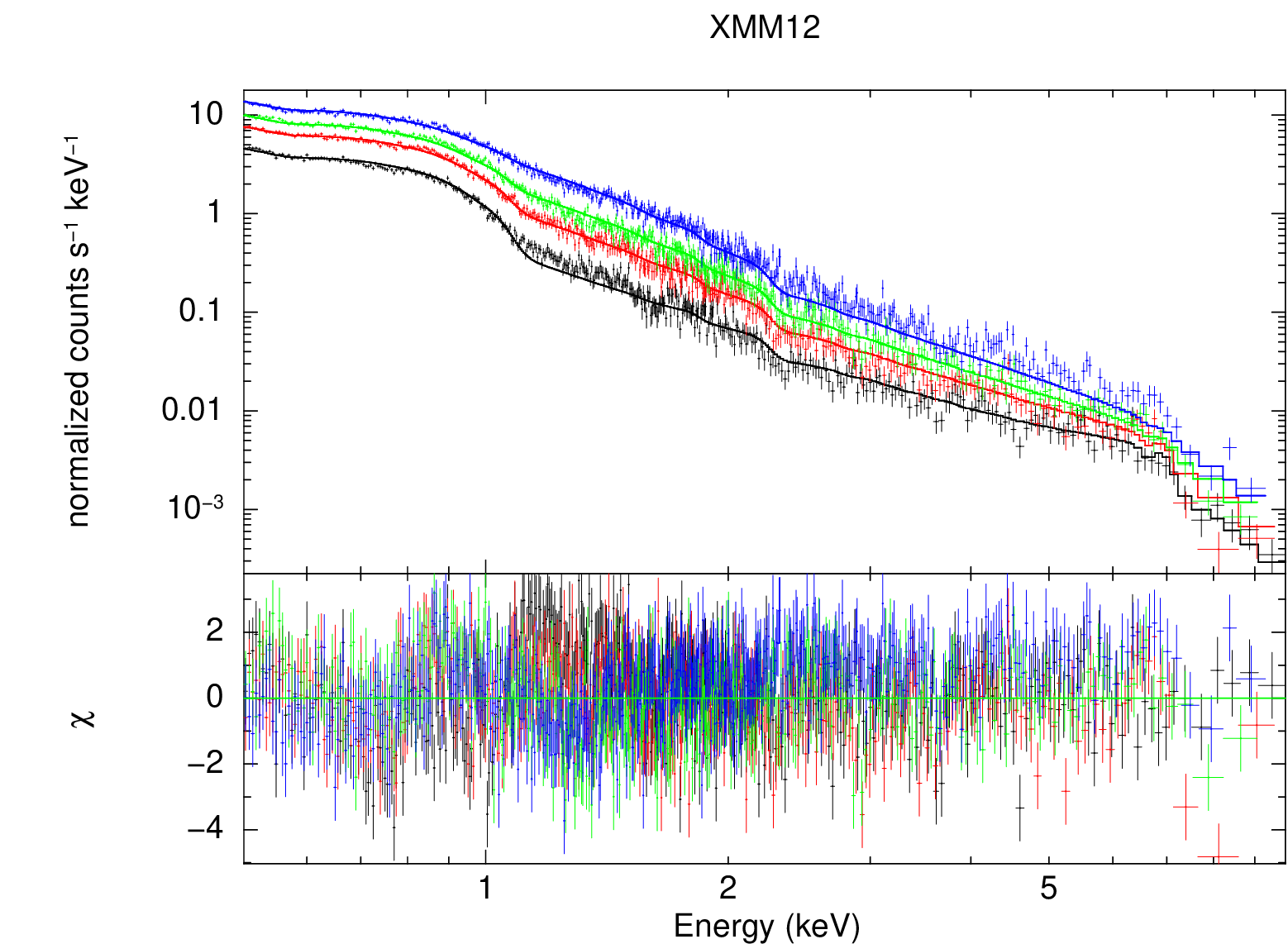}}
}
\caption{
        {\it Continued.}
}
\label{fig:slice:b}
\end{figure*}
\addtocounter{figure}{-1}
\begin{figure*}
\addtocounter{subfigure}{1}
\centering
\subfigure{
        \resizebox{8cm}{!}{\includegraphics[angle=270]{5m.eps}}
        \resizebox{8cm}{!}{\includegraphics[angle=270]{5n.eps}}
}
\subfigure{
        \resizebox{8cm}{!}{\includegraphics[angle=270]{5o.eps}}
        \resizebox{8cm}{!}{\includegraphics[angle=270]{5p.eps}}
}
\subfigure{
}
\caption{
        {\it Continued.}
}
\label{fig:slice:c}
\end{figure*}

\begin{figure}
 \begin{center}
  \includegraphics[width=8cm]{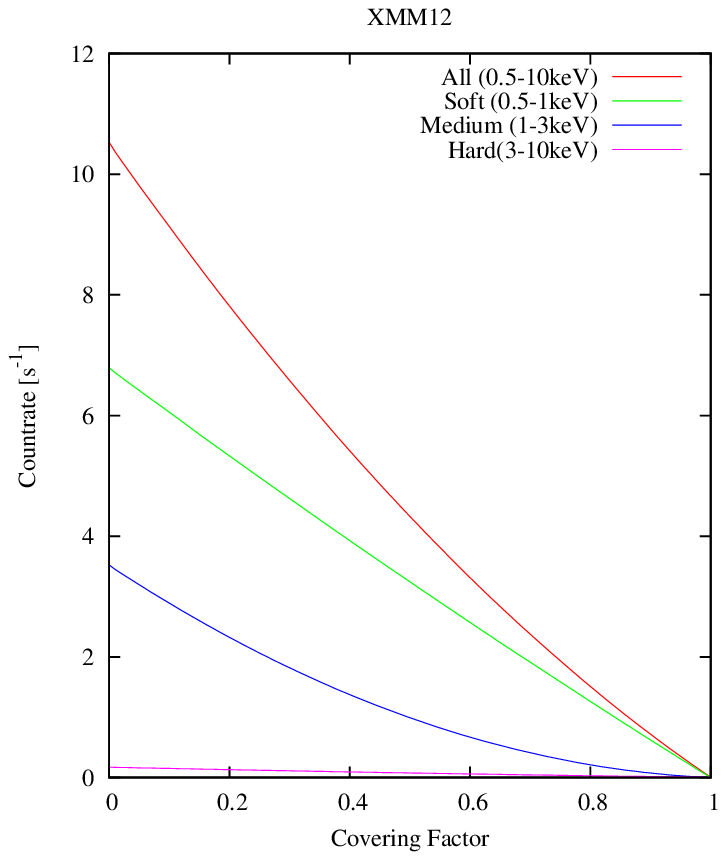}
 \end{center}
\caption{Model countrate of each energy band as a function of covering factor.
The figure is calculated for XMM12, which exhibits a wide range of the covering fraction variation. }\label{fig:cf}
\end{figure}


\begin{figure*}
\centering
\subfigure{
        \resizebox{15cm}{!}{\includegraphics[angle=270]{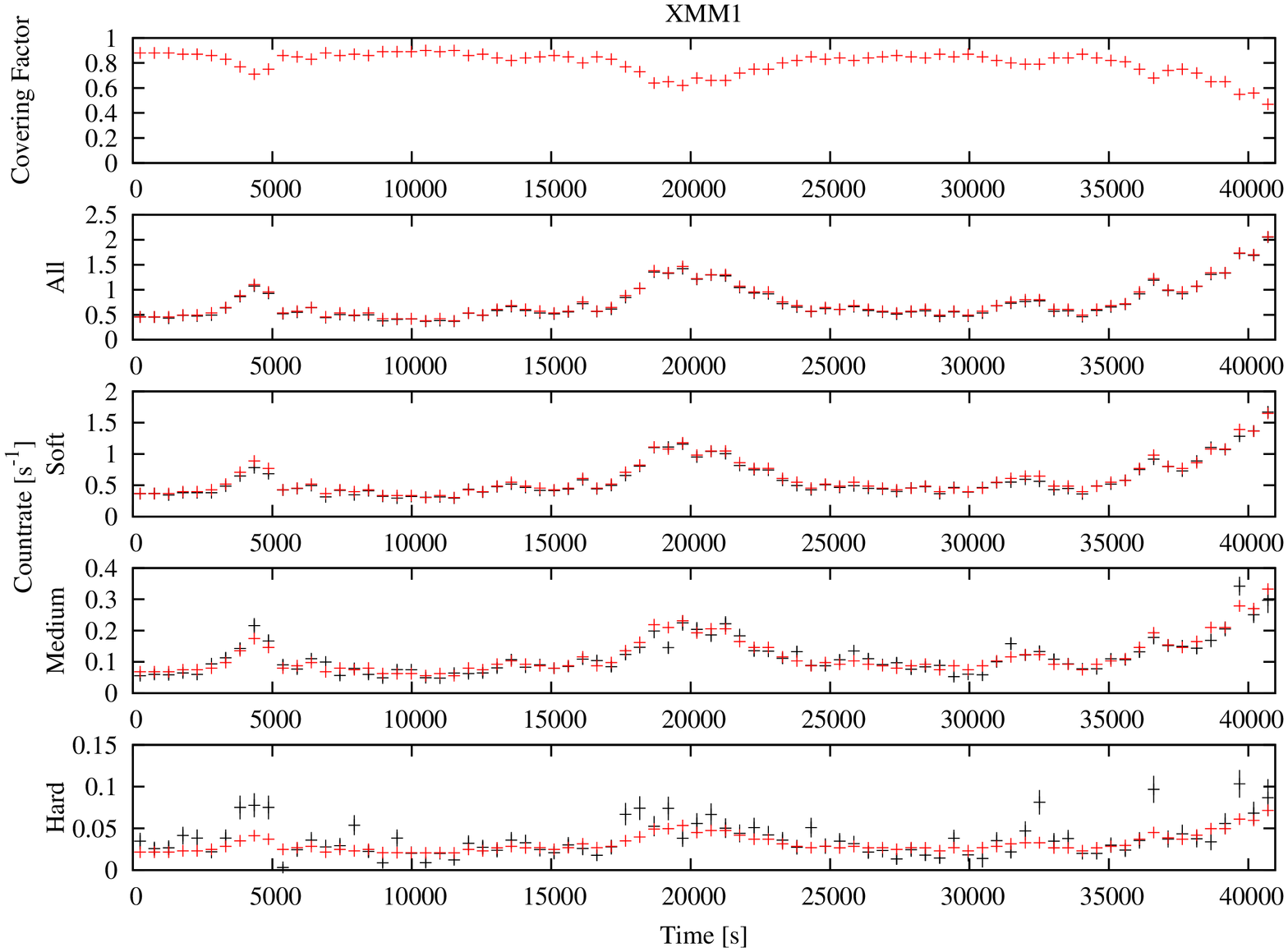}}
}
\subfigure{
        \resizebox{15cm}{!}{\includegraphics[angle=270]{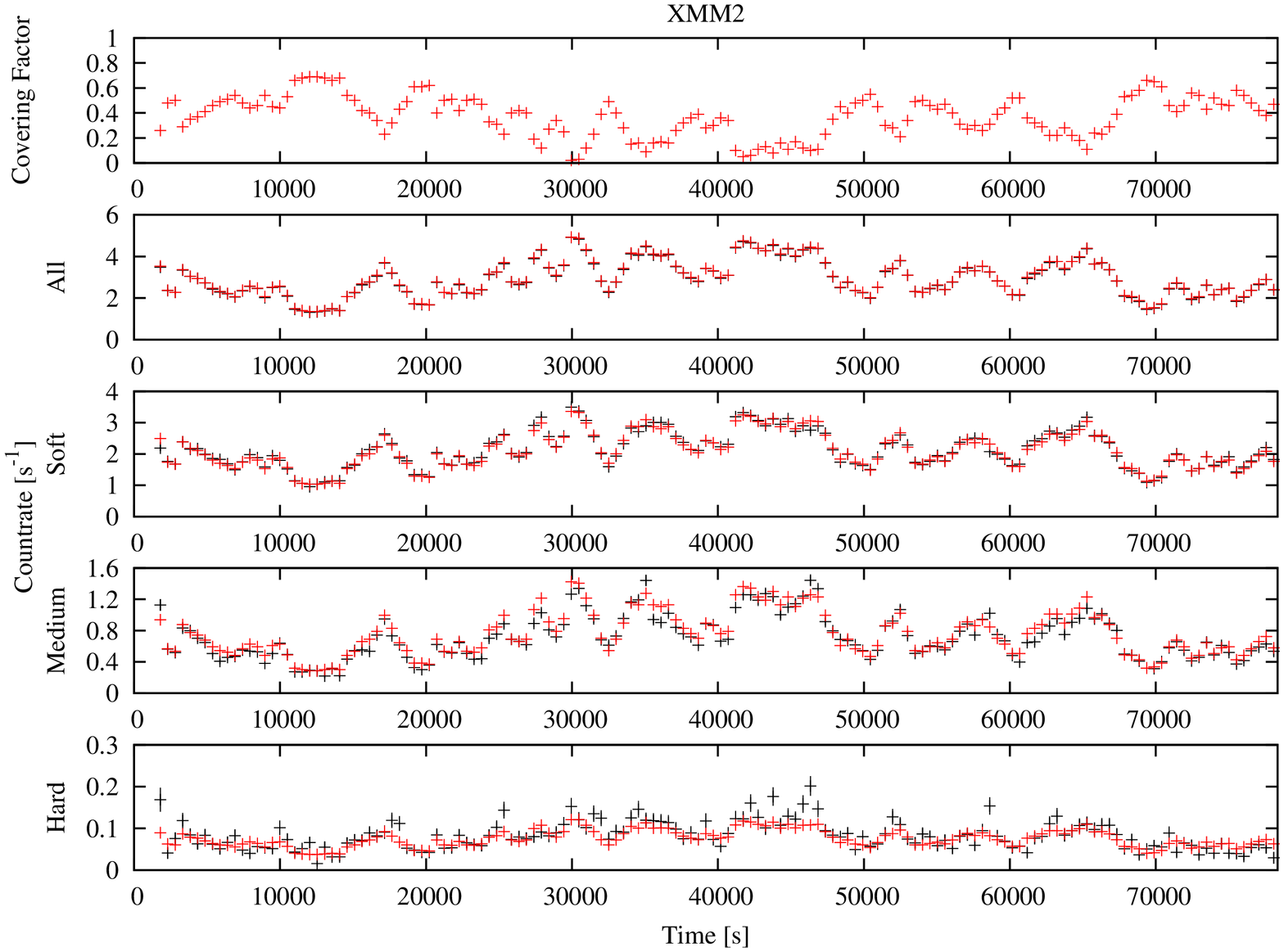}}
}
\caption{Top:~Variation of the partial covering fraction derived from the 0.5--12.0 keV light-curves assuming the VDPC model.
Upper Center:~The observed light-curve in 0.5--12.0 keV (black) and the model light-curve (red).
They should agree in definition.
Center:~The observed light-curve in 0.5--1.0 keV (black) and the model (red).
Lower Center:~The light-curve of 1.0--3.0 keV (black) and the model (red).
Bottom:~The light-curve of 3.0--10.0 keV (black) and the model (red).
}
\label{fig:modelltcrv:a}
\end{figure*}

\addtocounter{figure}{-1}
\begin{figure*}
\addtocounter{subfigure}{1}
\centering
\subfigure{
        \resizebox{15cm}{!}{\includegraphics[angle=270]{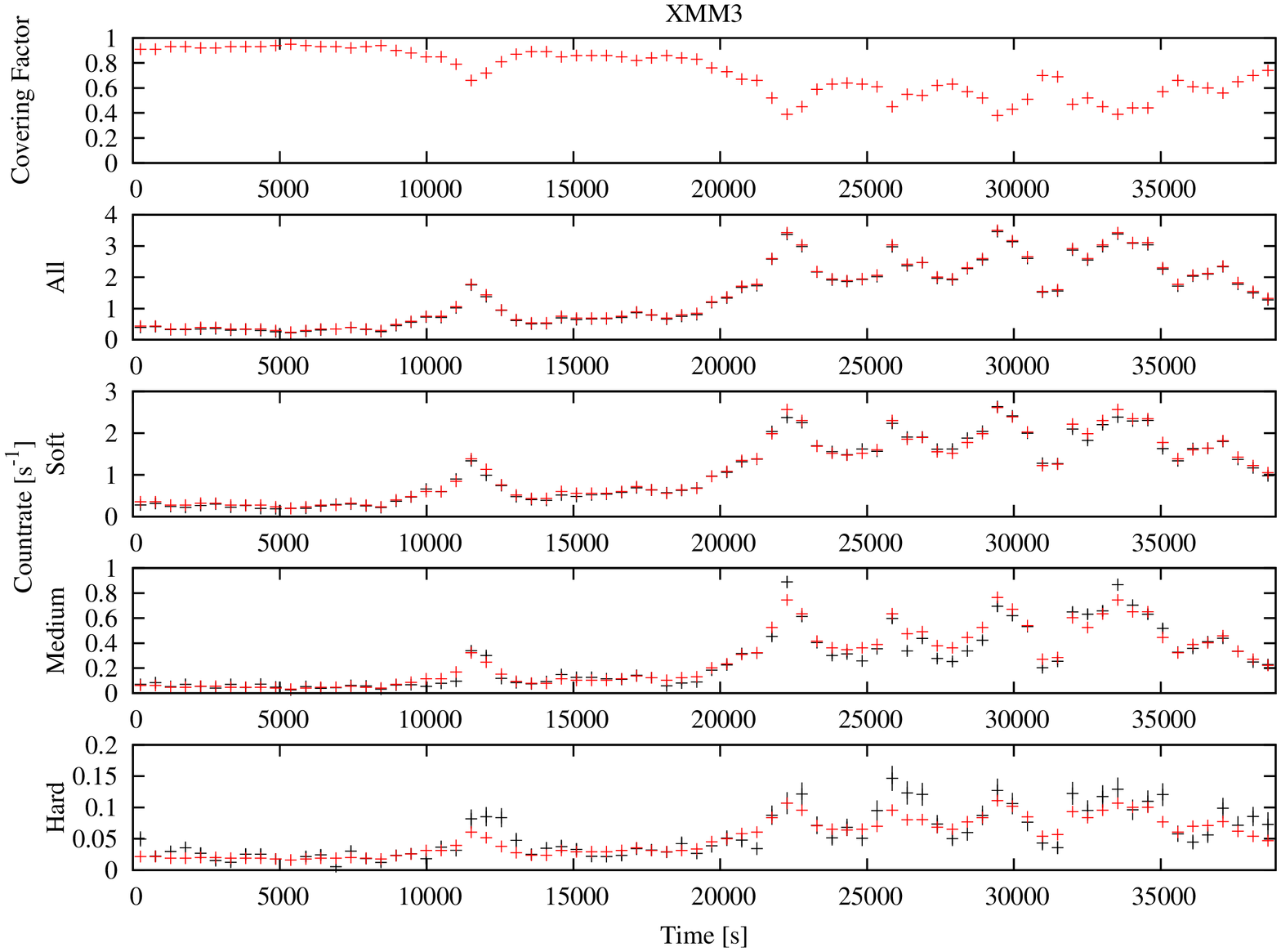}}
}
\subfigure{
        \resizebox{15cm}{!}{\includegraphics[angle=270]{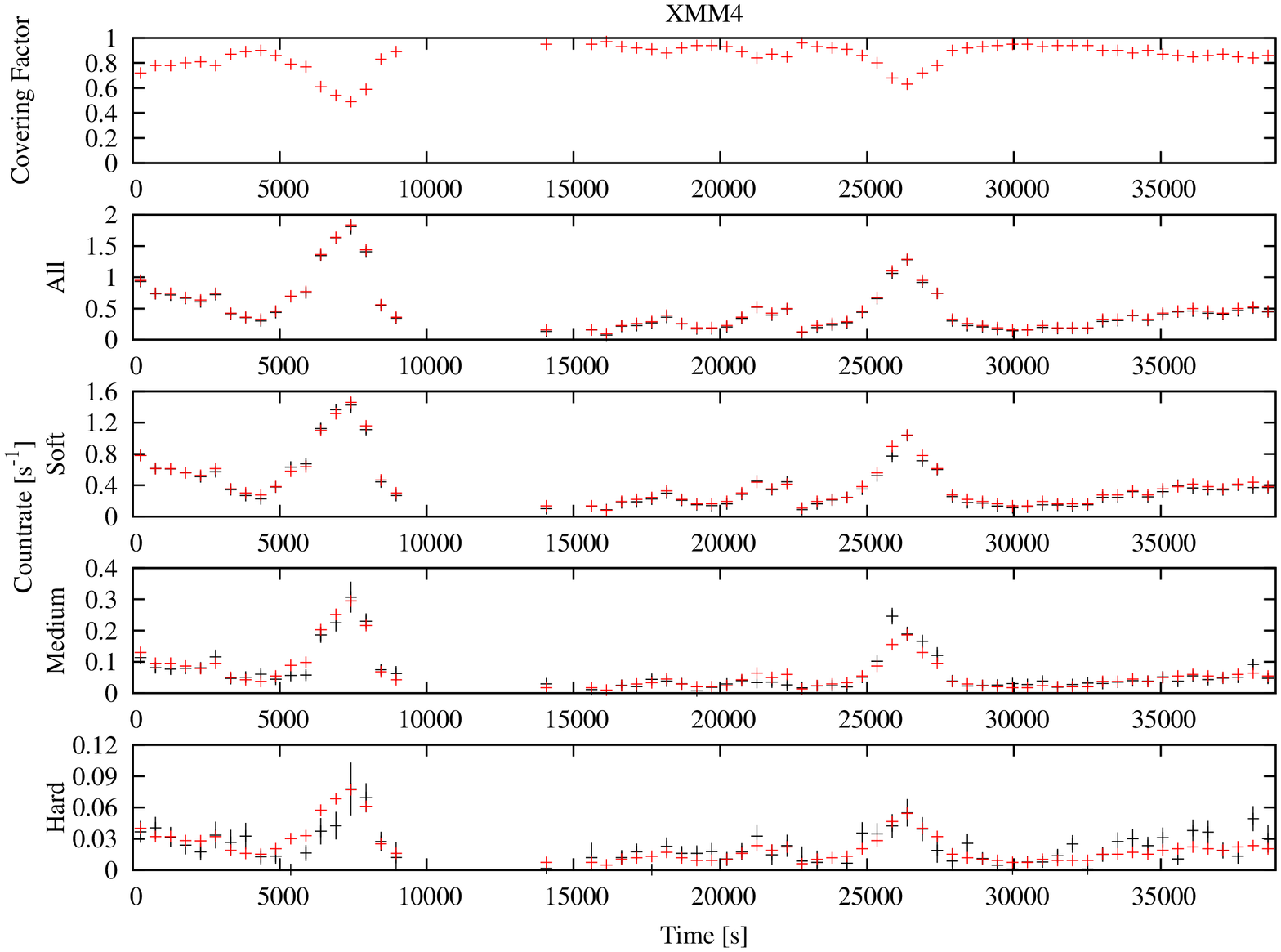}}
}
\caption{
        {\it Continued.}
}
\label{fig:modelltcrv:b}
\end{figure*}

\addtocounter{figure}{-1}
\begin{figure*}
\addtocounter{subfigure}{1}
\centering
\subfigure{
        \resizebox{15cm}{!}{\includegraphics[angle=270]{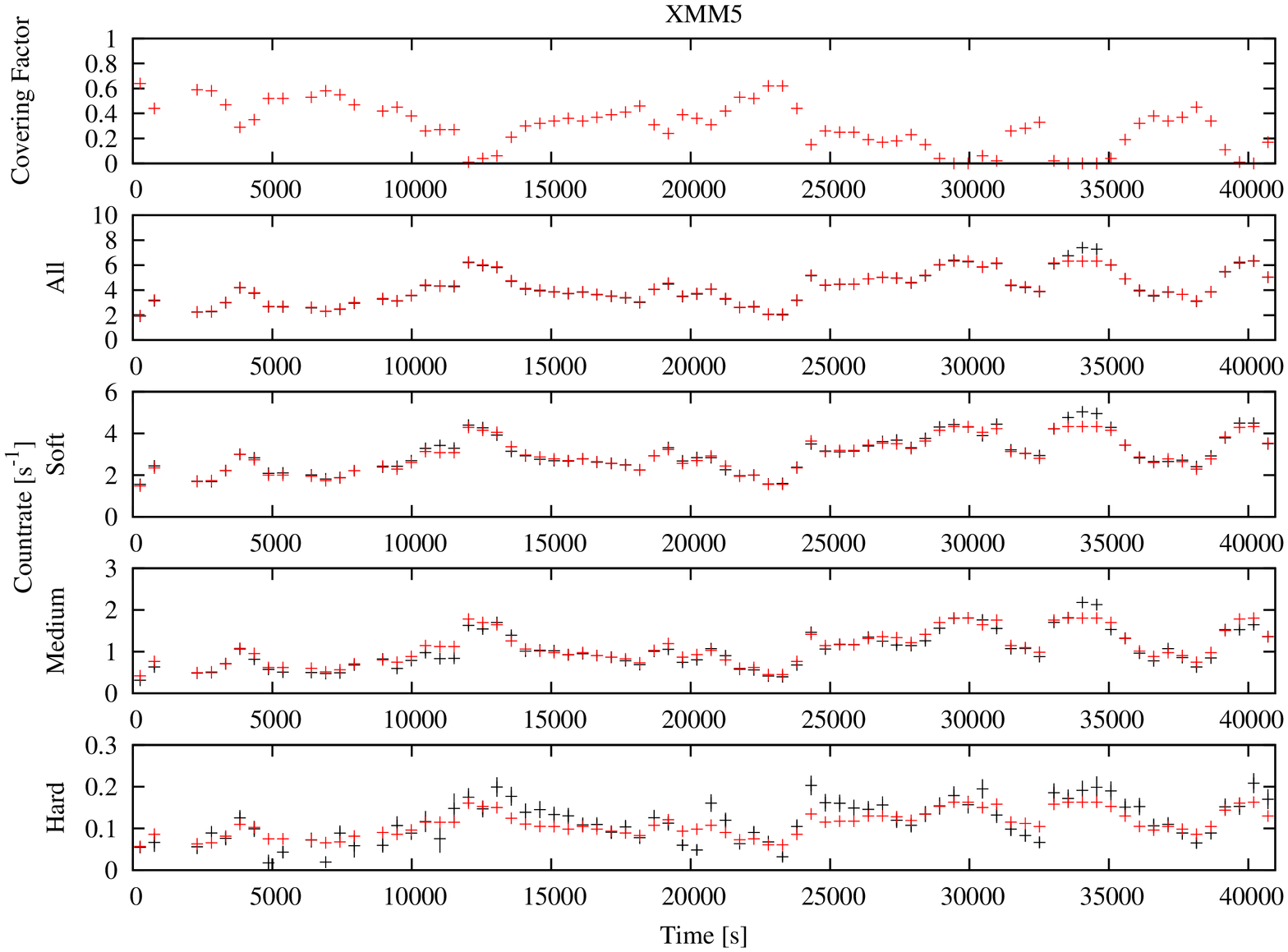}}
}
\subfigure{
        \resizebox{15cm}{!}{\includegraphics[angle=270]{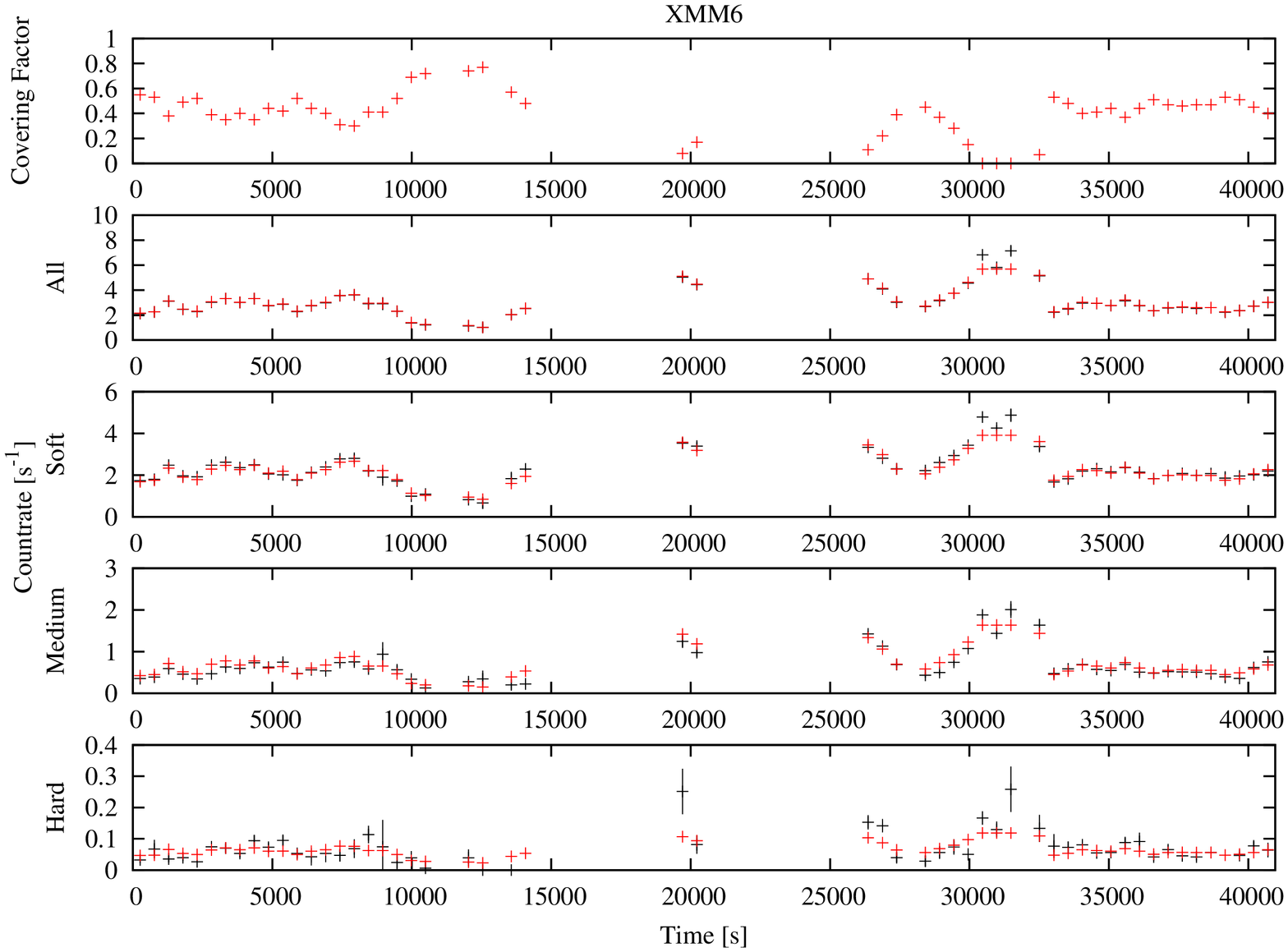}}
}
\caption{
        {\it Continued.}
}
\label{fig:modelltcrv:c}
\end{figure*}

\addtocounter{figure}{-1}
\begin{figure*}
\addtocounter{subfigure}{1}
\centering
\subfigure{
        \resizebox{15cm}{!}{\includegraphics[angle=270]{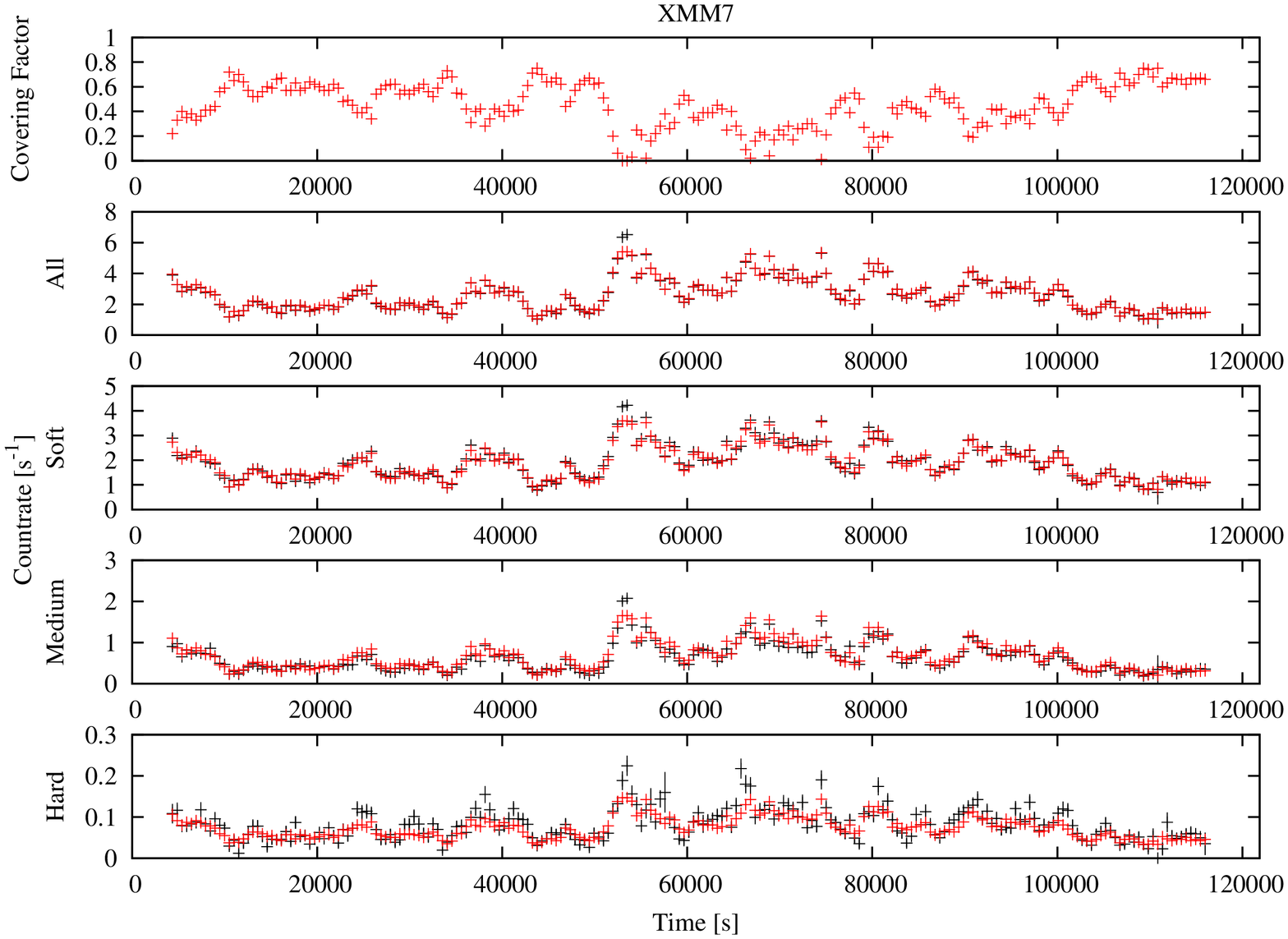}}
}
\subfigure{
        \resizebox{15cm}{!}{\includegraphics[angle=270]{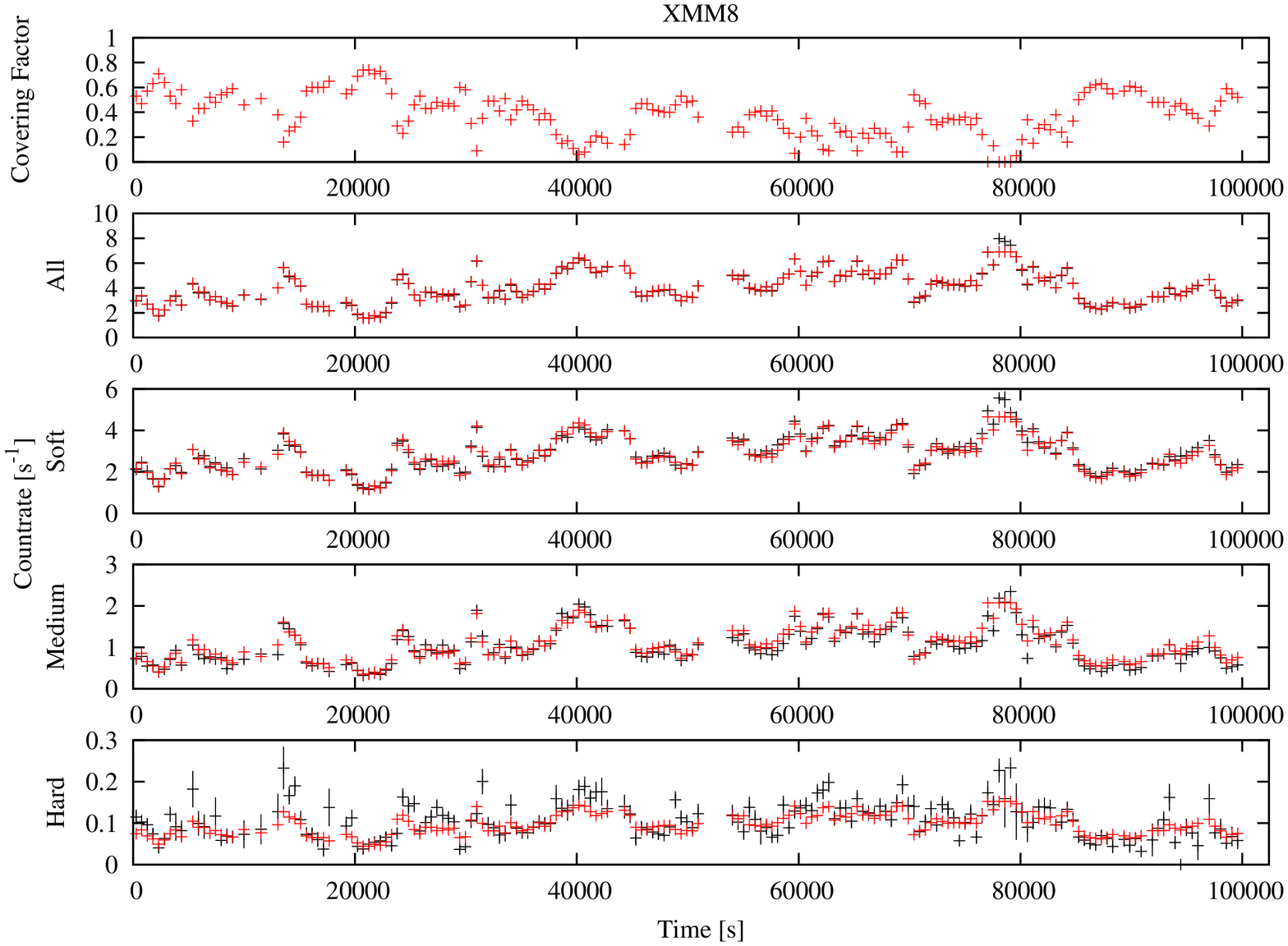}}
}
\caption{
        {\it Continued.}
}
\label{fig:modelltcrv:d}
\end{figure*}
\addtocounter{figure}{-1}
\begin{figure*}
\addtocounter{subfigure}{1}
\centering
\subfigure{
        \resizebox{15cm}{!}{\includegraphics[angle=270]{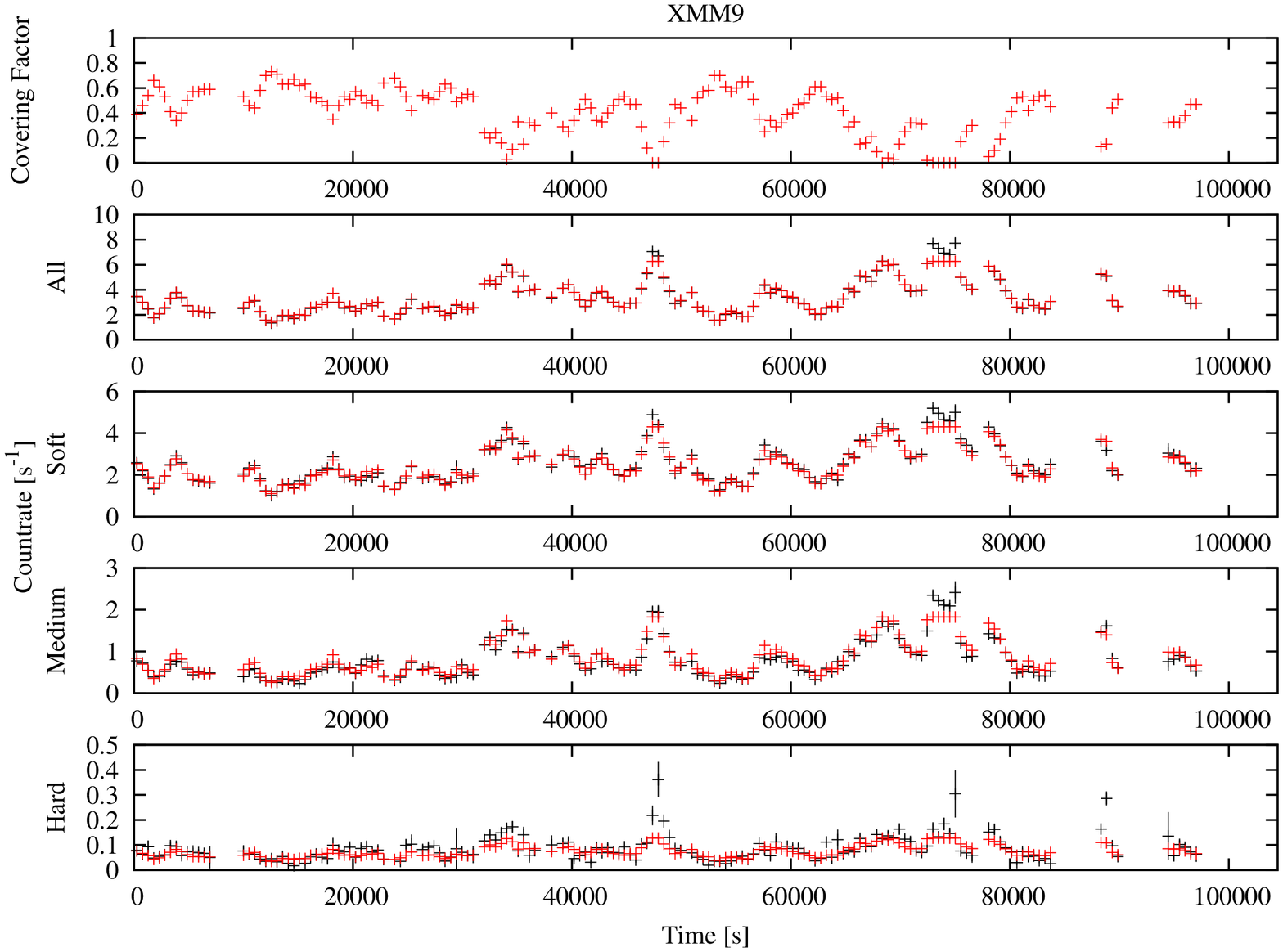}}
}
\subfigure{
        \resizebox{15cm}{!}{\includegraphics[angle=270]{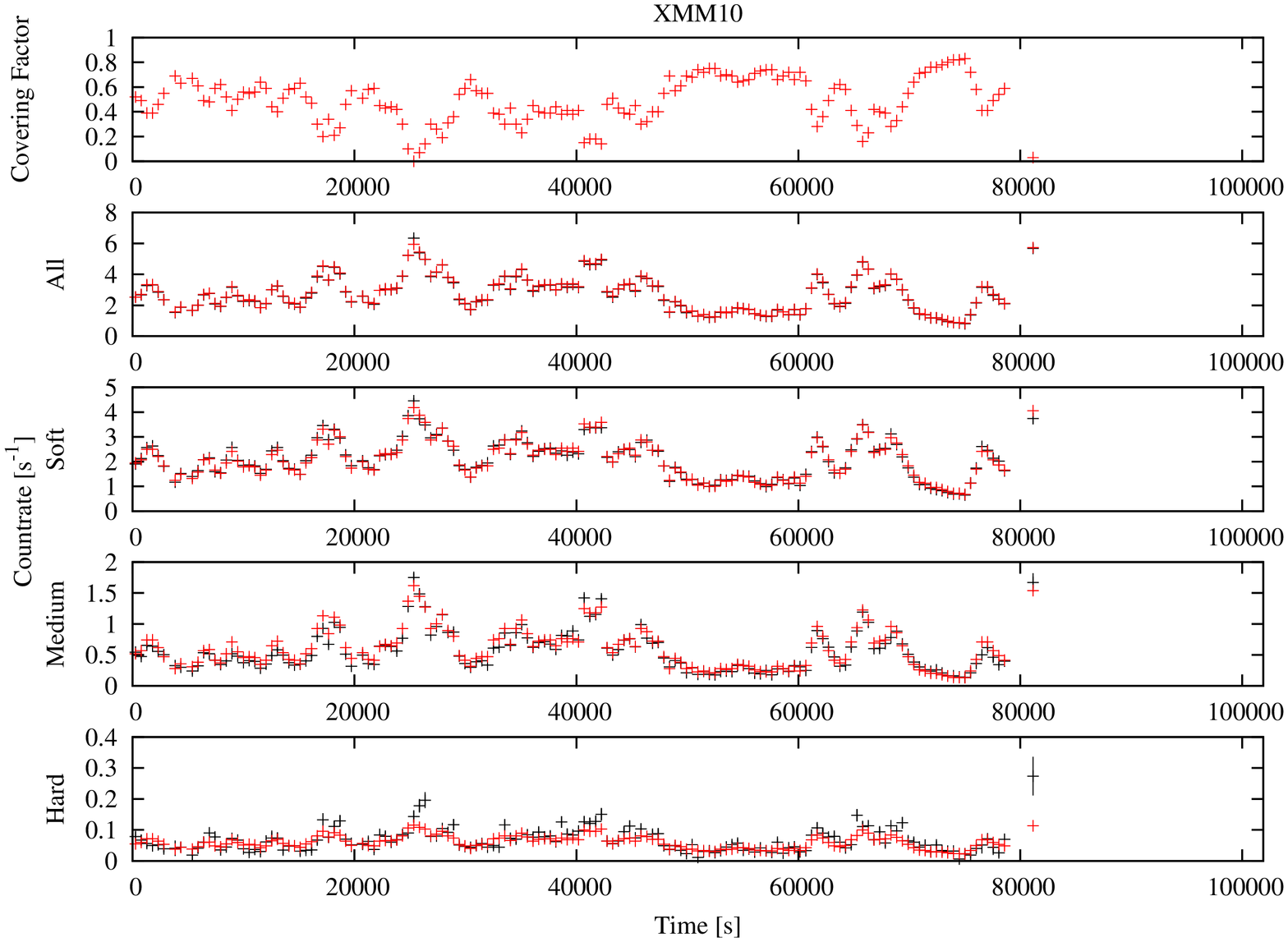}}
}
\caption{
        {\it Continued.}
}
\label{fig:modelltcrv:e}
\end{figure*}
\addtocounter{figure}{-1}
\begin{figure*}
\addtocounter{subfigure}{1}
\centering
\subfigure{
        \resizebox{15cm}{!}{\includegraphics[angle=270]{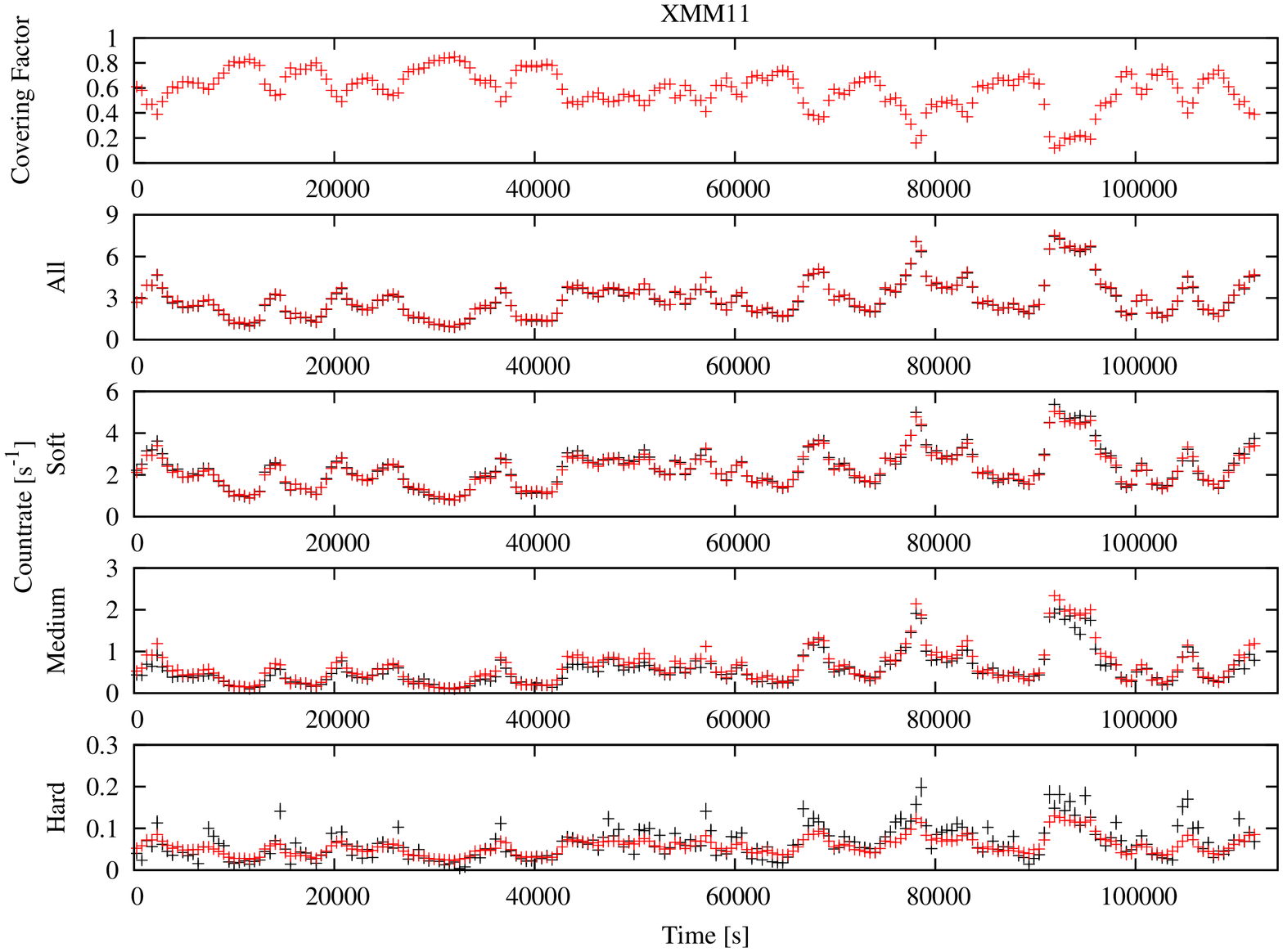}}
}
\subfigure{
        \resizebox{15cm}{!}{\includegraphics[angle=270]{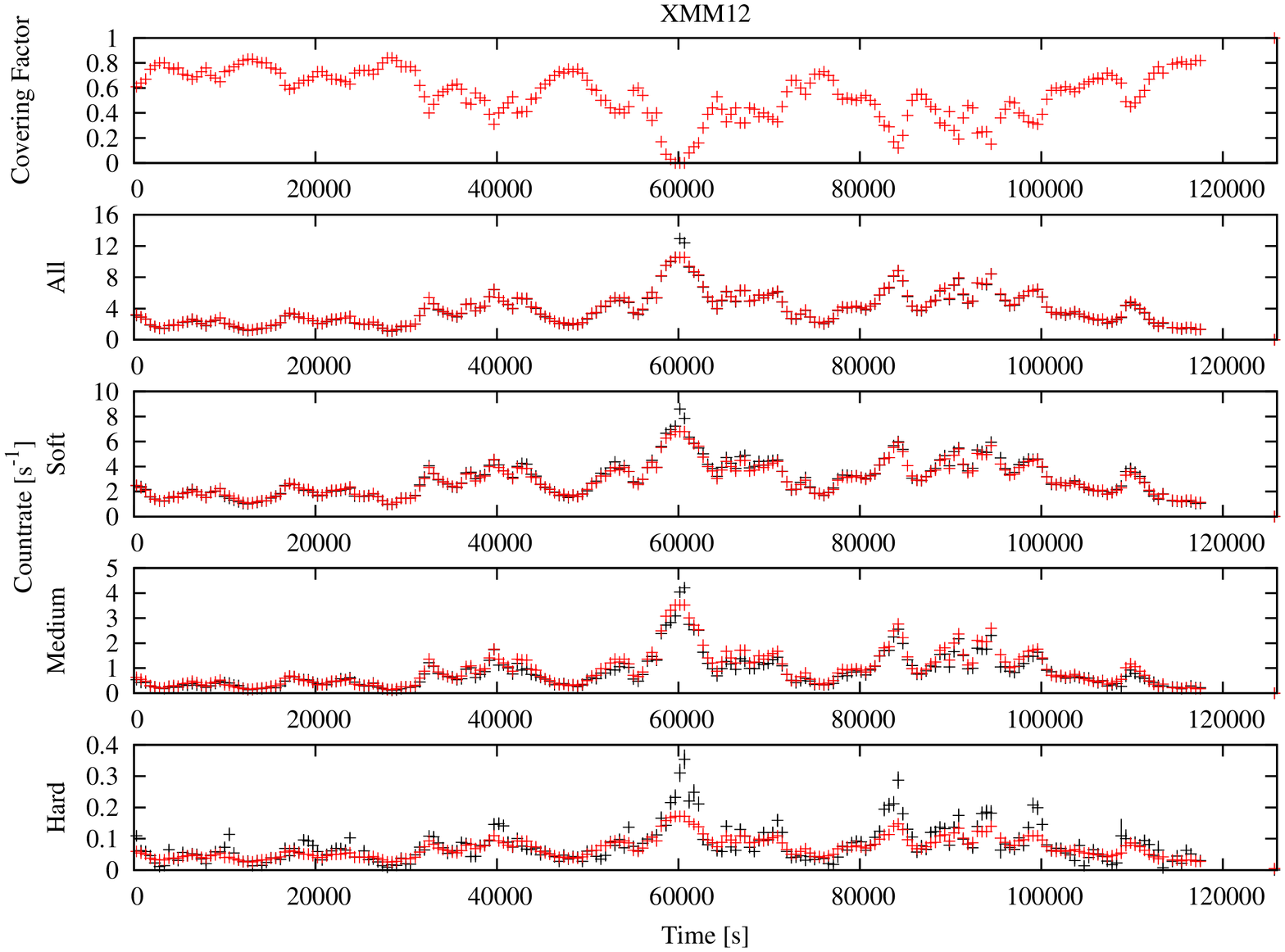}}
}
\caption{
        {\it Continued.}
}
\label{fig:modelltcrv:f}
\end{figure*}
\addtocounter{figure}{-1}
\begin{figure*}
\addtocounter{subfigure}{1}
\centering
\subfigure{
        \resizebox{15cm}{!}{\includegraphics[angle=270]{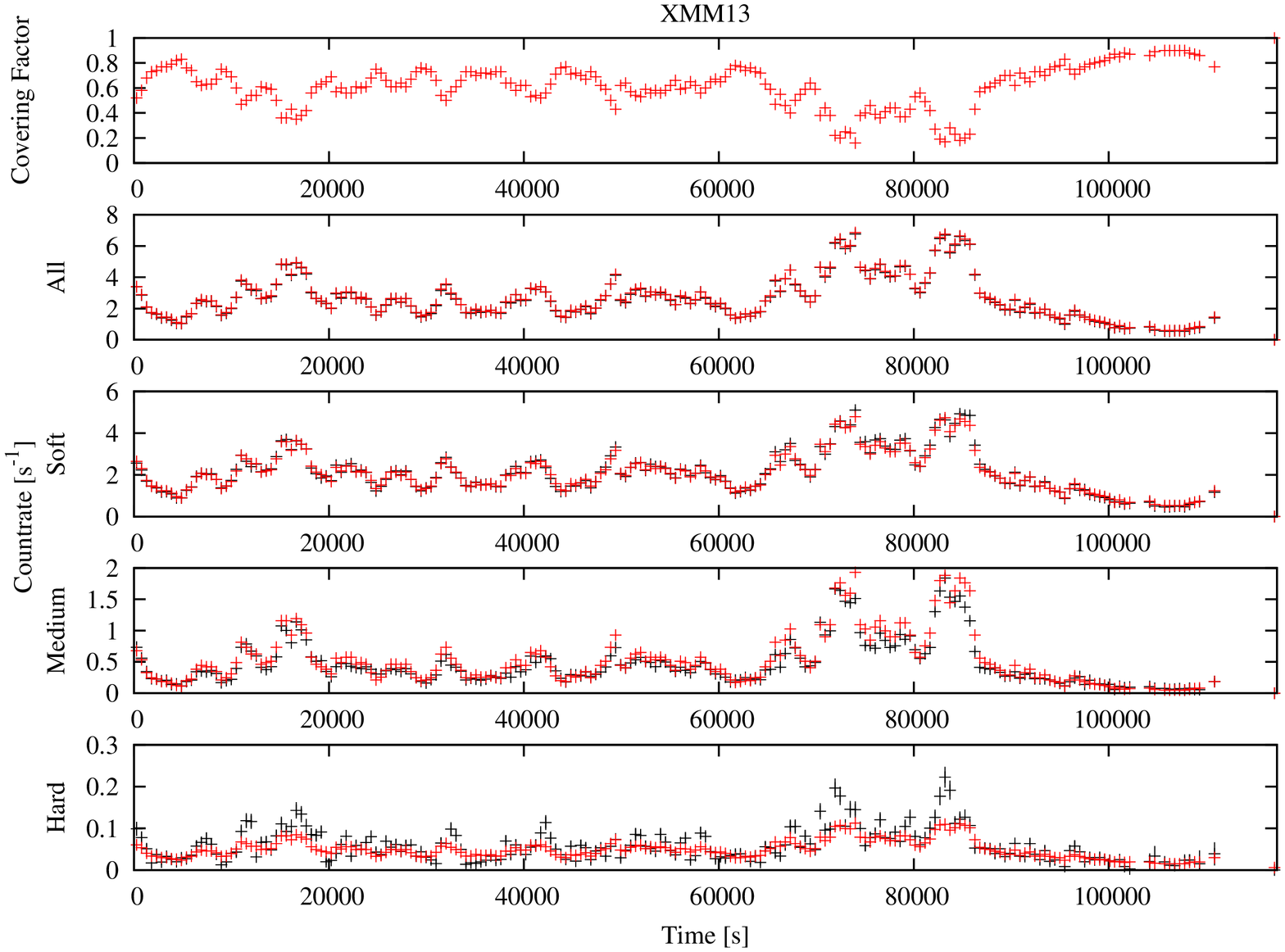}}
}
\subfigure{
        \resizebox{15cm}{!}{\includegraphics[angle=270]{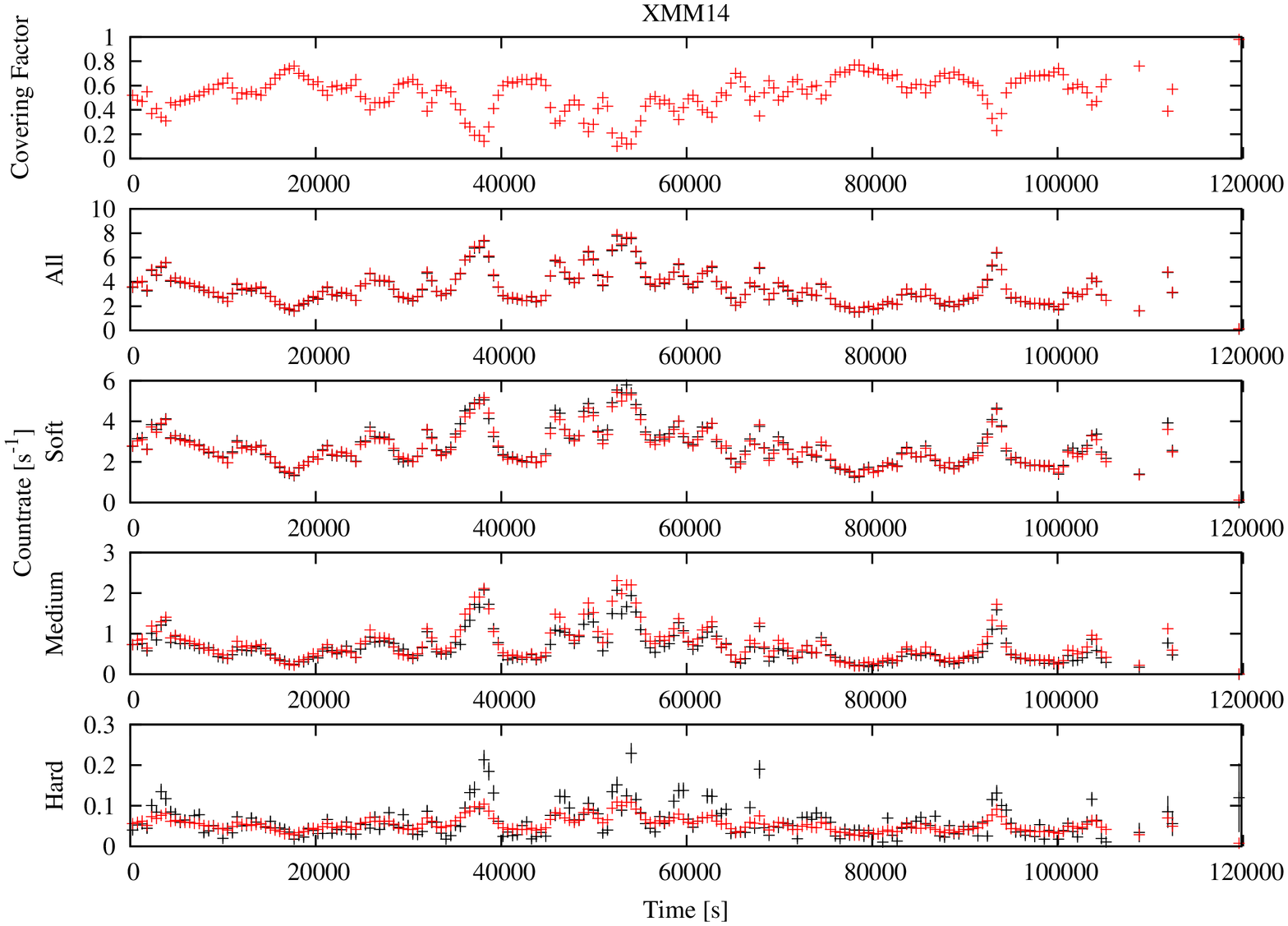}}
}
\caption{
        {\it Continued.}
}
\label{fig:modelltcrv:g}
\end{figure*}
\addtocounter{figure}{-1}
\begin{figure*}
\addtocounter{subfigure}{1}
\centering
\subfigure{
        \resizebox{15cm}{!}{\includegraphics[angle=270]{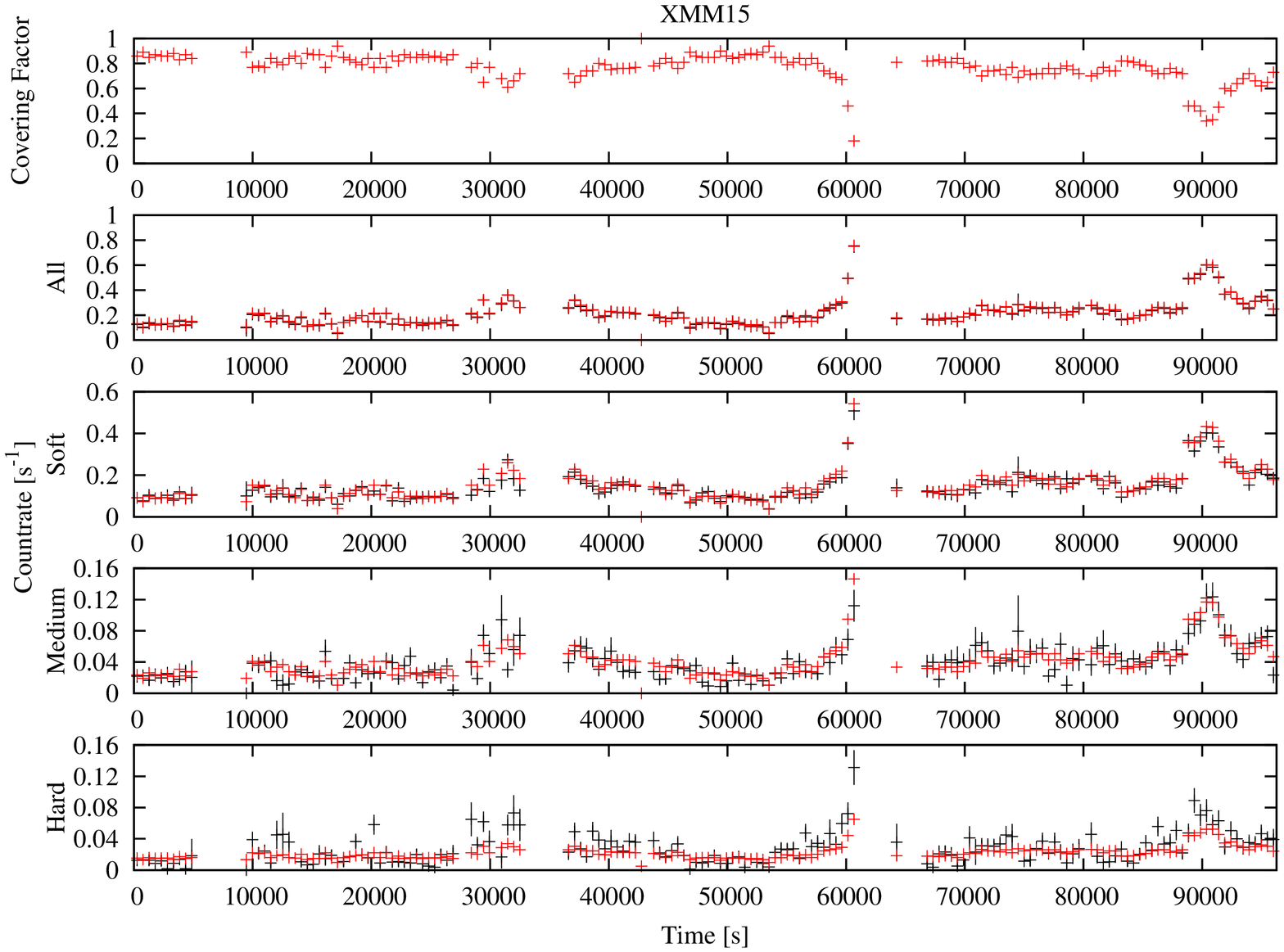}}
}
\subfigure{
        \resizebox{15cm}{!}{\includegraphics[angle=270]{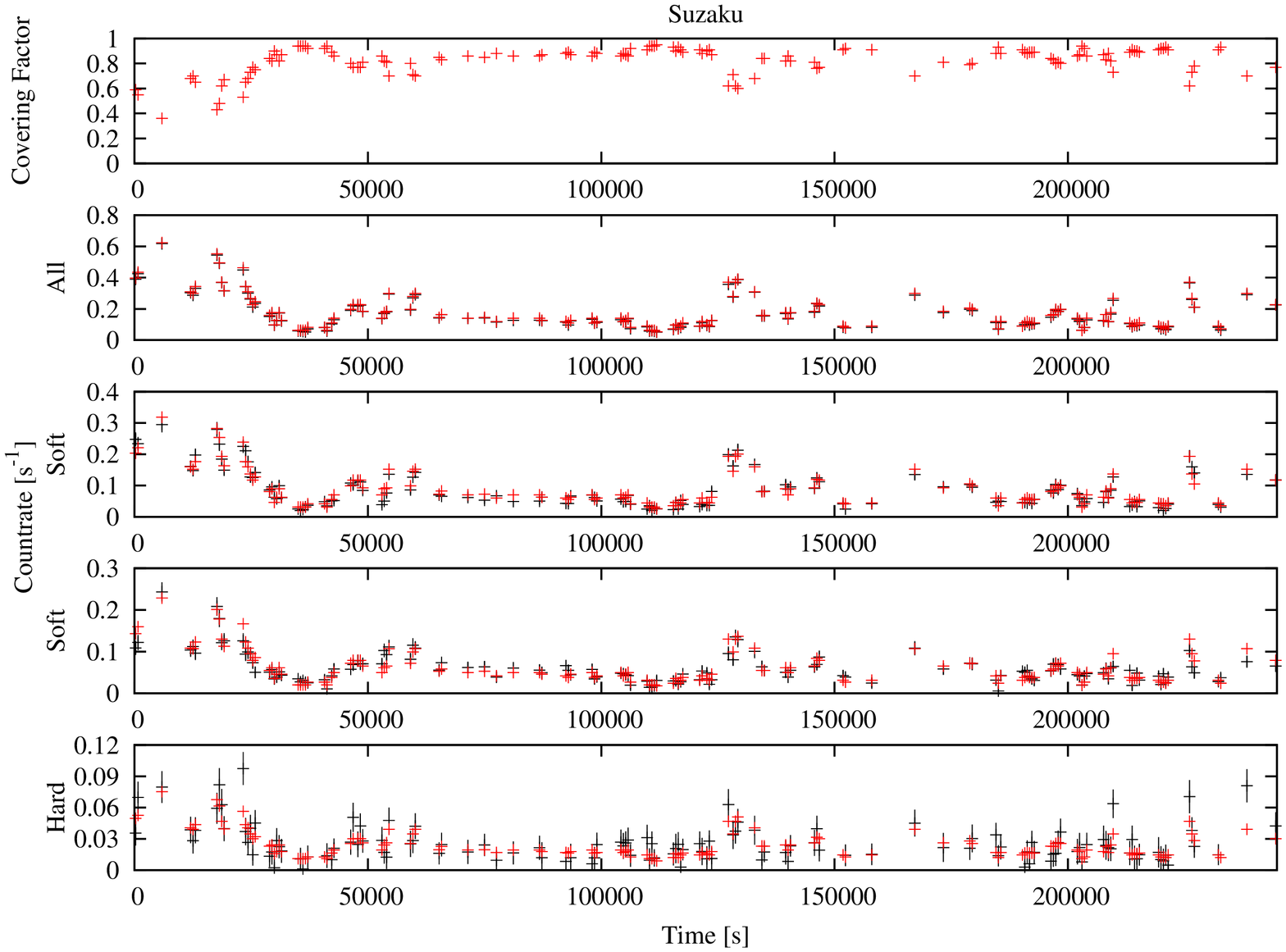}}
}
\caption{
        {\it Continued.}
}
\label{fig:modelltcrv:h}
\end{figure*}


\begin{figure*}
\centering
\subfigure{
        \resizebox{15cm}{!}{\includegraphics[angle=270]{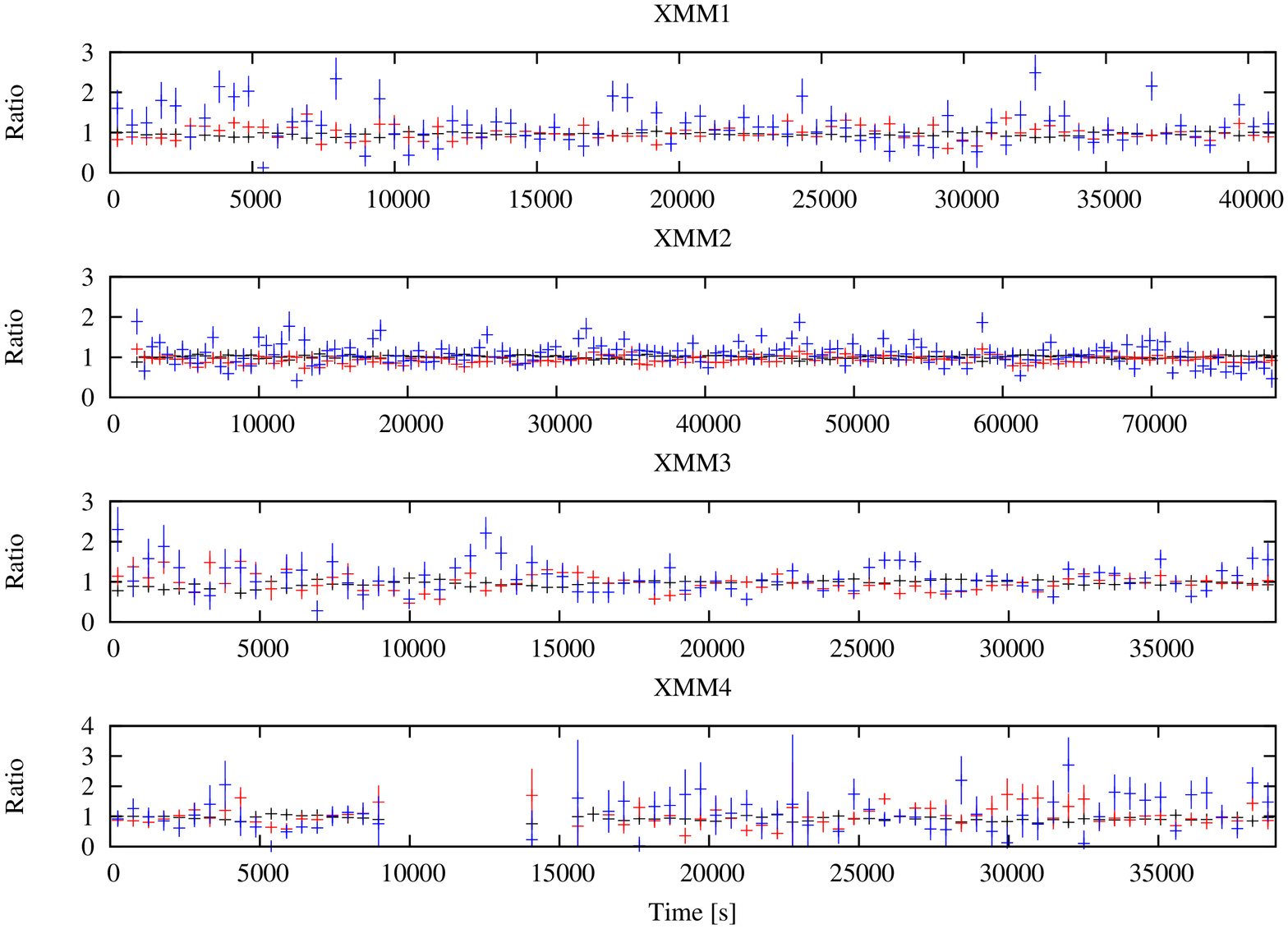}}
}
\subfigure{
        \resizebox{15cm}{!}{\includegraphics[angle=270]{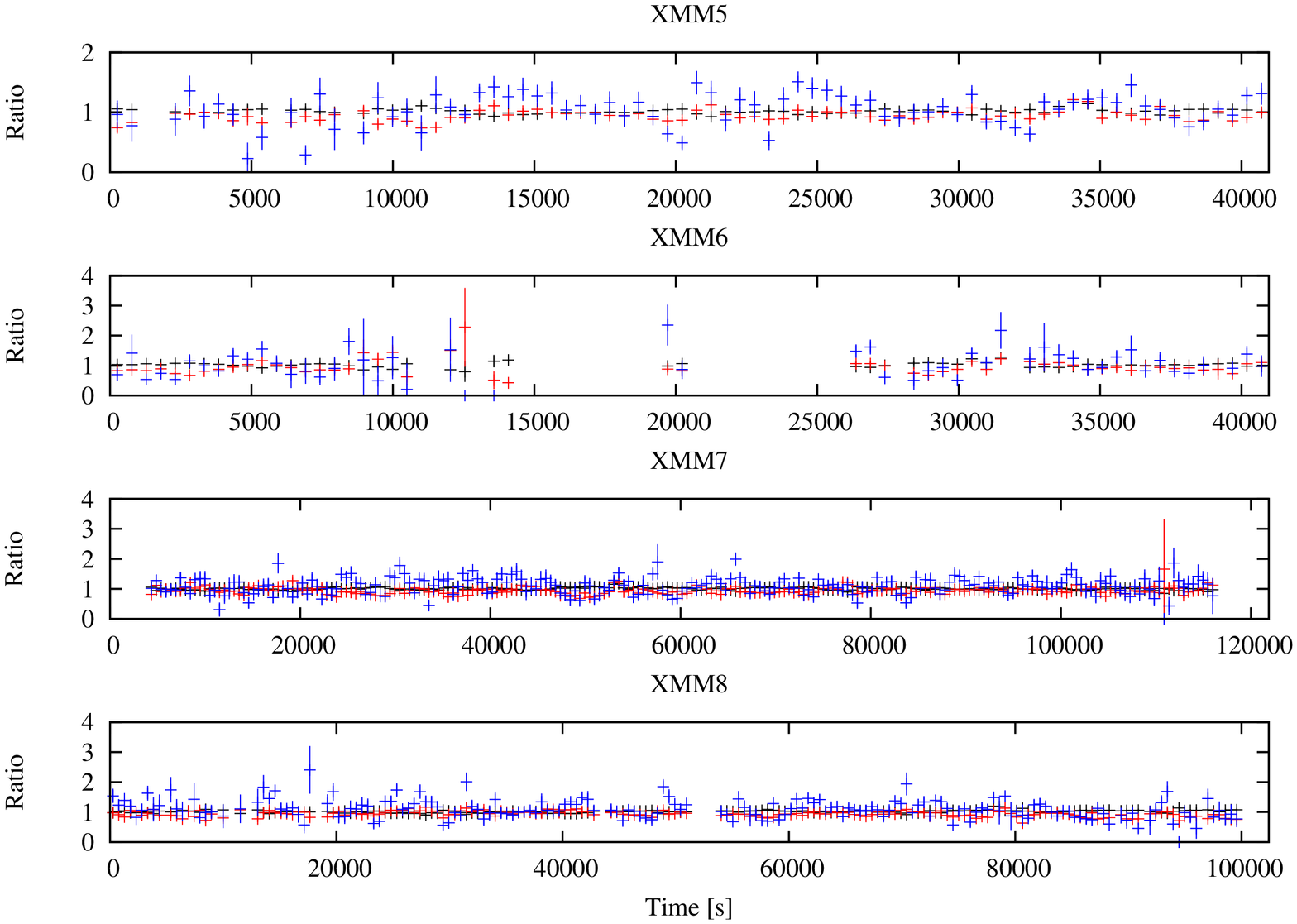}}
}
\caption{Ratios of the observed light-curves to the simulated ones.
The black, red, and blue points show the soft, medium, and hard band, respectively.
 }
\label{fig:ratio}
\end{figure*}

\addtocounter{figure}{-1}
\begin{figure*}
\addtocounter{subfigure}{1}
\centering
\subfigure{
        \resizebox{15cm}{!}{\includegraphics[angle=270]{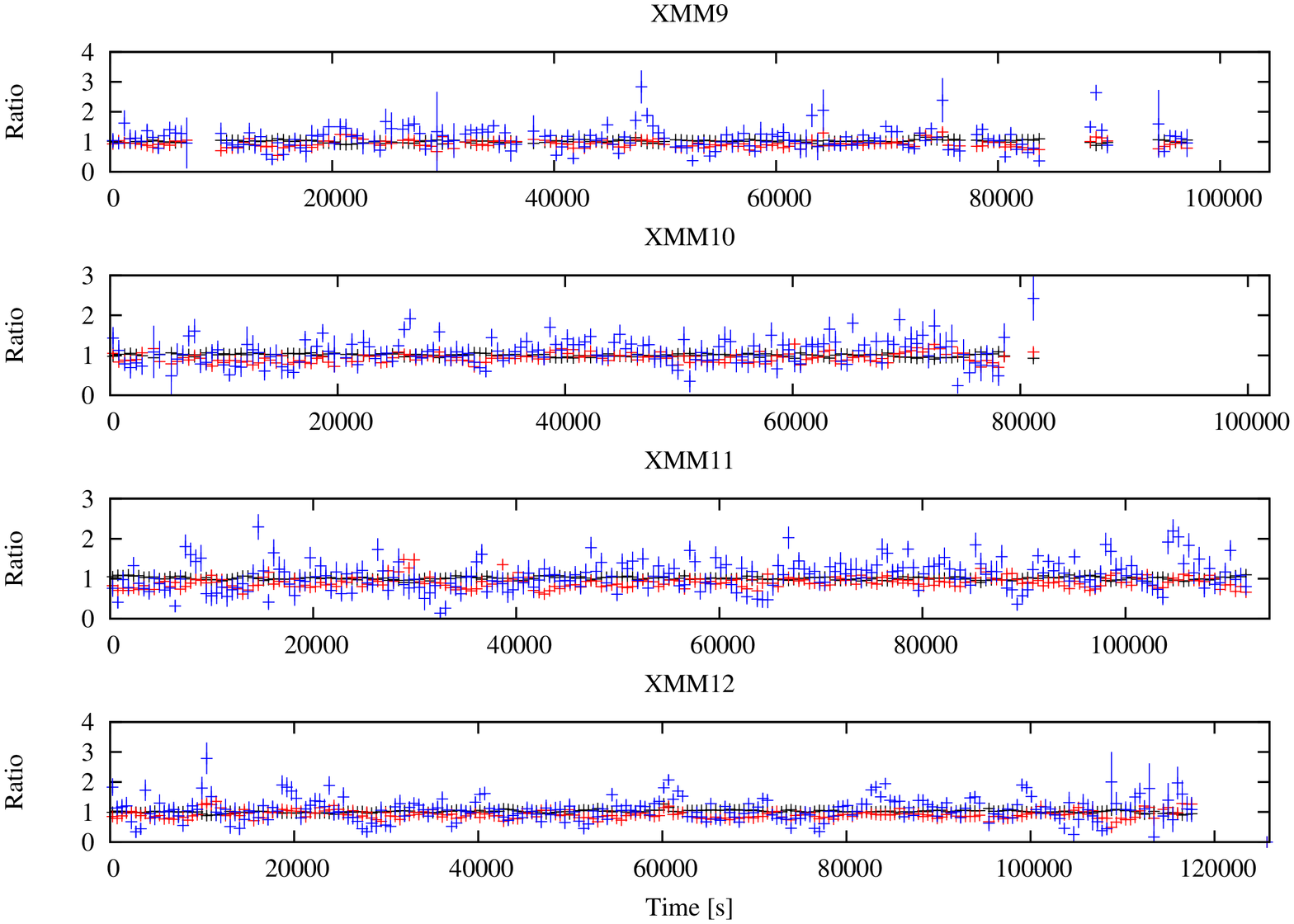}}
}
\subfigure{
        \resizebox{15cm}{!}{\includegraphics[angle=270]{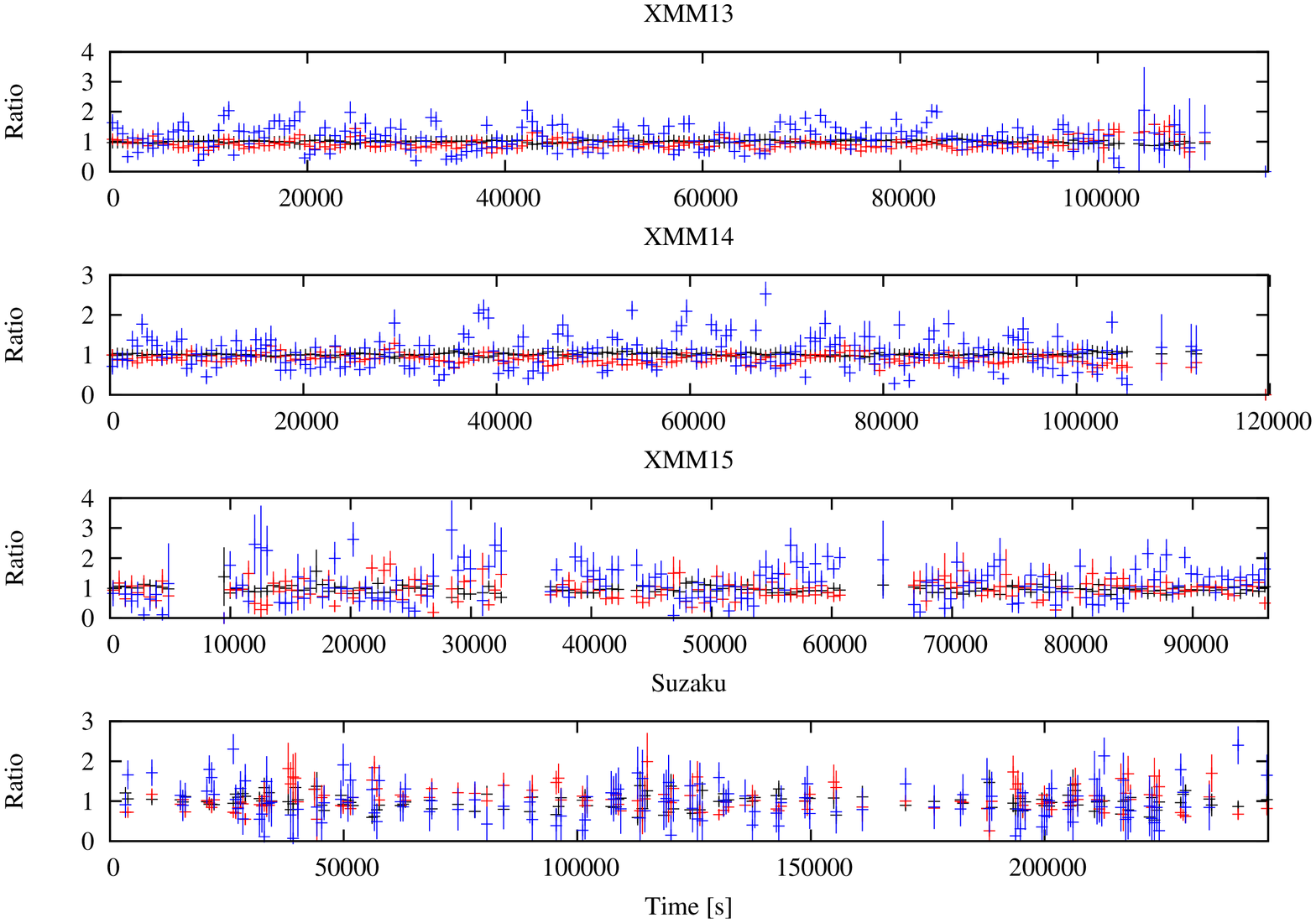}}
}
\caption{
        {\it Continued.}
}
\label{fig:ratio:b}
\end{figure*}

\begin{figure}
 \begin{center}
  \includegraphics[width=10cm,angle=270]{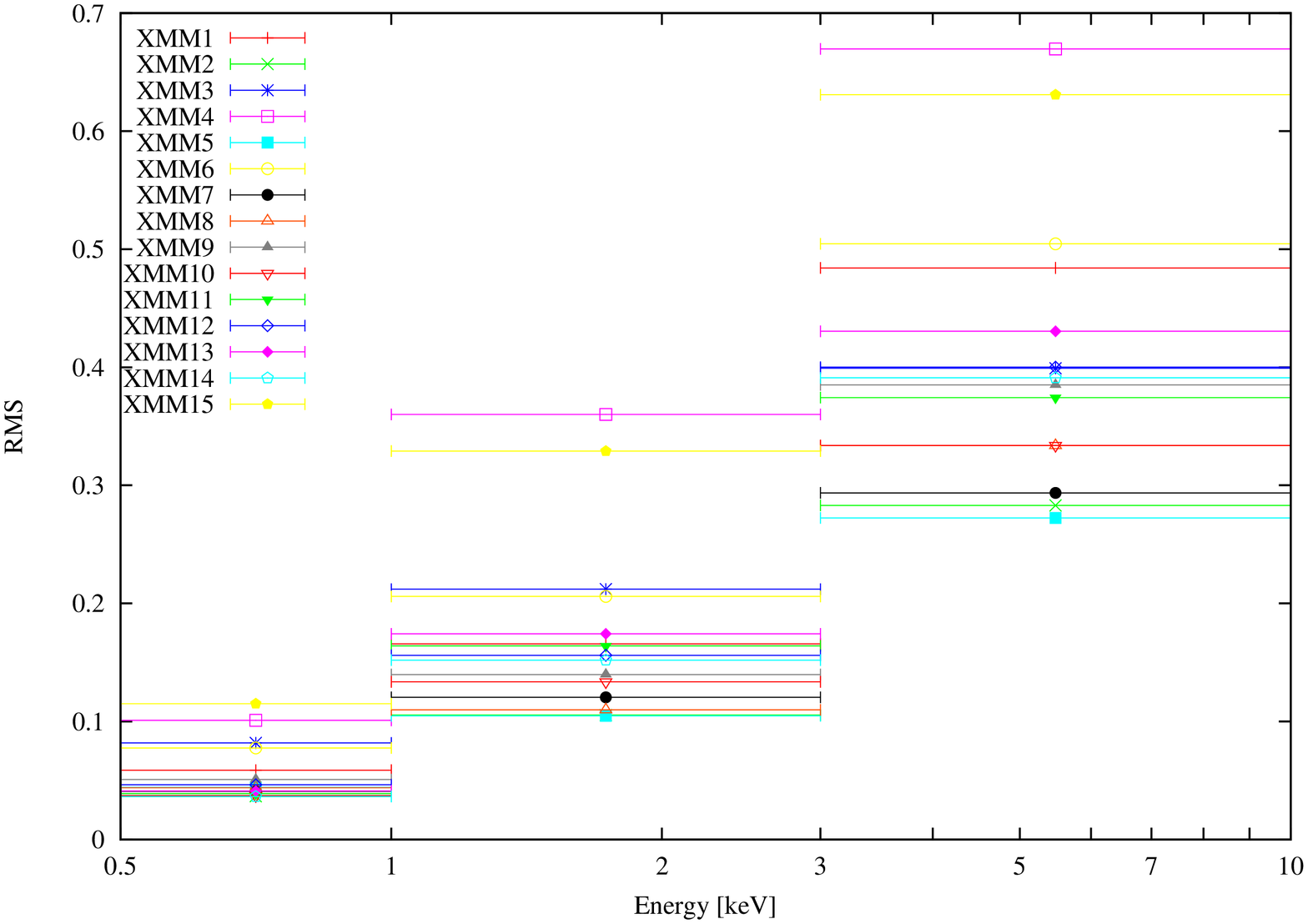}
 \end{center}
\caption{Root-mean-squares of the residual variations 
between the observed light-curves (XMM) and the model ones.
Suzaku data is not shown here because we cannot simply compare 
due to the difference of the effective area.}
\label{fig:rms}
\end{figure}


\clearpage

\appendix
\section*{Non-identity of the time-average of the variable partial covering model spectra and the model spectrum with the average partial covering fraction}

In the VDPC model we adopt in the present paper,
the spectral shape is expressed with the ``double partial covering'' as Equation (\ref{VPC}),
and the spectral variation is explained by only variation of the partial covering fraction $\alpha$.
In this appendix, we explain that 
the time-average of the VDPC model spectra where the partial covering fraction is variable 
and the VPDC model spectrum with the average partial covering fraction 
are not mathematically identical.

In Equation (\ref{VPC}), since the covering factor $\alpha$ is the only time-variable,
 the spectrum $F_i$ at a given time $t_i$ is written as 
\begin{eqnarray}
F_i&=&A_I\, N\, (1-\alpha_i + \alpha_i W_n \,e^{-\tau_1}) (1-\alpha_i + \alpha_i\, W_k \,e^{-\tau_2})(P+B)\nonumber \\
&=& A_I\,N\,\left((1\mathalpha{-}\alpha_i )^2 + (1\mathalpha{-}\alpha_i )\alpha_i  (W_n e^{-\tau_1}+W_k e^{-\tau_2})+\alpha_i^2\,W_n e^{-\tau_1}W_k e^{-\tau_2} \right)(P\mathalpha{+}B). \label{A1}
\end{eqnarray}
Thus the time-average spectrum $F$ is expressed as
\begin{eqnarray}
F&=&\left(\sum_i^n F_i\right) /n \nonumber   \\
&=& A_I\, N\left(\overline{(1-\alpha)^2} + \overline{(1-\alpha)\alpha} (W_n \,e^{-\tau_1}+W_k \,e^{-\tau_2})+\overline{\alpha^2}W_n \,e^{-\tau_1}W_k \,e^{-\tau_2} \right)(P+B)\label{alpha}
\end{eqnarray}
where $\overline{K} \equiv (\sum_i^n K_i)/n$.
On the other hand,
assuming that the time-average spectrum $F$ is expressed by the same double partial covering model with the covering factor $\beta$,
\begin{eqnarray}
F&=&A_I\, N\left((1-\beta)^2 + (1-\beta)\beta  (W_n \,e^{-\tau_1}+W_k \,e^{-\tau_2})+\beta^2W_n \,e^{-\tau_1}W_k \,e^{-\tau_2} \right)(P+B)  \label{beta}
\end{eqnarray}
should hold.

So that Equation (\ref{alpha}) and (\ref{beta}) agree, we should have
\begin{equation}
\left\{
\begin{array}{rll}
\overline{(1-\alpha)^2}&=&(1-\beta)^2 \\
\overline{(1-\alpha)\alpha}&=&(1-\beta)\beta \\
\overline{\alpha^2}&=&\beta^2.
\end{array}
\right.
\end{equation}
By putting $V(\{\alpha_i\})$ as the variance of $\{\alpha_i\}$,
we have
\begin{equation}
\overline{\alpha^2}=\overline{\alpha}^2+V(\{\alpha_i\})
\end{equation}
and 
\begin{equation}
\left\{
\begin{array}{rll}
1-2\overline{\alpha}+\overline{\alpha}^2+V(\{\alpha_i\})&=&1-2\beta+\beta^2 \\
\overline{\alpha}-\overline{\alpha}^2-V(\{\alpha_i\})&=&\beta-\beta^2  \\
\overline{\alpha}^2+V(\{\alpha_i\})&=&\beta^2.
\end{array}
\right.
\end{equation}
These three equations are only fulfilled when
\begin{eqnarray}
\overline{\alpha}=\beta \\
V(\{\alpha_i\})=0. \label{zero}
\end{eqnarray}
Equation (\ref{zero}) indicates that the time-averaged spectrum is identical with the VDPC model
only when $\alpha_i=\mathrm{const.}$ i.e.~there is no time variation.
This also means that some residual structures should appear  
when we attempt to fit the time-average of the VDPC model spectra with a VDPC model spectrum with the average partial covering fraction.
This is due to non-linearity of the double partial covering model with $\alpha$ (Equation \ref{A1}).
On the other hand, 
for example, a single-layer partial covering model given as
\begin{equation}
F=A_I((1-\alpha)+\alpha W)(P+B)
\end{equation}
does not make such residuals 
because the model is linear with $\alpha$; average of the time-variable single partial covering model spectra 
is mathematically identical to the same model with the covering factor $\beta=\overline{\alpha}$.
We also note that the residuals are not made
when the two covering fractions in a double-layer partial covering model are independent,
or the covariance is zero.

\begin{figure}[htbp]
 \centering
 \includegraphics[width=6cm,angle=270,clip]{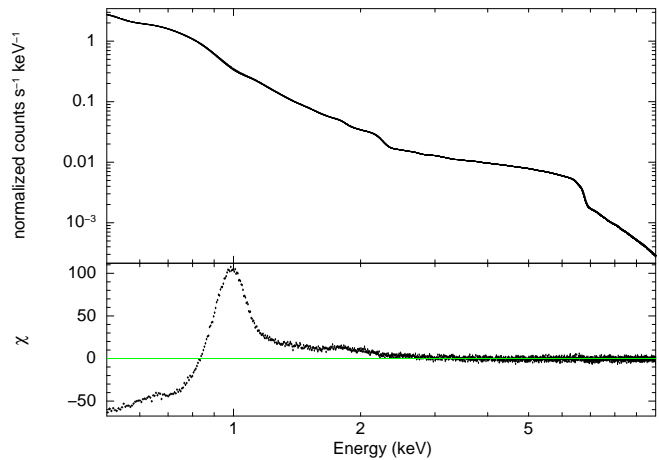}
 \caption{The simulated time-averaged spectrum and the fitted model.}
 \label{pic} 
\end{figure}

To demonstrate this effect, we carried out spectral simulations 
assuming the best-fit VDPC model of the data set XMM1 (Table 3).
We create the light-curve with a bin-width of 128 sec, 
where the best-fit covering fraction is determined for each light-curve bin.  
We create a simulated spectrum every 128 sec, and we integrated the simulated spectra.  
We fit the thus created time-averaged simulated spectrum with the VDPC model. 
For emphasizing the structure of simulated spectrum,
we take simulated exposure time of each light-curve bin 100 Msec.
The covering factors have $\overline{\alpha}=0.644$, $\overline{\alpha^2}=0.440$, and $V(\{\alpha_i\})=0.025$.
Figure \ref{pic} shows the simulated time-averaged spectrum and the best-fit VDPC model with the covering factor $\beta=0.642$.
A significant residual structure is seen below 2 keV.

Some residuals around 1 keV have been reported in the energy spectrum of 1H0707--495,
which were suggested as a blend of resonance absorption line of ionized iron L-shell and 
emission line arising from an extended warm medium \citep{gal04}.
Similar features have been also reported from IRAS 13224--3809 \citep{bol03} and other NLS1s \citep{nic99}.
Here, we suggest that these apparent residuals may be due to non-linearity of the model,
such that a time-averaged spectrum of the VDPC model is not identical to the
double partial covering model with the time-averaged partial covering fraction.

\begin{thebibliography}{}
\bibitem[Arnaud(1996)]{arn96} 
Arnaud, K.~A.\ 1996, Astronomical Data Analysis Software and Systems V, 101, 17 
\bibitem[Blustin \& Fabian(2009)]{blu09} 
Blustin, A.~J., \& Fabian, A.~C.\ 2009, \mnras, 399, L169 
\bibitem[Boller et al.(2002)]{bol02} 
Boller, T., et al.\ 2002, \mnras, 329, L1 
\bibitem[Boller et al.(2003)]{bol03} 
Boller, T., Tanaka, Y., Fabian, A., Brandt, W.~N., Gallo, L., Anabuki, N., Haba, Y., \& Vaughan, S.\ 2003, \mnras, 343, L89 
\bibitem[Boroson(2002)]{bor02} 
Boroson, T.~A.\ 2002, \apj, 565, 78
\bibitem[Brinkmann et al.(2007)]{bri07}
  Brinkmann, W., Papadakis, I.~E., \& Raeth, C.\ 2007, \aap, 465, 107 
\bibitem[Dauser et al.(2012)]{dau12} 
Dauser, T., et al.\ 2012, \mnras, 422, 1914 
\bibitem[den Herder et al.(2001)]{den01} 
den Herder, J.~W., et al.\ 2001, \aap, 365, L7 
\bibitem[Done et al.(2007)]{don07} 
Done, C., Sobolewska, M.~A., Gierli{\'n}ski, M., \& Schurch, N.~J.\ 2007, \mnras, 374, L15 
\bibitem[Fabian et al.(2002)]{fab02} 
Fabian, A.~C., Ballantyne, D.~R., Merloni, A., Vaughan, S., Iwasawa, K., \& Boller,~T.\ 2002, \mnras, 331, L35 
\bibitem[Fabian et al.(2004)]{fab04} 
Fabian, A.~C., Miniutti,~G., Gallo, L., Boller, T., Tanaka, Y., Vaughan, S., \& Ross, R.~R.\ 2004, \mnras, 353, 1071 
\bibitem[Fabian et al.(2009)]{fab09} 
Fabian, A.~C., et al.\ 2009, \nat, 459, 540 
\bibitem[Fabian et al.(2012)]{fab12} 
Fabian, A.~C., Zoghbi, A., Wilkins, D., et al.\ 2012, \mnras, 419, 116 
\bibitem[Gallo et al.(2004)]{gal04} 
Gallo, L.~C., Tanaka,~Y., Boller,~T., Fabian,~A.~C., Vaughan,~S., \& Brandt,~W.~N.\ 2004, \mnras, 353, 1064 
\bibitem[Grupe(2004)]{gru04} 
Grupe, D.\ 2004, \aj, 127, 1799 
\bibitem[Halpern \& Oke(1987)]{hal87} 
Halpern, J.~P., \& Oke, J.~B.\ 1987, \apj, 312, 91 
\bibitem[Ishisaki et al.(2007)]{ish07} 
Ishisaki, Y., et al.\ 2007, \pasj, 59, 113 
\bibitem[Jansen et al.(2001)]{jan01}
Jansen, F., et al.\ 2001, \aap, 365, L1
\bibitem[Kalberla et al.(2005)]{kal05} 
Kalberla, P.~M.~W., Burton, W.~B., Hartmann, D., et al.\ 2005, \aap, 440, 775  
\bibitem[Kallman et al.(2004)]{kal04} 
Kallman, T.~R., Palmeri, P., Bautista, M.~A., Mendoza, C., \& Krolik, J.~H.\ 2004, \apjs, 155, 675 
\bibitem[Koyama et al.(2007)]{koy07} 
Koyama, K., et al.\ 2007, \pasj, 59, 23 
\bibitem[Leighly(1999)]{lei99} 
Leighly, K.~M.\ 1999, \apjs, 125, 297 
\bibitem[Leighly(2004)]{lei04} 
Leighly, K.~M.\ 2004, Progress of Theoretical Physics Supplement, 155, 223
\bibitem[Longinotti et al.(2013)]{lon13} Longinotti, A.~L., 
Krongold, Y., Kriss, G.~A., et al.\ 2013, \apj, 766, 104 
\bibitem[Mathur(2000)]{mat00} 
Mathur, S.\ 2000, \mnras, 314, L17
\bibitem[McHardy et al.(2005)]{mch05}
McHardy, I.~M., Gunn,~K.~F., Uttley, P., \& Goad, M.~R.\ 2005, \mnras, 359, 1469 
\bibitem[Miller et al.(2010)]{mil10} 
Miller, L., Turner, T.~J., Reeves, J.~N., \& Braito, V.\ 2010, \mnras, 408, 1928 
\bibitem[Mineshige et al.(2000)]{min00} 
Mineshige, S., Kawaguchi, T., Takeuchi, M., \& Hayashida, K.\ 2000, \pasj, 52, 499 
\bibitem[Mitsuda et al.(2007)]{mit07} 
Mitsuda, K., et al.\ 2007, \pasj, 59, 1 
\bibitem[Miyakawa et al.(2009)]{miy09} 
Miyakawa,~T., Ebisawa,~K., Terashima,~Y., Tsuchihashi,~F., Inoue,~H., \& Zychi,~P.\ 2009, \pasj, 61, 1355
\bibitem[Miyakawa et al.(2012)]{miy12} 
Miyakawa,~T., Ebisawa,~K., \& Inoue,~H.\ 2012, \pasj, 64, 140
\bibitem[Nicastro et al.(1999)]{nic99} 
Nicastro, F., Fiore, F., \& Matt, G.\ 1999, \apj, 517, 108 
\bibitem[Nomura et al.(2013)]{nom13} 
Nomura,~M., Ohsuga,~K., Wada,~K., Susa,~H., \& Misawa,~T.\ 2013, \pasj, 65, 40 
\bibitem[Osterbrock \& Pogge(1985)]{ost85} 
Osterbrock, D.~E., \& Pogge, R.~W.\ 1985, \apj, 297, 166 
\bibitem[Proga et al.(2000)]{pro00}
Proga, D., Stone, J.~M., \& kallman, T.~R.\ 2000, \apj, 543, 686
\bibitem[Proga \& Kallman(2004)]{pro04} 
Proga, D., \& Kallman, T.~R.\ 2004, \apj, 616, 688
\bibitem[Puchnarewicz et al.(1992)]{puc92} Puchnarewicz, 
E.~M., et al.\ 1992, \mnras, 256, 589 
\bibitem[Stevens \& Kallman(1990)]{ste90} 
Stevens, I.~R., \& Kallman, T.~R.\ 1990, \apj, 365, 321 
\bibitem[Str{\"u}der et al.(2001)]{str01} 
Str{\"u}der, L., et al.\ 2001, \aap, 365, L18  
\bibitem[Sulentic et al.(2000)]{sul00} 
Sulentic, J.~W., Marziani, P., \& Dultzin-Hacyan, D.\ 2000, \araa, 38, 521 
\bibitem[Tanaka et al.(2004)]{tan04}
Tanaka,~Y., Boller,~T., Gallo,~L., \& Keil,~R.\ 2004, \pasj, 56, L9
\bibitem[Turner(2014)]{tur14}
Turner, T.\ 2014, The X-ray Universe 2014, ed. Jan-Uwe Ness, 198
\bibitem[Ulrich et al.(1997)]{ulr97} 
Ulrich, M.-H., Maraschi, L., \& Urry, C.~M.\ 1997, \araa, 35, 445
\bibitem[Zhou \& Wang(2005)]{zho05} Zhou, X.-L., \& Wang, J.-M.\ 2005, \apjl, 618, L83 
\bibitem[Zoghbi et al.(2010)]{zog10} 
Zoghbi, A., Fabian, A.~C., Uttley, P., Miniutti, G., Gallo, L.~C., Reynolds, C.~S., Miller, J.~M., \& Ponti, G.\ 2010, \mnras, 401, 2419 
\end{thebibliography}
\end{document}